*REVIEWS OF TOPICAL PROBLEMS*

# Local geometry of the Fermi surface and its effect on the electronic characteristics of normal metals

N A Zimbovskaya

## Contents

**Abstract. Fine features in the Fermi surface geometry, such as nearly cylindrical or nearly paraboloidal strips or local flattenings, are examined as regards their effect on the electronic, mainly galvanomagnetic, properties of metals. It is shown that under certain conditions, these features may significantly change the way a conventional normal metal or a layered structure with metallic conductivity responds to high-frequency external disturbances. All of the effects considered appear to be very sensitive to the disturbance propagation direction and/or to that of the external magnetic field. Experimental possibilities of observing the described effects are discussed.**

## 1. Introduction

Understanding the behavior of conduction electrons in conventional metals and other substances with a metallic type conductivity (including low-dimensional conductors) is very important because, among other things, the conduction electrons give rise to superconductivity and magnetism. Being an interacting many-particle system, conduction electrons are also exposed to crystalline fields, which crucially affect their behavior. The effects of these crystalline fields on electrons are summarized in the concept of the Fermi surface (FS), which is one of the most meaningful concepts in condensed matter physics providing excellent insight into the nature of the main physical properties of metals.

Active studies aimed at reconstructing the geometry of FSs of conventional metals started in the 1950s. These studies were based on experimental data concerning various phenomena responsive to the structure of electronic spectra in metals, combined with electron band structure computations. As a result of these efforts, the main geometric characteristics of conventional metal FSs, such as their connectivities, locations of open orbits, sizes, and arrangements of sheets, were studied rather well by the mid-1970s [1–5]. Further development of computational methods for electron structure calculations resulted in the impressive mapping of FSs of metals. Later, studies of FS geometries were revived due to the emergence of new conducting materials with remarkable properties such as high-temperature superconductivity. These materials include organic conductors belonging to the family of tetrathiafulvalene salts, dichalcogenides of transition metals, and graphite and its intercalates. As a whole, the above materials are characterized by the metallic type conductivity in the normal (nonsuperconducting) state and

**N A Zimbovskaya** Department of Physics and Electronics, University of Puerto Rico-Humacao, CUH Station, Humacao, Puerto Rico 00791, USA
Institute for Functional Nanomaterials, University of Puerto Rico, San Juan, Puerto Rico 00931, USA
E-mail: natalya.zimbovskaya@upr.edu

by the quasi-two-dimensional (Q2D) spectra of charge carriers. Significant effort has been and still is applied to the study of electron characteristics of these Q2D metals, including the geometries of their FSs [6–8]. Another extremely interesting group of novel materials are high-temperature superconducting cuprates. In addition to their transition temperatures, anomalies in the normal state transport properties of cuprates [9] are often referred to as one of the most formidable challenges in condensed matter physics [10]. In the first two decades of intensive research on cuprate (in particular, doped cuprate) properties, no truly convincing evidence of the existence of fermionic quasiparticles around coherent FSs was obtained. The failure to unambiguously observe such quasiparticles led to the theoretical search for novel strongly correlated electronic ground states appropriate for describing the unusual transport properties of doped cuprates, and this search continues up to the present. But in 2003, evidence of FS existence was revealed from the analysis of angle-dependent magnetoresistance data for overdoped $Tl_2Ba_2CuO_{6+\sigma}$ [11]. Another important advancement in cuprate fermiology was made by observation of quantum oscillations in the magnetoresistance of underdoped $YBa_2Cu_3O_{6+\sigma}$ [12]. This breakthrough was quickly followed by reports of similar oscillations observed in other cuprates, namely, underdoped $YBa_2Cu_4O_8$ [13, 14], overdoped $Tl_2Ba_2CuO_{6+\sigma}$ [15, 16], and the electron-doped compound $Nd_{2-x}Ce_xCuO_4$ [17]. Recently, de Haas–van Alphen oscillations were observed in overdoped $Tl_2Ba_2CuO_{6+\sigma}$ [10]. Unlike early observations of quantum oscillations in low-quality cuprate samples [18, 19], these recent experiments show a high signal-to-noise level, which provides better opportunities for interpretation of the experimental data. Another method successfully used to study the electronic characteristics of cuprates is angle-resolved photoemission spectroscopy [20]. Currently available experimental data give grounds to believe that charge carriers in doped cuprates in the normal (nonsuperconducting) state behave as fermionic quasiparticles associated with FSs. As the doping is reduced towards the insulating phase, these FSs reveal the tendency to break up into pieces, whose projections on the relevant planes in the quasimomenta space look like arcs. Also, it was suggested in [21] that cuprate FSs break apart at temperatures just above the critical temperature when these materials are in a special 'pseudogap' phase, which differs from the normal metallic-like or insulator state. Recent recognition of some layered iron-based compounds as high-temperature superconductors [22–24] spurred intensive research of their transport and magnetic properties, including intensive efforts aimed at reconstruction of FSs [25–30] using both experiments and electron band structure computations. This research continues up to the present.

Despite the very impressive achievements in FS mapping, comparatively little attention has been paid to fine features in their geometry related to some local characteristics of their curvature. These are points of flattening and nearly cylindrical strips, where the FS curvature vanishes. Also, points and lines can exist where the curvature is discontinuous, e.g., kink lines or conic points. These features can be easily missed in the electron structure calculations, including those that involve advanced computational methods.

When a FS incorporates points and/or lines of anomalous curvature, this leads to an enhancement/reduction in the contributions from the neighborhoods of these points/lines to the electron density of states (DOS) on the FS. Usually, these contributions are small compared to the main term in the DOS, which originates from the remaining major part of the FS. Therefore, they cannot produce noticeable changes in the response of the metal to an external perturbation, when all segments of the FS contribute to the response functions essentially equally. But at high frequencies of the external disturbance, the response is determined by 'efficient' charge carriers associated with small 'effective' segments of the FS. For instance, various kinds of magnetic oscillations, which are rightfully considered an important tool in experimental studies of the electronic properties of conductors with metallic-like conductivity [5], are formed by the response of charge carriers belonging to narrow strips or even vicinities of certain points on the FS. This last statement refers to commensurability oscillations (Pippard geometric resonances) that may appear when the conduction electrons are exposed to both uniform magnetic and nonuniform electric fields. The contribution to the DOS from 'efficient' charge carriers associated with the vicinities of the points and lines of zero curvature located at such segments can have the same, or even greater, magnitude as the contribution of the whole 'effective' segment. In other words, when the curvature vanishes at some part on an 'effective' segment of the FS, it can make a sensible change in the number of efficient electrons and, in consequence, in the response of the metal to the disturbance.

Some possible manifestations of local anomalies in the FS curvature were theoretically analyzed in earlier works [31–41]. An extraordinary attenuation rate for sound waves propagating in metals whose FSs incorporate flat or cylindrical segments was first predicted as early as 1973 [31]. Later, the effect of local flattenings and nearly cylindrical strips on the FSs on the frequency and angular dependences of sound attenuation and velocity shift were studied in more detail in [32]. Also, it was shown that such anomalies in the FS local geometry can give rise to noticeable changes in the conductivity and surface impedance of a metal under the anomalous skin effect [33–40]. A manifestation of a kink line, where the FS curvature has a discontinuity, was briefly discussed in [41].

The zero-curvature lines discussed above either indicate positions of nearly cylindrical strips on the FSs or are inflection lines separating concave parts of the FS from its convex segments. The presence of inflection lines can cause changes in the topology of the effective segments on the FS for some particular directions of propagation of the external disturbance [34]. As a result, noticeable changes in the response of a metal to the disturbance should occur. Some of the above fine features in the FS geometries are represented in Fig. 1.

The effects of FS curvature anomalies on the observables in metals and other metallic-like conductors were mostly studied in some particular cases, and only the simplest models for the FSs were used. The present work contains the principal results of advanced theoretical studies of possible manifestations of the FS local geometric structure in the experiments. The analysis is based on phenomenological models for the local shapes of the FSs. We do believe that adoption of phenomenological models is justified so far as these models are based on reasonable assumptions. Actually, phenomenological models were used to develop the theory of well-known effects, such as the de Haas–van Alphen effect. The suggested approach to fermiology does not contradict the standard approach based on electron band structure calcula-tions. On the contrary, it supplements the latter. Such supplementing analysis aims to bring new insight into the nature and origin of the considered physical effects. Also, our intention is to show the usefulness of these effects in further studies of FS geometries.

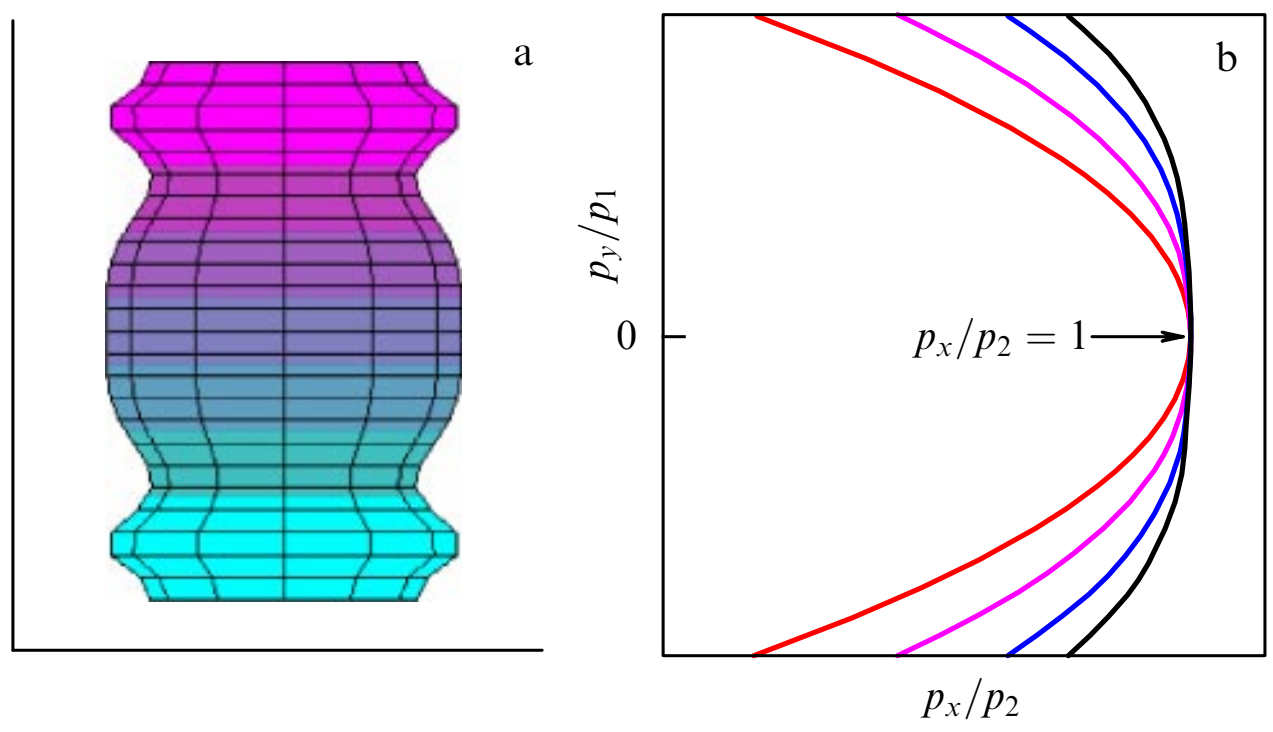

**Figure 1.** (a) Schematic plots of a FS with a complex profile including nearly cylindrical strips and kink lines and (b) profiles of the cross sections of an ellipsoid-like FS flattened at the points $(\pm p_2, 0, 0)$.

## 2. Basic concepts and equations

### 2.1 The effect of the Fermi surface curvature on the electron response of a metal

In studies of magnetotransport properties, a metal can be viewed as a Fermi sea of conduction electrons. These electrons interact with the ions of the crystalline lattice and with each other. The interactions can give rise to some radical changes in the ground state of the electron system and in the properties of charge-carrying states, and these changes can result in the metal transition from the normal to the superconducting state, the metal–insulator transition, and/or structural changes in the lattice. Putting aside these and other phase transitions, we can consider conduction electrons of a normal nonsuperconducting metal as a system of quasiparticles described by the Fermi–Dirac statistics and obeying the Pauli principle.

An equilibrium state of conduction electrons is described by the Fermi distribution function

$$f(E) = \left(\exp\left[\frac{E-\zeta(T)}{kT}\right]+1\right)^{-1}, \tag{1}$$

where $E$ is the quasiparticle energy, $\zeta$ is the chemical potential, $k$ is Boltzmann's constant, and $T$ is the temperature. In the low-temperature limit $T = 0$, conduction electrons fill all available states whose energies do not exceed the Fermi energy $E_F = \zeta(0)$. These states are densely packed in the interior of a certain surface in the quasimomentum space, which is called the Fermi surface. At $T = 0$, all quasiparticle states inside the FS are filled, whereas all states outside the FS are empty if the electron system is free from external disturbances and the interactions between quasiparticles are neglected.

It is especially easy to illustrate the concept of the Fermi surface using a model of an isotropic metal. Within this model, the crystalline potential is assumed to be uniformly spread over the region occupied by the metal. Then the FS takes a spherical shape. The radius $p_F$ of the Fermi sphere is determined by the relation

$$N = \frac{2V}{(2\pi\hbar)^3}\,\frac{4\pi}{3}\,p_F^3\,, \tag{2}$$

where $N/V$ is the quasiparticle density in the volume $V$ occupied by the metal.

The most essential characteristic of the FS local geometry is its Gaussian curvature

$$K(\mathbf{p}) = \pm\frac{1}{|R_1(\mathbf{p})R_2(\mathbf{p})|}\,, \tag{3}$$

where $R_{1,2}(\mathbf{p})$ are the principal radii of curvature.

At a certain point on the Fermi surface, its curvature is expressed in terms of the components of the velocity of electrons and their partial derivatives [40]. Any point on the FS where its curvature is finite can be specified as an elliptic, a hyperbolic, or a parabolic point depending on the curvature sign. At the elliptic points, we have $K(\mathbf{p}) > 0$, at the hyperbolic points, $K(\mathbf{p}) < 0$, and at the parabolic points, the Gaussian curvature of the surface becomes zero.

Points of zero curvature on a smooth surface are arranged into inflection lines, which separate 'hyperbolic' segments of the surface from its 'elliptic' segments. Such lines exist on the FSs of most metals. Some examples of inflection lines on surfaces that appear to be segments of the FSs of real metals are shown in the Fig. 2. Also, a zero-curvature line can belong to a segment of the FS where its Gaussian curvature form remains nonnegative. In such cases, the shape of the FS near the zero-curvature line is close to cylindrical.

The influence of zero-curvature lines on the FS of observables in metals originates from the enhancement of the electron density of states in the vicinities of such lines. The DOS on the Fermi surface is determined by the formula

$$N_\zeta = \frac{2V}{(2\pi\hbar)^3}\int\frac{\mathrm{d}A}{v}\,. \tag{4}$$

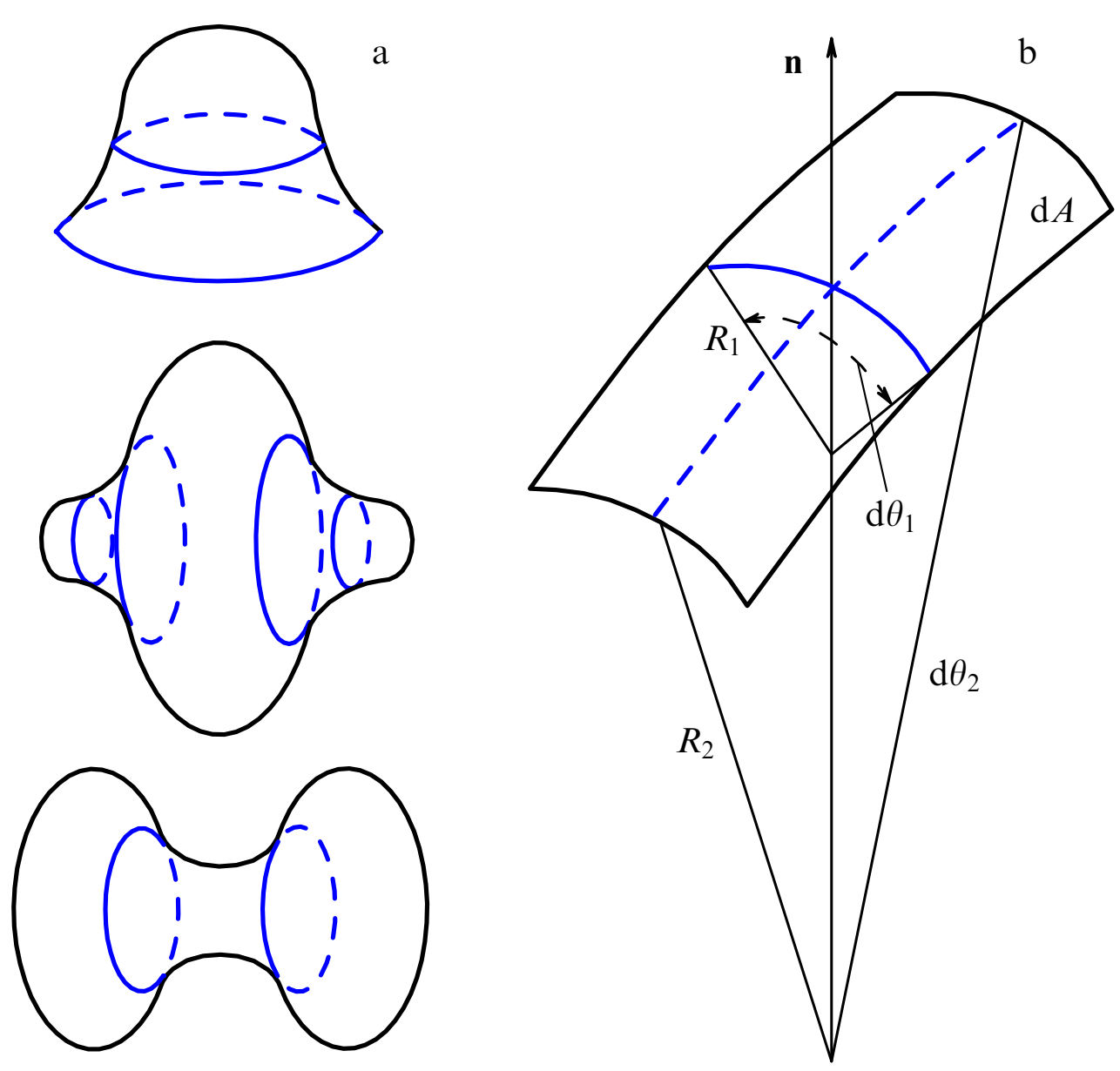

**Figure 2.** (a) Examples of zero-curvature lines on FSs of metals. (b) The surface element of a FS. The principal curvature radii of the shown element are $R_1$ and $R_2$; **n** is a normal vector to the surface.

Here, the integration is performed over the FS, $\mathrm{d}A$ is the surface area element, and $v$ is the magnitude of the electron velocity. We can transform the integral over the FS to the integral over the angles that determine the direction of the electron velocity vector $\mathbf{v}$ (that is, the direction of the normal to the surface). The surface element is shown in Fig. 2. The corresponding element of the surface area $\mathrm{d}A$ is $|R_1 R_2|\,\mathrm{d}\theta_1\,\mathrm{d}\theta_2$. Choosing polar coordinates $\theta, \varphi$, for which $\mathrm{d}\theta_1 = \mathrm{d}\theta$ and $\mathrm{d}\theta_2 = \sin\theta\,\mathrm{d}\varphi$, we have

$$\mathrm{d}A = |R_1 R_2| \sin\theta\,\mathrm{d}\theta\,\mathrm{d}\varphi \equiv \frac{\mathrm{d}\Omega}{|K(\theta,\varphi)|}\,, \tag{5}$$

where $K(\theta, \varphi)$ is the curvature of the FS. We can then write

$$N_\zeta = \frac{2V}{(2\pi\hbar)^3} \int \frac{\mathrm{d}\Omega}{|K(\theta,\varphi)|}\,. \tag{6}$$

It follows from this expression that contributions to the electron DOS from those parts of the FS that correspond to the vicinities of zero-curvature lines ($K(\theta, \varphi) = 0$) exceed contributions from the remaining FS segments. In other words, when the curvature of the FS vanishes at some point (or line), the number of electrons associated with the vicinity of this point (or line) increases.

The occurrence of zero-curvature lines on a FS can affect high-frequency phenomena in metals. This happens when such lines intersect effective segments on the surface or belong to these segments. In the vicinity of a zero-curvature point, the FS is close in shape to a cylinder (if one of the principal curvature radii tends to infinity) or to a plane (so-called points of flattening, where both principal curvature radii tend to infinity). Therefore, when a zero-curvature line is located on an effective segment of the FS or even intersects it, the number of effective electrons increases. In turn, this brings changes to the observable properties for certain directions of propagation of the external perturbation.

The dependence of the attenuation rate of a sound wave on the direction of its propagation through a metal with a nearly planar and/or cylindrical FS was first predicted in Ref. [31]. The effect of zero-curvature lines and points on the frequency and angular dependences of the dispersion and attenuation of ultrasound, conductivity, and surface impedance of a metal under the anomalous skin effect was theoretically analyzed in Refs [31–38] using very simple models of the FS shape near the points/lines of zero curvature. Some results in Refs [31–38] were confirmed in experiments on the propagation of ultrasound waves in metals (see, e.g., Refs [36, 37, 39]).

Along with the points and lines of zero curvature, points and lines can exist where the curvature of the surface reveals discontinuities or diverges. A very simple illustration of such a case is a kink line on a piecewise smooth surface. If a FS has kink lines, the effective segments can be missed for appropriate directions of propagation of electromagnetic or sound waves, which causes anomalies in the high-frequency response of the metal. It is shown in Refs [41, 42] that under these conditions, electromagnetic waves of a special kind can appear in metals.

In the last two decades, various quasi-two-dimensional (Q2D) materials with metallic-type conductivity have been synthesized. Galvanomagnetic phenomena and quantum oscillatory effects in these materials were intensely studied (see, e.g., Refs [6–8, 43, 44]). A common feature of Q2D metals is their layered structure with a pronounced anisotropy of the electrical conductivity. In such materials, the electron energy only weakly depends on the quasimomentum projection $p = \mathbf{pn}$ onto the normal $\mathbf{n}$ to the layer planes. A simple model of a Q2D Fermi surface originates from the energy–momentum relation of the form

$$E(\mathbf{p}) = \frac{\mathbf{p}_\perp^2}{2m_\perp} - 2w\cos\left(\frac{p_z d}{\hbar}\right), \tag{7}$$

where $\mathbf{p}_\perp$ is the quasimomentum projection onto the layer plane and $m_\perp$ is the effective mass corresponding to the motion of quasiparticles in this plane. The parameter $w$ in expression (7) is the interlayer transfer integral, whose value determines how much the FS is warped. When $w$ tends to zero, the FS becomes perfectly cylindrical. However, there are grounds to believe that FSs of some realistic Q2D conductors have more complex geometries than those described by Eqn (7). For instance, such a conclusion follows from experiments on magnetic quantum oscillations in the overdoped cuprate $TlBa_2CuO_{6+\sigma}$ [10] and in the layered perovskite oxide $Sr_2RuO_4$ [45, 46]. It may be conjecture that FSs of some Q2D conductors can have various profiles, including cross sections with maximum/minimum external areas where the curvature becomes zero or discontinuous, as shown in Fig. 3.

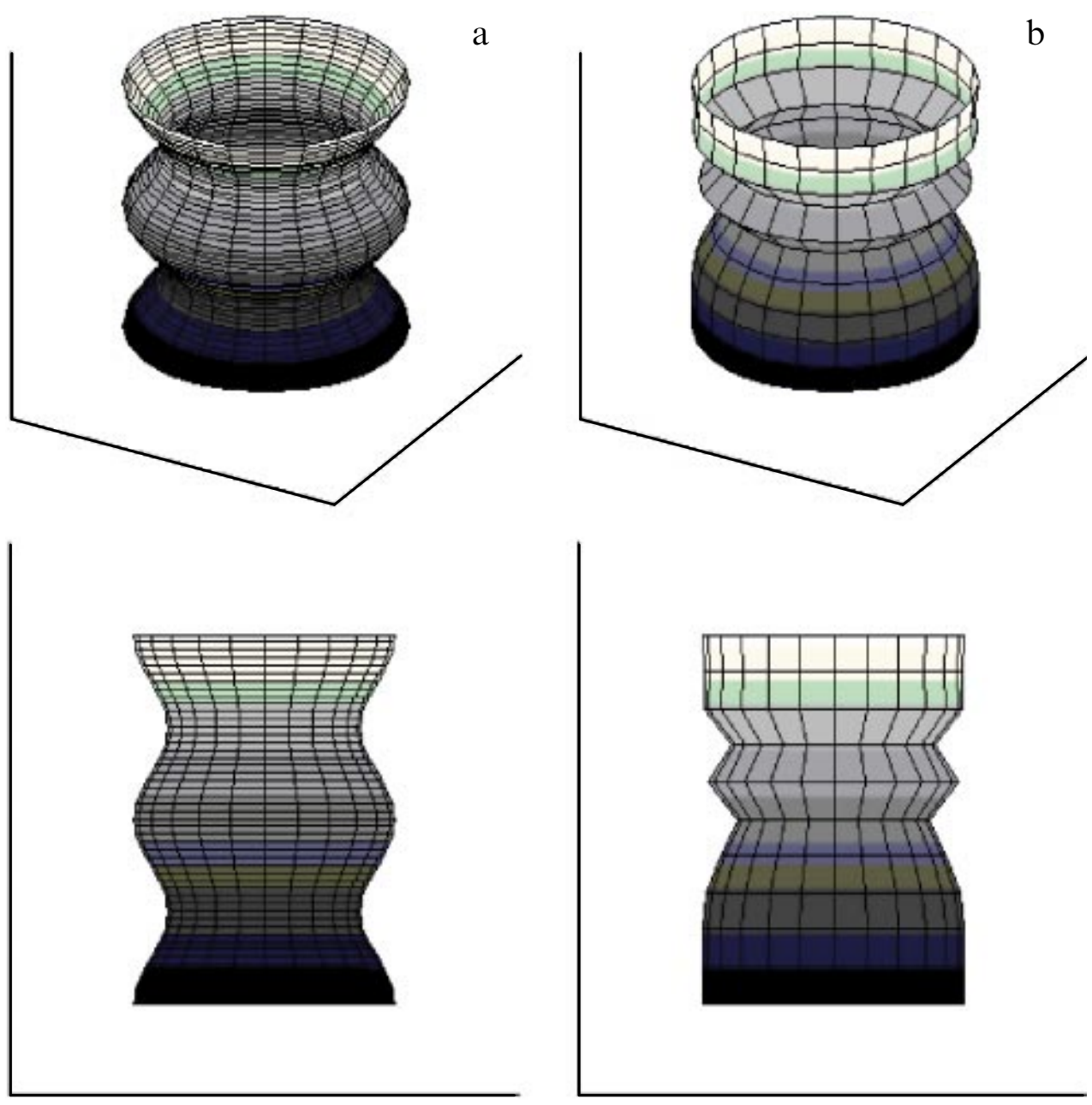

**Figure 3.** Schematic plots of FSs shaped as corrugated cylinders (a) with a cosine warping corresponding to energy–momentum relation (7) and (b) with complex profiles.

## 2.2 Main equations of the Fermi-liquid theory

Studies of the effects arising due to electron–electron interactions within the general many-body quantum field theory approach started in the early 1960s in the work of Luttinger [47], and have continued through the next four decades. Nevertheless, one of the oldest methods for dealing with electron–electron interactions, namely the Landau Fermi liquid (FL) theory [4, 48, 49], is still useful. It is important to realize that the *phenomenological* FL theory and the microscopic many-body perturbation theory (and, when applicable, the exact density functional theory) by definition lead to the same observable quantities such as the response to an external field, while both the zeroth approximation and its renormalization depend on the approach chosen. A discussion of this effect applied to the dielectric response of metals can be found, e.g., in [50]. Recognizing the value of the many-body perturbation theory, we nevertheless emphasize that the FL theory has provided important insights in areas such as high-frequency collective modes in metals [51–57] and oscillations of various thermodynamic observables in quantizing magnetic fields [58–60].

An advantage of this phenomenological theory is that it enables describing the effects of quasiparticle interactions in such a way that the interpretation of the results is rather transparent, as compared to field theory methods. It is known that the many-body approach gives general but cumbersome results, and it usually takes great calculational efforts and/or significant simplifications to extract suitable expressions for comparison with experimental data. More importantly, the adopted simplifications may lead to omission of some qualitative effects of electron–electron interactions.

Within the phenomenological Landau FL theory, single quasiparticle energies are renormalized, and the renormalization is determined by the distribution of excited quasiparticles. Accordingly, the energy of a 'bare' (noninteracting) quasiparticle $E_0(\mathbf{p})$ moving in the effective crystal potential is replaced by the renormalized energy defined by the relation [49]

$$E_\sigma(\mathbf{p},\mathbf{r},t) = E_0(\mathbf{p}) + \sum_{\mathbf{p}',\mathbf{s}'} F(\mathbf{p},\mathbf{s};\mathbf{p}',\mathbf{s}')\,\rho(\mathbf{p}',\mathbf{s}',\mathbf{r},t)\,. \qquad (8)$$

We note that $E_0(\mathbf{p})$ is here the energy spectrum in the absence of any excited quasiparticles (that is, at equilibrium and at zero temperature), whereas $\rho(\mathbf{p}',\mathbf{s}',\mathbf{r},t)$ represents the nonequilibrium part of the electron distribution function, which may depend on both the quasiparticle position $\mathbf{r}$ and time $t$. Also, $\mathbf{s}$ and $\mathbf{s}'$ are quasiparticle spins ($\sigma$ is the spin quantum number), and $F(\mathbf{p},\mathbf{s};\mathbf{p}',\mathbf{s}')$ is the FL kernel (Landau correlation function), which describes additional renormalization of the quasiparticle spectrum due to interaction with other excited quasiparticles. With spin–orbit interactions, the Landau correlation function can be written as

$$F(\mathbf{p},\mathbf{s};\mathbf{p}',\mathbf{s}') = \varphi(\mathbf{p},\mathbf{p}') + 4\psi(\mathbf{p},\mathbf{p}')(\mathbf{s}\,\mathbf{s}')\,. \qquad (9)$$

As follows from Eqn (8), the quasiparticle velocity $\mathbf{v} = \nabla_{\mathbf{p}} E$ differs from the bare velocity $\mathbf{v}_0 = \nabla_{\mathbf{p}} E_0$.

The gradient of $E$ in the coordinate space describes the average force exerted by the ambient quasiparticles on a given electron. It can be regarded as a 'diffusion' force, which tends to push the quasiparticle toward regions of minimum energy. The 'diffusion' force has to be included in the net force acting on an electron. When both an electric field $\mathbf{E}$ and a magnetic field $\mathbf{B}$ are applied to the electron system, we can write

$$\frac{\mathrm{d}\mathbf{p}}{\mathrm{d}t} = \frac{e}{c}[\mathbf{v}_p \times \mathbf{B}] + e\mathbf{E} - \sum_{\mathbf{p}',\mathbf{s}'} F(\mathbf{p},\mathbf{s};\mathbf{p}',\mathbf{s}')\,\frac{\partial g}{\partial \mathbf{r}}\,, \qquad (10)$$

where the third term represents the 'diffusion' force. To analyze the conduction electron response to an external disturbance, we must use a transport equation for the nonequilibrium distribution function. Here, we restrict our consideration to the linearized transport equation for the nonequilibrium distribution function $g(\mathbf{p},\mathbf{r},t) = \mathrm{Tr}_s\left(\rho(\mathbf{p},\mathbf{r},\hat{\mathbf{s}},t)\right)$. Accordingly, the transport equation takes the form suggested in Ref. [61]:

$$\frac{\partial g}{\partial t} + \frac{\partial g^e}{\partial \mathbf{r}}\,\mathbf{v}_{\mathbf{p}} + \frac{e}{c}[\mathbf{v}_{\mathbf{p}} \times \mathbf{B}]\,\frac{\partial g^e}{\partial \mathbf{p}} + \frac{\partial f_{\mathbf{p}}}{\partial E_{\mathbf{p}}}\,e\mathbf{v}_{\mathbf{p}}\mathbf{E} = I[g]\,, \qquad (11)$$

where $I[g]$ is the collision integral. Again, the distribution function $g$ describes a deviation of the system of quasiparticles from the equilibrium ground state, and the function $g^e$ is defined by the relation

$$g^e(\mathbf{p},\mathbf{r},t) = g(\mathbf{p},\mathbf{r},t) - \frac{\partial f_{\mathbf{p}}}{\partial E_{\mathbf{p}}}\sum_{\mathbf{p}',\mathbf{s}'} F(\mathbf{p},\mathbf{s};\mathbf{p}',\mathbf{s}')\,g(\mathbf{p}',\mathbf{r},t)\,. \qquad (12)$$

This function corresponds to a deviation from the 'local equilibrium' state (see Ref. [49]), which is described by Fermi distribution function (1) where $E$ is the 'renormalized' energy determined by Eqn (8).

Exact expressions for the functions $\varphi(\mathbf{p},\mathbf{p}')$ and $\psi(\mathbf{p},\mathbf{p}')$ are of course unknown. The model of an extremely short-range (contact) Coulomb interaction between quasiparticles is often used when applying the many-body theoretical approach to study the effects of electron–electron interactions (see, e.g., Ref. [62]). Within the phenomenological FL theory, this model results in the approximation of the functions $\varphi(\mathbf{p},\mathbf{p}')$ and $\psi(\mathbf{p},\mathbf{p}')$ by constants $\varphi_0$ and $\psi_0$. Such an approximation allows obtaining the Stoner renormalization of the paramagnetic susceptibility and a suitable approximation for the electron compressibility [4, 49]. But it misses other FL effects, such as the renormalization of cyclotron masses of the conduction electrons.

The phenomenological theory of an electronic liquid is based on the assumption that weakly excited states of the electronic system of a normal metal are single-particle states. Beyond the semiclassical approximation, it suggests the existence of a single-particle Hamiltonian $H_0$ with a set of quantum numbers $\nu$ and energies $E_\nu$. Weakly excited states have energies close to the Fermi energy $\zeta$ and are represented by a single-particle density matrix, which is the sum of an equilibrium part $f$ ($f_{\nu\nu'} = f_\nu \delta_{\nu\nu'}$, where $f_\nu$ is the Fermi distribution function for the energy $E_\nu$) and a small nonequilibrium correction $\rho$. The density matrix $\rho$ satisfies the equation

$$\mathrm{i}\hbar\,\frac{\partial \rho}{\partial t} = [H,\rho]\,, \qquad (13)$$

where the Hamiltonian $H$ includes the term $H_0$ and corrections describing interactions of quasiparticles with external disturbances (e.g., electromagnetic fields), their scattering on the defects of the crystalline lattice, and the interactions of quasiparticles. For a disturbance harmonic in time ($\sim \exp(-\mathrm{i}\omega t)$), the matrix elements $\rho_{\nu\nu'}$ satisfy the transport equation

$$-\mathrm{i}\omega\rho_{\nu\nu'} + \frac{1}{\mathrm{i}\hbar}(E_{\nu'} - E_\nu)\,\rho_{\nu\nu'} + \frac{1}{\mathrm{i}\hbar}(f_\nu - f_{\nu'})\,W^*_{\nu\nu'} = I_{\nu\nu'}[\rho]\,, \qquad (14)$$

where $I[\rho]$ is the quantum collision integral and the operator $W^*$ describes the interaction of a quasiparticle with the disturbance and the interactions between the quasiparticles, which cause the deviation from the equilibrium state.

In the framework of the FL theory, $W^*$ is a linear functional of $\rho$:

$$W^*_{\nu\nu'} = W_{\nu\nu'} + \sum_{\nu_1,\,\nu_2} F^{\nu_1\nu_2}_{\nu\nu'} \rho_{\nu_1\nu_2}\,, \tag{15}$$

where $W_{\nu\nu'}$ represents the interactions with the perturbing fields and the integral term describes the interactions between quasiparticles. For disturbances whose frequencies are small compared to $\zeta/\hbar$ and wavelengths are greater than the electron wavelength on the FS (such as electromagnetic fields or ultrasound waves), matrix elements of the FL kernel $F^{\nu_1\nu_2}_{\nu\nu'}$ can be computed assuming that the electron system remains in the ground state.

To study the electromagnetic response of electrons, we can write the operator $W$ as

$$W_{\nu\nu'} = \int \mathrm{d}\mathbf{r} \left( e n_{\nu\nu'}(\mathbf{r}) \Phi(\mathbf{r}) - \frac{1}{c}\, \mathbf{j}_{\nu\nu'}(\mathbf{r}) \mathbf{A}(\mathbf{r}) \right), \tag{16}$$

where $\Phi(\mathbf{r})$ and $A(\mathbf{r})$ are the scalar and vector potentials of the alternating electromagnetic field and $en(\mathbf{r})$ and $\mathbf{j}(\mathbf{r})$ are the operators of the charge and current density of electrons. When we are dealing with perturbations of a different kind (e.g., originating from a sound wave propagating through a metal), the form of $W_{\nu\nu'}$ changes accordingly.

Transport equation (14) can be converted into a form more convenient for further calculations and for comparison with the semiclassical transport equation. To transform the equation, we introduce the operator of effective current density $\mathbf{j}^*(\mathbf{r})$ that determines the form of the correction $\rho_A$ to the equilibrium electronic density matrix in a weak static magnetic field $\mathbf{b}(\mathbf{r}) = \mathrm{rot}\,\mathbf{A}(\mathbf{r})$:

$$\rho^A_{\nu\nu'} = -\frac{1}{c}\, \frac{f_\nu - f_{\nu'}}{E_\nu - E_{\nu'}} \int \mathrm{d}\mathbf{r}\, \mathbf{j}^*_{\nu\nu'}(\mathbf{r})\, \mathbf{A}(\mathbf{r})\,, \tag{17}$$

where the matrix elements $\mathbf{j}^*_{\nu\nu'}(\mathbf{r})$ are defined by the relations [63]

$$\mathbf{j}^*_{\nu\nu'}(\mathbf{r}) = \mathbf{j}_{\nu\nu'}(\mathbf{r}) + \sum_{\nu_1,\,\nu_2} \frac{f_{\nu_1} - f_{\nu_2}}{E_{\nu_1} - E_{\nu_2}}\, F^{\nu_1\nu_2}_{\nu\nu'}\, \mathbf{j}^*_{\nu_1\nu_2}(\mathbf{r})\,. \tag{18}$$

In the semiclassical limit, this expression turns into the well-known relation between 'bare' ($\mathbf{v}_0$) and renormalized ($\mathbf{v}$) velocities of a quasiparticle suggested by Pines and Noziers [49]. Then we can separate the contribution $\rho_A$ from the nonequilibrium part of the density matrix:

$$\rho_{\nu\nu'} = \rho^A_{\nu\nu'} + g_{\nu\nu'}\,. \tag{19}$$

When the disturbance arises due to an external electromagnetic field, the matrix elements $g_{\nu\nu'}$ are determined by the equation

$$\begin{gathered} -\mathrm{i}\omega g_{\nu\nu'} + \frac{1}{\mathrm{i}\hbar}(E_{\nu'} - E_\nu) g^e_{\nu\nu'} + \frac{f_\nu - f_{\nu'}}{E_\nu - E_{\nu'}} \\ \times \int \mathrm{d}\mathbf{r}\, \mathbf{j}^*_{\nu\nu'}(\mathbf{r})\, E_\omega(\mathbf{r}) = I_{\nu\nu'}[g]\,, \\ g^e_{\nu\nu'} = g_{\nu\nu'} - \frac{f_\nu - f_{\nu'}}{E_\nu - E_{\nu'}} \sum_{\nu_1\nu_2} F^{\nu_1\nu_2}_{\nu\nu'} g_{\nu_1\nu_2}\,, \end{gathered} \tag{20}$$

where

$$\mathbf{E}_\omega(\mathbf{r}) = \frac{\mathrm{i}\omega}{c}\, \mathbf{A}_\omega(\mathbf{r}) - \nabla_{\mathbf{r}} \Phi_\omega(\mathbf{r})$$

is the amplitude of the electric field. This result is the quantum analog of semiclassical transport equation (11), and it can be converted into the latter in the semiclassical limit.

### 2.3 Symmetry of the crystalline lattice and approximations of the FL kernel

Expressions for the Landau correlation functions $\varphi(\mathbf{p}, \mathbf{p}')$ and $\psi(\mathbf{p}, \mathbf{p}')$ beyond the limit of the short-range interactions between the quasiparticles strongly depend of the crystalline lattice symmetries. Using a simple isotropic model of a metal with a spherical FS allows easily concluding that these functions depend only on the angle between the quasimomenta $\mathbf{p}$ and $\mathbf{p}'$ corresponding to a pair of quasiparticles on the FS. Accordingly, the correlation functions can be expanded in series in spherical harmonics [64]:

$$\begin{aligned} \varphi(\mathbf{p}, \mathbf{p}') &= \frac{(2\pi\hbar)^3 v_{\mathrm{F}}}{2p_{\mathrm{F}}^2} \sum_{j=0} \sum_{|m| \leqslant j} A_j Y_{j,m}(\theta, \Phi)\, Y_{j,-m}(\theta', \Phi')\,, \\ \psi(\mathbf{p}, \mathbf{p}') &= \frac{(2\pi\hbar)^3 v_{\mathrm{F}}}{2p_{\mathrm{F}}^2} \sum_{j=0} \sum_{|m| \leqslant j} B_j Y_{j,m}(\theta, \Phi)\, Y_{j,-m}(\theta', \Phi')\,, \end{aligned} \tag{21}$$

where $v_{\mathrm{F}}$ is the Fermi velocity and the vectors $\mathbf{p}$ and $\mathbf{p}'$ are represented using their spherical coordinates: $\mathbf{p} = (p_{\mathrm{F}}, \theta, \Phi)$. Dimensionless coefficients $A_j$ and $B_j$ describe the effects of electron–electron correlations. Assuming that all coefficients $A_j$, $B_j$ with $j > 0$ vanish, we reduce expressions (21) to the approximations corresponding to the limit of short-range interactions. Usually, the FL coefficients become smaller as $j$ increases, and the terms with $j \leqslant 2$ must be kept in the expansions for the correlation functions to satisfactorily describe the main correlation effects in the electronic characteristics of metals.

Considering practical metals with nonspherical FSs, we can decompose the FL functions with respect to basis functions determined by the crystal symmetry, such as Allen's FS harmonics [65]:

$$\begin{aligned} \varphi(\mathbf{p}, \mathbf{p}') &= \sum_{j=1}^{d} \sum_{m=1}^{d_j} \varphi_j(p, p')\, R_{jm}(\theta, \Phi)\, R^*_{jm}(\theta', \Phi')\,, \\ \psi(\mathbf{p}, \mathbf{p}') &= \sum_{j=1}^{d} \sum_{m=1}^{d_j} \psi_j(p, p')\, R_{jm}(\theta, \Phi)\, R^*_{jm}(\theta', \Phi')\,. \end{aligned} \tag{22}$$

Here, $d$ is the order of the point symmetry group of the lattice, the index $j$ labels irreducible representations of the group, $d_j$ is the dimension of the $j$th irreducible representation, and $\{R_{jm}(\theta, \Phi)\}$ is a basis of the $j$th irreducible representation including $d_j$ functions.

Keeping various Q2D metals in mind, it is interesting to consider suitable approximations for the correlation functions for an axially symmetric FS. In this case, the functions $\varphi(\mathbf{p}, \mathbf{p}')$ and $\psi(\mathbf{p}, \mathbf{p}')$ do not vary with identical changes in directions of the projections $\mathbf{p}_\perp$ and $\mathbf{p}'_\perp$ of the quasimomenta $\mathbf{p}$ and $\mathbf{p}'$ onto the plane $p_z = 0$ (assuming that the $z$ axis of the chosen coordinate system is directed along the FS symmetry axis). These functions actually depend only on the cosine of the angle $\theta$ between the vectors $\mathbf{p}_\perp$ and $\mathbf{p}'_\perp$ and on the longitudinal components of the quasimomenta $p_z$ and $p'_z$.

On these grounds, we can separate the even and odd parts of the FL functions of $\cos\theta$. Then the function $\varphi(\mathbf{p}, \mathbf{p}')$ can be presented in the form [66]

$$\varphi(\mathbf{p}, \mathbf{p}') = \varphi_0(p_z, p'_z, \cos\theta) + (\mathbf{p}_\perp \mathbf{p}'_\perp)\, \varphi_1(p_z, p'_z, \cos\theta)\,, \tag{23}$$

where $\varphi_0$ and $\varphi_1$ are even functions of $\cos\theta$. Due to the invariance of the FS under the replacement $\mathbf{p} \to -\mathbf{p}$ and $\mathbf{p}' \to -\mathbf{p}'$, the functions $\varphi_0$ and $\varphi_1$ should not change under a simultaneous change of the signs of $p_z$ and $p_z{}'$. Using this, we can split the functions $\varphi_0$ and $\varphi_1$ into the parts that are even and odd in $p_z$, $p_z{}'$, and rewrite Eqn (23) as

$$\varphi(\mathbf{p}, \mathbf{p}') = \varphi_{00} + p_z p_z' \varphi_{01} + (\mathbf{p}_\perp \mathbf{p}_\perp')(\varphi_{10} + p_z p_z' \varphi_{11}) \,. \quad (24)$$

The function $\psi(\mathbf{p}, \mathbf{p}')$ can be presented similarly. In Eqn (24), the functions $\varphi_{00}$, $\varphi_{01}$, $\varphi_{10}$, and $\varphi_{11}$ are even in all their arguments, $p_z$, $p_z{}'$, and $\cos\theta$. Each term in Eqn (24) corresponds to a particular part in the expansion of the FL functions in spherical harmonics, which is appropriate for an isotropic metal. Comparing Eqns (24) and (21), we see that the function $\varphi_{00}$ matches the part of expansion (21) that includes all terms with even values of both $j$ and $m$. The product $p_z p_z' \varphi_{01}$ represents the sum of all terms in Eqn (21) with odd values of $j$ and even values of $m$. The third term in Eqn (24) corresponds to that part of expansion (21) containing terms with odd values of both indices. Finally, the terms in Eqn (21) labeled with even $j$ and odd $m$ are matched with the expression $(\mathbf{p}_\perp \mathbf{p}_\perp') p_z p_z' \varphi_{11}$. Again, the main FL effects in the response of a metal to an external disturbance can be adequately analyzed while keeping the terms with $j \leqslant 2$ in expansion (21). Omission of all terms with $j > 2$ from this expansion agrees with the assumption that the functions $\varphi_{00}$, $\varphi_{01}$, $\varphi_{10}$, and $\varphi_{11}$ are constants. Therefore, we can adopt approximation (24) assuming $\varphi_{00}$, $\varphi_{01}$, $\varphi_{10}$, and $\varphi_{11}$ to be material constants for an arbitrary axially symmetric FS.

## 3. Local features of the Fermi surface curvature and the linear response of a metal to a high-frequency electromagnetic disturbance

### 3.1 Anomalous skin effect

It is well known that electromagnetic waves incident on a metal surface cannot penetrate deeply inside. Actually, the field inside the metal vanishes at distances of the order of $\delta$ from the surface. This effect is called the skin effect, and the characteristic depth $\delta$ is called the skin depth. The suppression of the electromagnetic field inside the metal originates from the response of conduction electrons, and it occurs when the frequency $\omega$ of the incident wave is smaller than the electron plasma frequency $\omega_{\mathrm{p}}$. The latter is the characteristic frequency for the response of the electrons to an external disturbance. When $\omega > \omega_{\mathrm{p}}$, the electrons are too slow to respond, and the electromagnetic field penetrates into the metal without decay. At moderately high frequencies $\tau^{-1} \ll \omega \ll \omega_{\mathrm{p}}$ ($\tau$ is the scattering time for conduction electrons) and low temperatures, $\delta$ can become smaller than the electron mean free path $l$. When the condition $\delta < l$ is satisfied, the effect is referred to as the anomalous skin effect. Under the anomalous skin effect, the response of a metal to an incident electromagnetic wave is determined by the electrons moving in the skin layer nearly parallel to the surface of the metal sample. These 'efficient' electrons are associated with 'effective segments' on the FS. The remaining electrons stay in the skin layer only very briefly, which prevents them from responding to the electromagnetic field.

A theory of the anomalous skin effect in metals was first proposed more than five decades ago by Pippard [67], Reuter, and Zondheimer [68], and Dingle [69] using an isotropic model for a metal. The theory was further developed to make it applicable to realistic metals with anisotropic FSs [70–78]. It became clear that the response of conduction electrons to an external electromagnetic field under the anomalous skin effect depends on the FS geometry and especially on its Gaussian curvature. Some effects of the FS geometry on the metal response under the anomalous skin effect were analyzed in [4, 70, 71, 75], adopting some simplified models for the FS. Here, we perform a general analysis whose results are independent of particularities in energy–momentum relations and can be applied to a broad class of metals.

We consider a metal filling the half-space $z < 0$. A plane electromagnetic wave is incident on the metal surface at a right angle. To analyze the response of the metal to the wave, we calculate the surface impedance

$$Z_{\alpha\beta} = \frac{E_\alpha(0)}{\int_0^{-\infty} j_\beta(z)\,\mathrm{d}z} \,, \quad (25)$$

where $\alpha, \beta = x, y$, and $E_\alpha(z)$ and $j_\beta(z)$ are the components of the electric field $\mathbf{E}$ and the electric current density $\mathbf{j}$.

Due to the high density of conduction electrons in good metals, the skin depth $\delta$ can be very small. Assuming the electron density to be of the order of $10^{30}$ m$^{-3}$ and the mean free path $l \sim 10^{-3}$ m (a clean metal), we estimate the skin depth at the disturbance frequency $\omega \sim 10^9$ s$^{-1}$ as $\delta \sim 10^{-6}$ m. Therefore, at high frequencies $\omega$, the skin effect in good metals becomes extremely anomalous, with $\delta/l \sim 10^{-2} - 10^{-3}$ or even smaller. Under these conditions, electrons must move nearly parallel to the metal surface to remain in the skin layer for a sufficiently long time. The effect of the surface roughness on such electrons is rather small. As shown in [79], electron reflection from the metal surface under the extremely anomalous skin effect can be treated as nearly specular, because the corrections originating from the diffuse scattering have the order $\delta/l$. Therefore, we can limit our analysis to the case of specular reflection of electrons from the surface. Then the surface impedance tensor has the form

$$Z_{\alpha\beta} = \frac{8\mathrm{i}\omega}{c^2} \int_0^\infty \left( \frac{4\pi\mathrm{i}\omega\sigma}{c^2} - q^2 I \right)^{-1}_{\alpha\beta} \mathrm{d}q \,, \quad (26)$$

where $\mathbf{q}$ is the wave vector of the incident wave ($\mathbf{q} = (0, 0, q)$), $\sigma$ is the electron conductivity tensor, and $I_{\alpha\beta} = \delta_{\alpha\beta}$.

We split each sheet of the FS into segments such that a one-to-one correspondence is established between the quasimomentum $\mathbf{p}$ and the electron velocity $\mathbf{v}$ over each segment. The segments may coincide with the FS sheets. Also, it can happen that some sheets include several segments. This depends on the FS shape. In calculating the conductivity, we integrate over each segment using spherical coordinates in the velocity space, which include the velocity magnitude at the $j$th segment $v_j$ and the spherical angles $\theta$ and $\varphi$. Therefore, the surface area element is $\mathrm{d}A_j = \sin\theta\,\mathrm{d}\theta\,\mathrm{d}\varphi / |K_j(\theta, \varphi)|$, where $K_j(\theta, \varphi)$ is the Gaussian curvature of the $j$th FS segment. Adding the contributions from all these segments, we obtain

$$\sigma_{\alpha\beta}(\omega, q) = \frac{\mathrm{i}e^2}{4\pi^3\hbar^3 q} \sum_j \int \mathrm{d}\varphi \times \int \frac{n_\alpha n_\beta \sin\theta\,\mathrm{d}\theta}{|K_j(\theta, \varphi)| \left[ (\omega + \mathrm{i}/\tau)/(q v_j) - \cos\theta \right]} \,, \quad (27)$$

where $n_{\alpha,\beta} = v_{j\alpha,\beta}/v_j$, and the limits in the integrals over $\theta$ and $\varphi$ are determined by the shape of the segments. The leading contribution to the surface impedance under the anomalous skin effect comes from the region of large $q$, where $\omega/qv \ll 1$. To calculate the corresponding asymptotic expressions for the conductivity components, we expand the integrand in Eqn (27) in powers of $\omega/qv$. Then we can write the well-known result for the leading term in the expansion of the conductivity component $\sigma_{xx}(\omega, q)$:

$$\sigma_0(q) = \frac{e^2}{4\pi^3\hbar^3 q}\sum_l \int \mathrm{d}\varphi \, \frac{\cos^2\varphi}{|K_l(\pi/2,\varphi)|} \equiv \frac{e^2}{4\pi\hbar^3 q}\, p_0^2 \,. \quad (28)$$

The same asymptotic form can be obtained for $\sigma_{yy}$, and we therefore omit the indices here and in the following expressions for simplicity. Summation over $l$ ranges all segments of the FS containing effective strips that correspond to $\theta = \pi/2$ ($v_z = 0$), and the curvature $K_l(\pi/2, \varphi)$ is supposed to take a finite and nonzero value at any point of any effective strip. For a spherical FS, $p_0$ is equal to the Fermi momentum $p_{\mathrm{F}}$. In realistic metals, the two are not equal, but have the same order of magnitude. We can then calculate the next term in the expansion of conductivity in powers of the small parameter $u = \omega/qv_{\mathrm{F}}$. For a FS whose curvature is everywhere finite and nonzero, we arrive at the result [73]

$$\sigma_1(\omega, q) = \sigma_0(q)\, \frac{\mathrm{i}\omega}{qv_0} \equiv \mathrm{i}\sigma_0(q)\, \frac{v_{\mathrm{F}}}{v_0}\, u \,, \quad (29)$$

where the velocity $v_0$ has the order of the Fermi velocity $v_{\mathrm{F}}$.

When the curvature vanishes at an effective line, the asymptotic form of the conductivity changes. First, we assume that the curvature vanishes when a whole effective line passes through one of the assigned segments of the FS. The effective line is determined by the condition $v_z = 0$. Therefore, the curvature $K(\theta, \varphi)$ in the vicinity of this line can be approximated by the expression

$$K(\theta, \varphi) = W(\theta, \varphi)(\cos\theta)^{-\beta} \,. \quad (30)$$

In this expression, the function $W(\theta, \varphi)$ is everywhere finite and nonzero and the exponent $\beta$ takes negative values, and therefore the curvature vanishes at $\theta = \pi/2$. In the close vicinity of this line, the FS is nearly cylindrical in shape. The closer $\beta$ is to $-1$, the closer the effective strip on the FS is to a cylinder. The contribution to the conductivity from the nearly cylindrical segment on the FS has the form

$$\sigma_{\mathrm{a}}(\omega, q) = \eta\sigma_0(q)\left(\frac{\omega}{qv_{\mathrm{a}}}\right)^{\beta}\left[1 - \mathrm{i}\tan\left(\frac{\pi\beta}{2}\right)\right], \quad (31)$$

$$\eta \approx \frac{1}{\pi p_0^2}\int \frac{\mathrm{d}\varphi\, \cos^2\varphi}{|W(\pi/2,\varphi)|}\,, \quad (32)$$

where $v_{\mathrm{a}}$ is the maximum speed of a conduction electron on the effective line. Comparing Eqn (32) with the definition of $p_0^2$ in (28), we see that the dimensionless factor $\eta$ is determined by the relative number of 'effective' electrons concentrated in the nearly cylindrical effective segment.

The contribution to the conductivity from the 'anomalous' effective strip strongly depends on the relative number of effective electrons concentrated there. This is illustrated by Fig. 4. In this figure, we display plots of $|\sigma(\omega,q)/\sigma_0(q)| \equiv |1 + \sigma_{\mathrm{a}}(\omega,q)/\sigma_0(q)|$ versus $\omega/qv$. When the parameter $\eta$ takes values of the order of or greater than 0.1 (the number of effective electrons associated with the anomalous sections on the FS is comparable to the total number of effective electrons), the term $\sigma_{\mathrm{a}}(\omega,q)$ can dominate over $\sigma_0(q)$ and determine the conductivity value at large $q$. This occurs when the shape parameter $\beta$ takes values not too close to zero, and the curvature anomaly on the effective line is well pronounced. When either $\eta$ or $\beta$ or both are very small, the conductivity in the leading approximation is described by Eqn (28), just as for a metal whose FS curvature is everywhere nonzero. Nevertheless, in such cases, the term $\sigma_{\mathrm{a}}(\omega,q)$ remains important because it gives the first correction to the leading term in the expression for the conductivity.

Also, the anomalous contribution to the conductivity can appear when the FS is flattened at some points belonging to an effective segment. To avoid lengthy calculations, we illustrate the effect of such points on the conductivity using a simple expression representing the energy–momentum

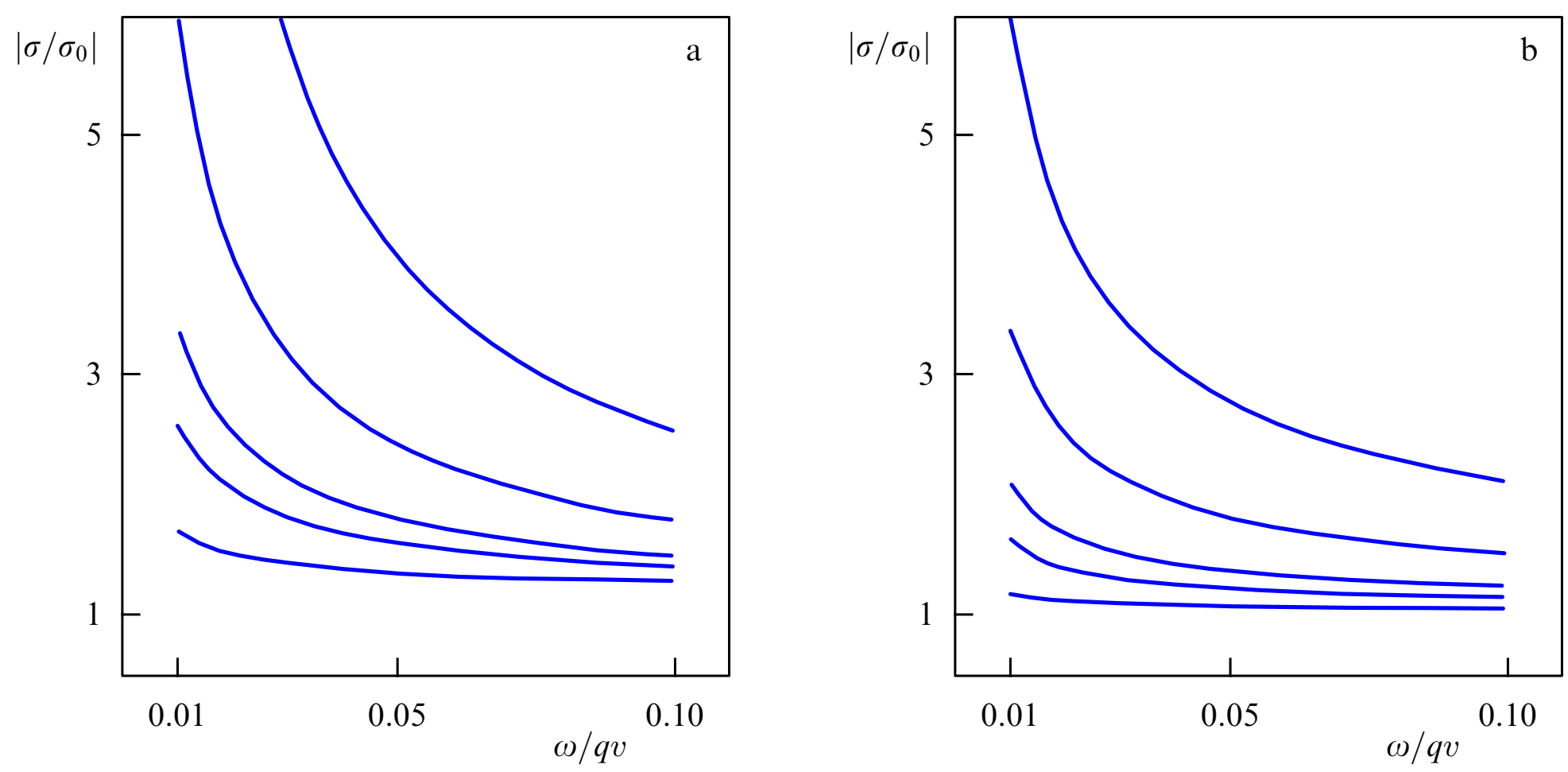

**Figure 4.** The conductivity component $\sigma_{xx}(\omega,q)$ including the contribution from a zero-curvature segment on the FS at large $q$ ($\omega/qv \ll 1$): (a) the curves are plotted at $\eta = 0.2$ and $\beta = -0.8, -0.7, -0.6, -0.4$, and $-0.2$ from top down. (b) The curves are plotted at $\beta = -0.6$ and $\eta = 0.2, 0.1, 0.05, 0.03, 0.01$ from top down.

relation near the point of flattening $\mathcal{M}_0(p_1, 0, 0)$,

$$E(\mathbf{p}) = \frac{p_1^2}{2m_1}\left(\frac{p_x^2}{p_1^2}\right) + \frac{p_2^2}{m_2}\left(\frac{p_y^2+p_z^2}{p_2^2}\right)^k, \tag{33}$$

where $p_1$ and $p_2$ have the dimension of momentum. When $k=1$, this expression corresponds to an ellipsoidal FS, and $m_1$ and $m_2$ are the eigenvalues of the effective mass tensor. The curvature of the FS associated with energy–momentum relation (33) is given by [60]

$$K(\mathbf{p}) = \frac{k}{m_2 v^4}\left(\frac{p_y^2+p_z^2}{p_2^2}\right)^{k-1} \times \left[\frac{1}{m_1}(v_y^2+v_z^2) + v_x^2\,\frac{k(2k-1)}{m_2}\left(\frac{p_y^2+p_z^2}{p_2^2}\right)^{k-1}\right]. \tag{34}$$

For $k>1$, the curvatures of both principal cross sections of the FS vanish at the points $(\pm p_1, 0, 0)$, indicating local flattening of the FS.

The 'anomalous' contribution to the conductivity originating from the flattened segments of the FS has a form similar to that in Eqn (31), namely,

$$\sigma_{\rm a}(\omega, q) = \mu\sigma_0(q)\left(\frac{\omega}{qv(\pi/2,0)}\right)^{\beta}\left[1-{\rm i}\tan\left(\frac{\pi\beta}{2}\right)\right] \tag{35}$$

with the shape parameter $\beta = -1 + 1/(2k-1)$, where $\mu$ is a small dimensionless factor proportional to the relative number of conduction electrons associated with the flattened part of the FS. Due to the smallness of $\mu$, term (35) can be significant only when $\beta < 0$ $(k>1)$. Otherwise, it can be neglected.

We now proceed with the calculations of the surface impedance given by expression (26). Under the anomalous skin effect conditions, the impedance can be represented as an expansion in inverse powers of the anomaly parameter $(\xi \gg 1)$. Representing the conductivity as the sum of terms (28) and (29), we can calculate the first two terms of this expansion:

$$Z \equiv R - {\rm i}H = -\frac{8{\rm i}\omega}{c^2}\,\delta\int_0^\infty {\rm d}t\,\frac{1}{1-{\rm i}t^3(1+{\rm i}t/\xi)} \approx \frac{Z_0}{\xi}\left[1-{\rm i}\sqrt{3}-\frac{2}{3\xi}\left(1+{\rm i}\sqrt{3}\right)\right], \tag{36}$$

where $\delta = (c^2\hbar^3/e^2p_0^2\omega)^{1/3}$ is the skin depth,

$$Z_0 = \frac{8\pi}{3\sqrt{3}}\frac{v_0}{c^2}; \qquad \xi = \frac{v_0}{\omega\delta} \equiv \left(\frac{\omega}{\omega_0}\right)^{-2/3}$$

is the anomaly parameter, and

$$\omega_0 = \left(\frac{v_0}{\hbar}\right)^{3/2}\frac{ep_0}{c} \tag{37}$$

is the frequency. Keeping in mind that $v_0 \sim v_{\rm F}$ and $p_0 \sim p_{\rm F}$, we can roughly estimate the characteristic frequency $\omega_0$. In good metals, $\omega_0 \sim 10^{12} - 10^{13}$ s$^{-1}$. This is significantly smaller than the plasma frequency $\omega_{\rm p}$, which is of the order of $10^{15}-10^{16}$ s$^{-1}$ in good metals. As could be expected, the inequality $\omega \ll \omega_0$ $(\xi \gg 1)$ agrees with the general requirement for frequencies $\omega \ll \omega_{\rm p}$ and can be satisfied for $\omega \sim 10^{10}-10^{11}$ s$^{-1}$.

An expression like (36) was first obtained by Dingle [69] within the isotropic model of metals. It was subsequently generalized to be applied to realistic metals, assuming that their FSs do not include zero-curvature segments [74]. For such FSs, the frequency dependence of the surface impedance has the same character as for a Fermi sphere. The main approximation of the surface impedance is proportional to $\omega^{2/3}$, and the first correction to it is proportional to $\omega^{4/3}$.

We can expect that in realistic metals, either $\eta$ or $\beta$ or both take small values (a zero-curvature segment on the effective part of the FS is small and/or the curvature anomaly is only moderately pronounced). In this case, the anomalous contribution to the conductivity of form (31) and/or (35) is the first correction to the leading approximation and determines the first correction to the approximation for the surface impedance [80]:

$$\Delta Z \equiv \Delta R - {\rm i}\Delta H \approx -Z_0\eta(\beta)\left(\frac{\omega}{\omega_0}\right)^{2(\beta+1)/3} \times\left\{\cot\left[\frac{\pi(\beta+1)}{3}\right]+{\rm i}\right\}, \tag{38}$$

where $\eta(\beta)$ is a small dimensionless parameter of the order of $\eta$ or $\mu$. We can use the result in (38) to describe the contribution to the surface impedance from a narrow or weakly developed, nearly cylindrical strip and also from a point of flattening located on the effective segment of the FS. The correction to the leading approximation of the surface impedance is now proportional to $\omega^{2(\beta+1)/3}$, as shown in Fig. 5. The obtained asymptotic expression indicates that a curvature anomaly on an effective line changes the frequency dependence of the surface impedance. This follows from the above relation between the curvature of the FS and the number of effective electrons.

### 3.2 Cyclotron resonance in metals in a normal magnetic field

It is well known that a periodic motion of conduction electrons in a magnetic field can cause a resonance with the electric field of an incident electromagnetic wave. This resonance occurs when the cyclotron frequency of electrons coincides with the frequency of the electromagnetic field. There are two geometries providing the resonance features in the surface impedance to be displayed. First, strong resonance arises when the magnetic field is directed nearly parallel to the surface of a metal [81]. Conduction electrons spiral around the magnetic field and at each revolution some of them return to the skin layer, where they gain energy from the electromagnetic wave. Another possibility of a resonance appears when the magnetic field is perpendicular to the surface. In this case, some conduction electrons remain within the skin layer for a long time, absorbing the energy. As a rule, the cyclotron resonance in a normal magnetic field is not manifested in good metals. The standard explanation is that the skin layer in metals is very thin at high frequencies. Therefore, the percentage of conduction electrons moving within the skin layer parallel to the metal surface is too small to provide a distinguishable resonance feature at $\omega = \Omega$ (where $\Omega$ is the cyclotron frequency of conduction electrons) in either the surface impedance of the metal or the absorbed power. This scenario was supported with thorough calculations carried out assuming the FS of a metal to be a smooth surface of a nonzero curvature (see, e.g., Refs [82, 83]). Nevertheless, cyclotron resonance in a normal magnetic field was observed in a polycrystalline sample of potassium [84] and in monocrystalline samples of cadmium and zinc [85, 86] forty years ago. During the next two decades, no detailed theory was suggested to explain these old experiments. Then a qualitative explanation of the experiments on potassium was offered, based on the assumption that the FS of this metal includes some cylindrical segments [87].

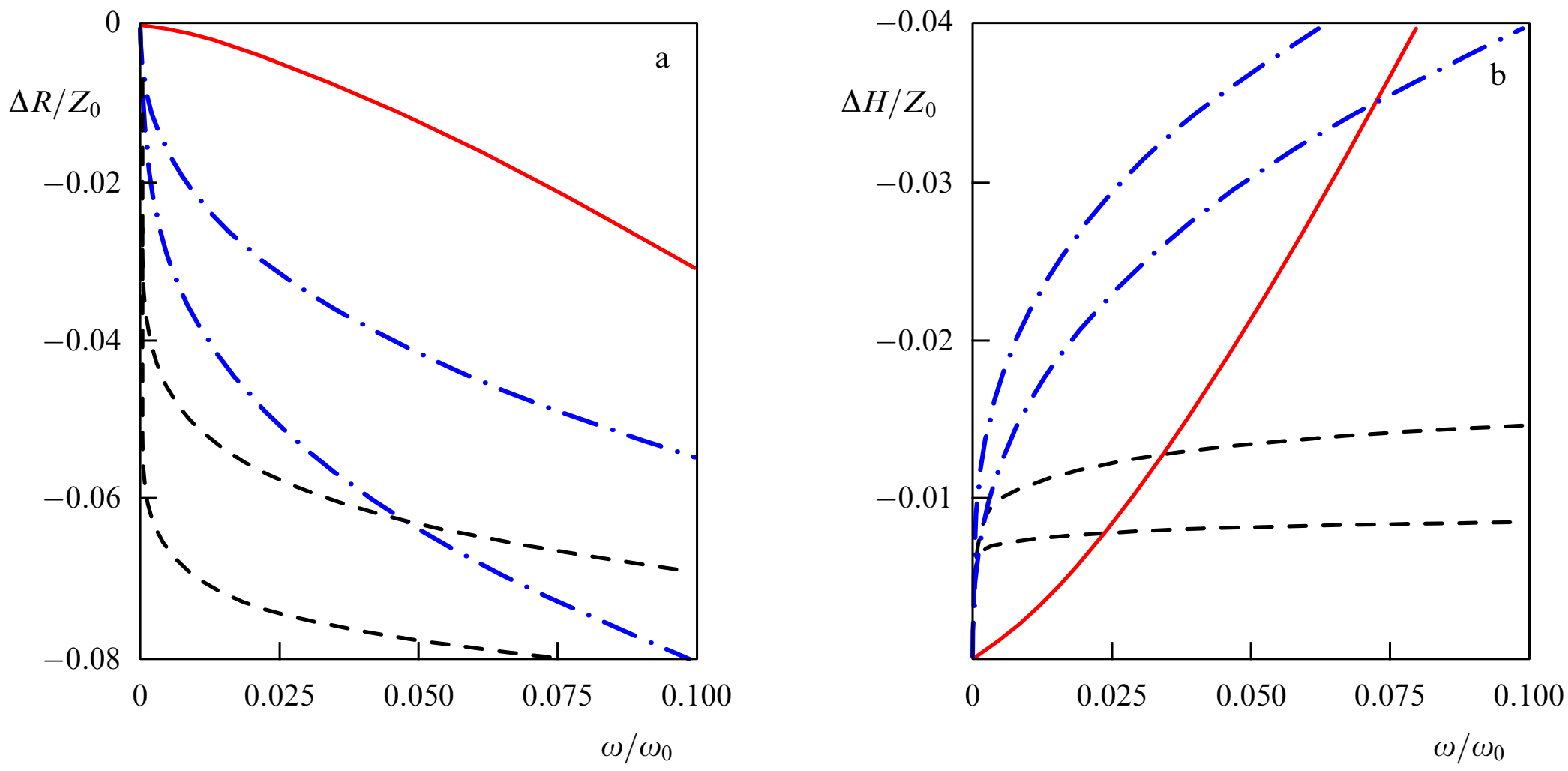

**Figure 5.** Frequency dependences of the real ($\Delta R$) and imaginary ($\Delta H$) parts of the first correction to the leading term in the surface impedance expansion in inverse powers of the anomaly parameter. The curves are plotted at $\eta = 0.01$, $\beta = -0.9$; $\eta = 0.02$, $\beta = -0.8$ (dashed lines) and $\eta = 0.1$, $\beta = -0.5$; $\eta = 0.1$, $\beta = -0.4$ (dashed-dotted lines). Solid lines represent the real and imaginary parts of the first correction to the leading approximation for the impedance of a metal whose FS does not include nearly cylindrical or flattened segments.

This assumption agrees with the results concerning the FS shape of potassium that follow from the charge density wave theory [88, 89] and with the experimental results reported in Refs [90, 91]. Another attempt to explain the resonance in potassium was undertaken in Ref. [92], assuming that the FS of potassium has inflection lines where its curvature vanishes. This is a realistic assumption, because even a slightly distorted Fermi sphere has small lumps arranged at the segments closest to the Brillouin zone boundaries. Zero-curvature lines separate these lumps from the main body of the FS [93].

Here, we further develop the approach suggested in [92, 93], and we show that the cyclotron resonance in the normal magnetic field can appear in metals when their FSs have some special local geometric characteristics, such as local flattenings and/or nearly cylindrical segments, which induce changes in the frequency dependences of the surface impedance considered in the previous section. As before, we consider a semi-infinite metal that fills the half-space $z < 0$. We assume that the external magnetic field $\mathbf{B}$ is directed perpendicularly to the surface of the metal. The response of the conduction electrons to an electromagnetic disturbance with a frequency $\omega$ and wave vector $\mathbf{q}$ (the latter being parallel to the $z$ axis) can be expressed in terms of the electron conductivity components. For a metal with a closed single-sheet FS, we can write the expression [73]

$$\sigma_{\alpha\beta} = \frac{2\mathrm{i}e^2}{(2\pi\hbar)^3}\sum_r \int_0^{2\pi} \mathrm{d}\psi \int \mathrm{d}p_z \times \int_{-\infty}^{0} \mathrm{d}\epsilon\, e^{\epsilon}\, \frac{m_\perp(p_z) v_\mathrm{a}(p_z,\psi) v_{n\beta}(p_z)\exp(-\mathrm{i}n\psi)}{\omega + \mathrm{i}/\tau - r\Omega(p_z) - qv_z\big[p_z, \psi + \delta\psi(\epsilon)\big]}\,, \quad (39)$$

where $m_\perp(p_z)$ is the cyclotron mass and the angular variable $\psi = \Omega t$ is related to the time $t$ of electron motion along the cyclotron path. For a metal with a complex multisheet FS (for example, cadmium or zinc), integration with respect to $p_z$ in Eqn (39) is carried out within the limits determined by the form of each sheet, and extra summation is to be carried out over all sheets of the FS.

To analyze the cyclotron resonance, we need an appropriate asymptotic expression for the conductivity components at large $q$, when the parameter $u$ takes values that are small compared to unity. At $u \ll 1$, the correction $\delta\psi(\epsilon)$ is nearly proportional to $\epsilon$, and we can use the approximation $\delta\psi(\epsilon) \approx -\mathrm{i}\epsilon\Omega/qv_z(p_z,\psi)$. To proceed, we expand the last term in the denominator in Eqn (39) in a series in $\delta\psi$, and keep the first two terms:

$$qv_z\big(p_z, \psi + \delta\psi(\epsilon)\big) \approx qv_z(p_z,\psi) - \mathrm{i}\epsilon\Omega\,\frac{\partial v_z(p_z,\psi)/\partial\psi}{v_z(p_z,\psi)} \equiv qv_z(p_z,\psi) - \mathrm{i}\epsilon\Delta(p_z,\psi)\,. \quad (40)$$

We then change variables in Eqn (39) and pass to integration in the velocity space. As before, we split the FS into segments such that a one-to-one correspondence between $\mathbf{v}$ and $\mathbf{p}$ holds over each segment.

The leading terms in the expansions of the conductivity components in powers of the small parameter $u$ are independent of the magnetic field and keep the same form as in its absence. For instance, the leading approximation for the component $\sigma_{xx}$ is described by expression (28). Also, in a metal with the FS whose curvature is everywhere finite and nonzero, the next terms in the expansions of the conductivity components appear to be of the order of $\sigma_0(q)u$ [93].

We now proceed with the analysis of the effect of the FS local geometry on the electron conductivity near the cyclotron resonance. First, we assume that the FS curvature vanishes at a whole effective line. We suppose that the curvature anomaly is attributed to the curvature radius directed perpendicularly to the effective line. This means that the FS includes a nearly cylindrical segment whose curvature can be approximated by Eqn (30). As a result, a special term emerges in the asymptotic expression for the conductivity at large $q$. To avoid tedious calculations, we use the model of an axially symmetric FS in computing this term.

Assuming that the magnetic field is directed along the symmetry axis, neither the velocity magnitude nor the FS curvature depend on the angle $\varphi$. Under these conditions, the conductivity tensor is diagonalized in circular components $\sigma_{\pm}(\omega, q) = \sigma_{xx}(\omega, q) \pm \mathrm{i}\sigma_{yx}(\omega, q)$. Using approximation (30) for the curvature [the factor $W(\theta, \varphi)$ now reduces to $W(\theta)$], we obtain the following contribution to the conductivity from the nearly cylindrical strip on the FS:

$$\sigma_{\pm}^{\mathrm{a}}(\omega, q) \approx \sigma_0 \eta \left[1 - \mathrm{i} \tan\left(\frac{\pi\beta}{2}\right)\right] (u\chi_{\pm})^{\beta} . \quad (41)$$

Here, $\chi_{\pm} = 1 \mp \Omega/\omega + \mathrm{i}/\omega\tau$ and $\eta$ is a dimensionless parameter whose value is determined by the relative number of the conduction electrons concentrated at the nearly cylindrical effective cross section. The term $\sigma_{\pm}^{\mathrm{a}}(\omega, q)$ can dominate if the parameter $\eta$ takes values of the order of unity. Otherwise, this term corresponds to the first correction to the leading approximation for the conductivity.

We remark that Eqn (41) gives a general expression for the contribution to the transverse conductivity originating from a nearly cylindrical effective strip at any axially symmetric FS, which is not associated with a particular form of the energy–momentum relation near the effective line. As shown above [see Eqn (35)], local flattenings of the FS also cause the occurrence of an 'anomalous' contribution to the electron conductivity at large $q$. In the presence of an external magnetic field, the conduction electrons associated with locally flat segments of the FS can give rise to resonance features in the surface impedance of a metal sample and its derivative with respect to the field magnitude [93]. Assuming that the relevant segment of the FS is shaped like a lens flattened near its vertices, as is determined by energy–momentum relation (33), we can express the correction $\sigma_{\mathrm{a}}(\omega, q)$ in the first approximation as

$$\sigma_{\mathrm{a}}(\omega, q, B) = \sigma_{\mathrm{a}}(\omega, q, 0) \times \sum_r \int_{-\infty}^{\infty} e^{\epsilon} \left(1 - \frac{r\Omega(\pi/2)}{\omega} + \frac{\mathrm{i}\Delta(\pi/2)\epsilon}{\omega} + \frac{\mathrm{i}}{\omega\tau}\right)^{\beta} \mathrm{d}\epsilon , \quad (42)$$

where $\sigma_{\mathrm{a}}(\omega, q, 0)$ is defined in (35).

The analysis in Refs [83, 93] revealed no qualitative difference between the expressions for the leading terms of the surface impedance computed for the axially symmetric FSs and those not having that symmetry, if **B** is directed along a high-order symmetry axis of the Fermi surface, such that the transverse conductivity is diagonalized in circular components. This gives grounds to expect the model of an axially symmetric FS to reflect the main features in the electronic response of a metal whose FS includes nearly cylindrical strips. We use this model, and we have

$$Z_{\pm} = \frac{8\mathrm{i}\omega}{c^2} \int_0^{\infty} \frac{\mathrm{d}q}{4\pi\mathrm{i}\omega\sigma_{\pm}/c^2 - q^2} . \quad (43)$$

We consider the case where some effective cross sections of the FS are zero-curvature lines: this can be the single central nearly cylindrical cross section; otherwise, the effective cross sections are combined in pairs, which are arranged symmetrically with respect to the plane $p_z = 0$. The contribution of the nearly cylindrical cross sections to the surface impedance depends on the relative number of the effective electrons associated with them. When a considerable part of the conduction electrons is concentrated in the nearly cylindrical effective segments of the FS ($\eta \sim 1$), the anomalous contribution to the conductivity $\sigma_{\mathrm{a}}$ can be given by the leading term in the expansion of the conductivity in powers of $u$, and it strongly contributes to the surface impedance. In this case, the leading term of the impedance is

$$Z_{\pm} = Z_0 \zeta(\beta) \left(\frac{\omega}{\omega_0}\right)^{2/(3+\beta)} \chi_{\mp}^{-\beta/(\beta+3)} . \quad (44)$$

This expression shows a very peculiar dependence of the surface impedance on the magnetic field. Near the cyclotron frequency, the real ($R$) and imaginary ($H$) parts of the component $Z_-$ rapidly increase. The increase in $R(B)$ is no smaller in order than $R(0)$ at the same frequency. In strong magnetic fields ($\omega \ll \Omega$), the value of $R(B)$ increases proportionally to $(\Omega/\omega)^{-\beta/(\beta+3)}$. When the parameter $\beta$ takes values close to zero, $R(B)$ almost levels off in the presence of strong magnetic fields, whereas in the limit $\beta \to -1$, $R(B) \sim \sqrt{\Omega/\omega}$. No experimental observations of such impedance behavior have been reported to date. The reason is that we can hardly expect the parameter $\eta$ to take values of the order of unity in real metals. We must rather expect that the relative number of electrons associated with the nearly cylindrical cross section is small, $\eta \ll 1$. Then the first correction to the surface impedance is described by the expression

$$\Delta Z_{\pm} = -Z_0 \eta(\beta) \left(\frac{\omega}{\omega_0}\right)^{2(\beta+1)/3} \chi_{\mp}^{\beta} . \quad (45)$$

This correction describes resonance features at the cyclotron frequency in both real and imaginary parts of the impedance component $Z_-$ (Fig. 6a). The shape of the peak in the real part of the impedance resembles that recorded in potassium [84]. The peak height depends on the value of $\eta$. For $\eta \sim 10^{-2}$, $\omega\tau \sim 10$, and $\xi^3 \sim 10^4$, the resonance amplitude in Re $Z$ is approximately $10^{-2}$ of the leading term, which agrees with the experiments in Ref. [84], as well as with experiments on organic metals [94, 95]. Even better agree-ment with the experimental result reported in Ref. [84] can be achieved under the assumption that the efficient electrons in potassium are associated with the vicinities of inflection lines separating the main, nearly spherical body of the FS from bulbs located near the boundaries of the Brillouin zone [93]. Within the nearly free electron approximation appropriate for alkali metals, the FS is flattened near the inflection lines. Due to the polycrystallinity of the potassium sample used in the experiments, the FS flattening points are quite widely distributed and fall on the cyclotron orbits of the efficient electrons over a rather broad interval of the magnetic field orientations. Using Eqn (42), it can be shown that these small flattened pieces on an effective strip of the FS can yield an additional term to the expression for the surface impedance [92], similarly to the case where one of the effective strips on the FS is nearly cylindrical. The emergence of this contribution is caused by the increase in the number of efficient electrons associated with the effective strip due to the presence of flattened segments of the FS included there. Both the height and shape of the resonance feature can be brought into good quantitative agreement with the experimental result, as is demonstrated in Fig. 6b.

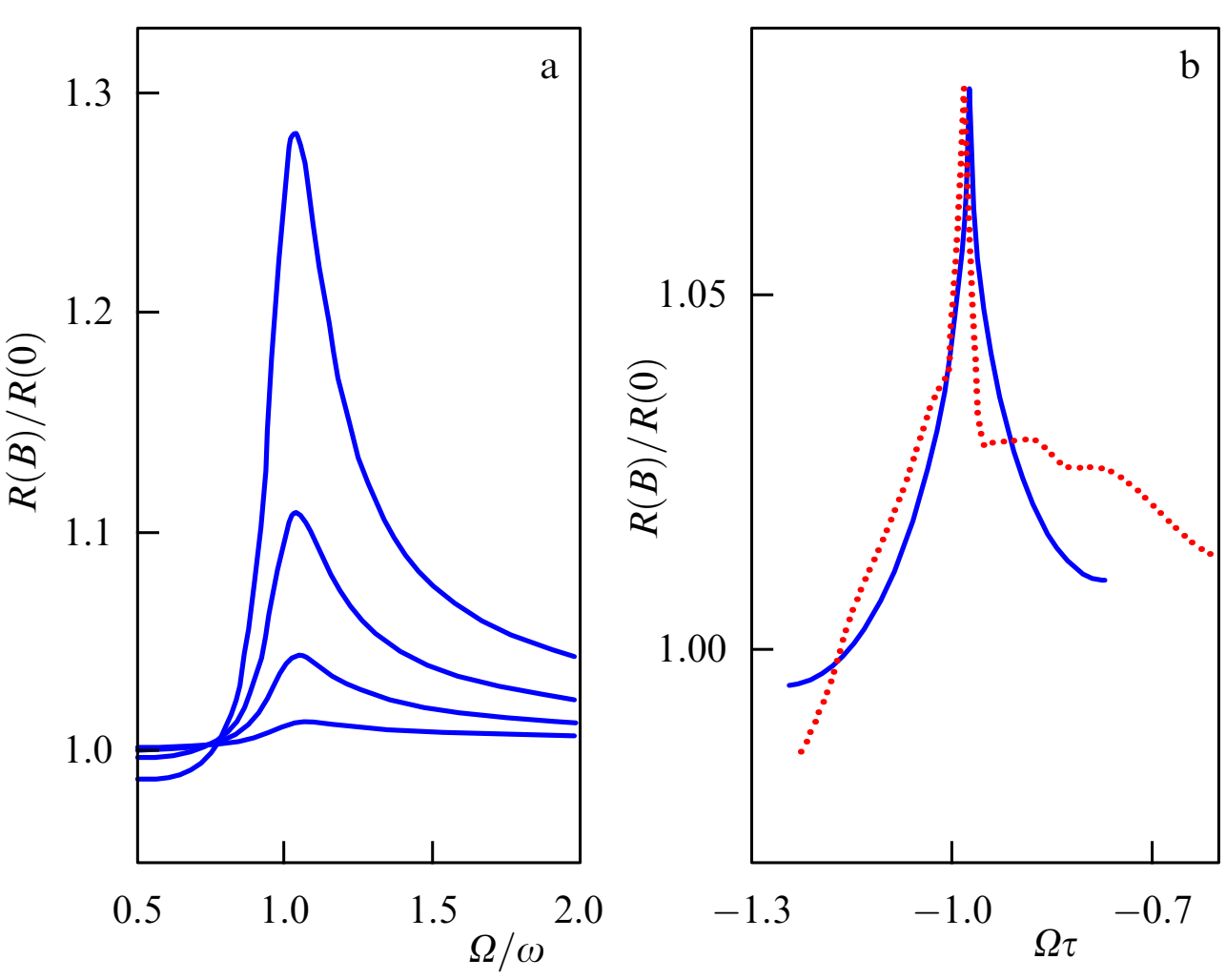

**Figure 6.** Magnetic field dependences of the real part of the surface impedance near the cyclotron resonance. (a) The curves are plotted assuming that the FS has the axial symmetry and the magnetic field is directed along the symmetry axis for $\omega\tau = 10$, $\xi^3 = 10^4$, $\beta = -0.8, -0.7, -0.6, -0.4$ (from top down) and $\eta = 0.01$. (b) The curves represent the experimental result reported for the real part of surface impedance in potassium [84] (dotted line) compared to the theory proposed in Ref. [92].

The extent to which locally flat regions of the FS can affect the surface impedance is determined by the relative number of the conduction electrons associated with locally flat regions and the FS shape in the vicinity of flattening points. We analyze these effects using energy–momentum relation (33). It is worth mentioning that electron lenses similar to the FS corresponding to this relation are incorporated in the FSs of cadmium and zinc, and the cyclotron resonance in a normal magnetic field was observed in both metals. When the parameter $k$ in Eqn (33), which characterizes the shape of the lens near its vertices, considerably differs from unity and $\eta$ is not too small, the resonance contribution to the surface impedance can be significant, which leads to the appearance of noticeable resonance-type features in the frequency/field dependences of those features. When $k > 3/2$, the resonance peak can be observed in the impedance itself, while for a less pronounced flattening of the FS ($1 < k < 3/2$), the resonance singularity is manifested only in the field and frequency dependences of the impedance derivative with respect to the magnetic field amplitude.

The derivative of the real part of the contribution to the surface impedance from locally flat neighborhoods of the lens vertices has the form [93]

$$\frac{\mathrm{d}R_-}{\mathrm{d}B} = a(k)\; Y_k(\chi_+)\,, \tag{46}$$

where $a(k)$ is a dimensionless coefficient and the function $Y_k(\chi_+)$ describes the shape of the resonance curve. The top panels in Fig. 7 display the field dependence of the function $Y_k(\chi_+)$ for several values of the parameter $k$. A comparison with the recordings of $\mathrm{d}R/\mathrm{d}B$ obtained for cadmium and zinc and reported in Refs [85, 86], which are shown in the bottom panels, indicates that the shape of the resonance lines is in fairly good agreement with the experimental results. Variations of the amplitude and shape of the resonance lines with decreasing $k$ resemble the variations of the experimental resonance lines upon an increase in the angle $\Phi$ between the magnetic field and the $[11\bar{2}0]$ axis of the crystalline lattice of these metals. The latter lies in the plane perpendicular to the symmetry axis of the electron lens included in the FSs of cadmium and zinc. The above similarity occurs because as the magnetic field deviates from the $[11\bar{2}0]$ axis, the effective cross section of the electron lens no longer passes through the flattening points at the vertices of the lens, but still remains quite close to them for small deviation angles. It can therefore be assumed that if the angle $\Phi$ does not exceed a certain critical value $\Phi_0$ characterizing the size of the flattened region on the FS, then the parameter $k$ decreases with increasing $\Phi$, but remains greater than unity. The value $k = 1$ corresponds to angles $\Phi \geqslant \Phi_0$, when the cyclotron orbit of effective electrons no longer passes through locally flat regions on the FS.

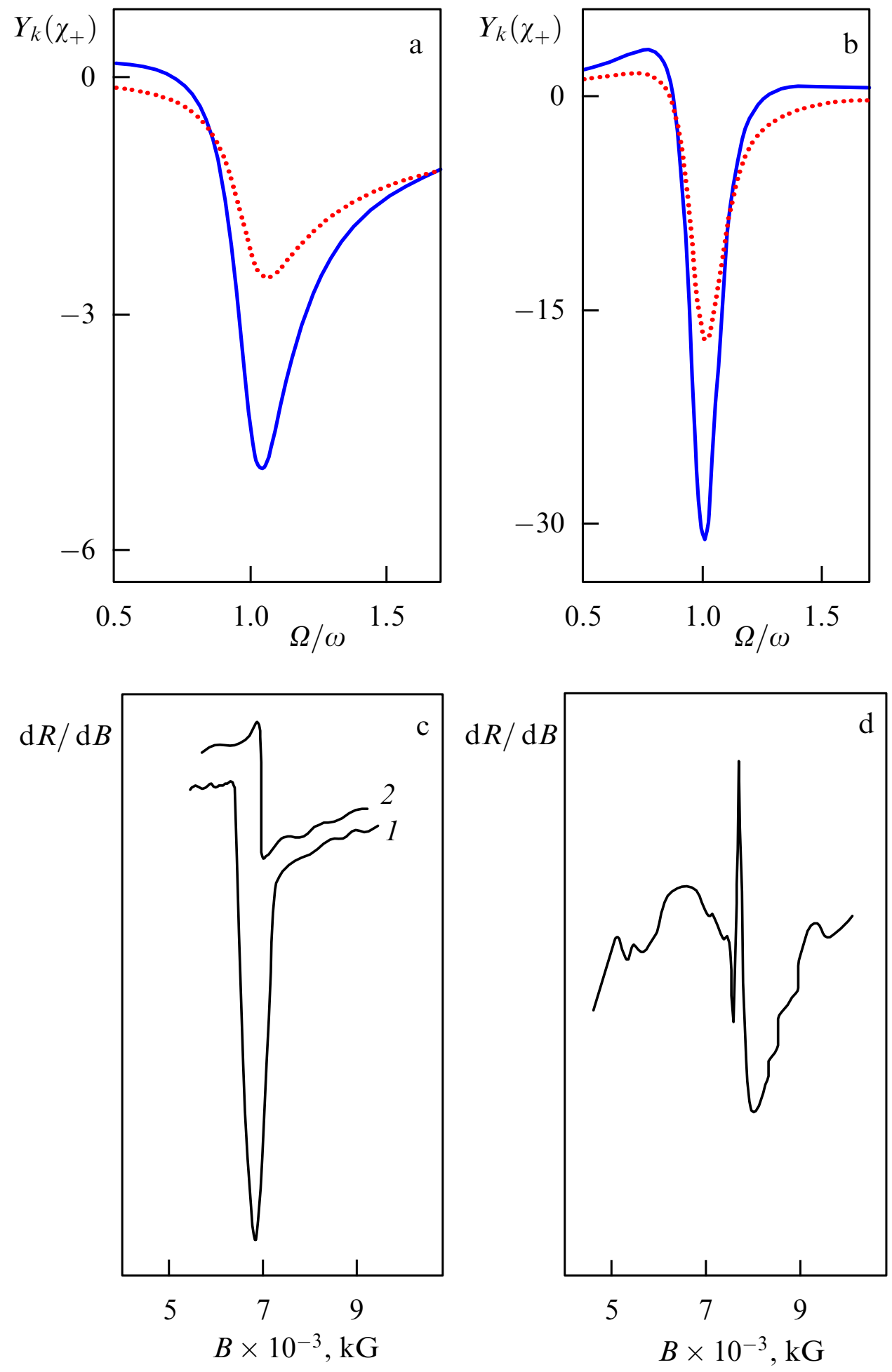

**Figure 7.** (a, b) Plots of the function $Y_k(\chi_+)$ vs the magnetic field for (a) moderate and (b) strong flattening of the FS near the vertices of a lens determined by Eqn (33). Curves are plotted assuming that $\Delta(\pi/2, 0) = 0$, $\omega\tau = 10$ for $k = 1.2$ and 1.4 (fig. a, dashed line and solid lines, respectively) and $k = 1.6$ and 2.0 (fig. b, dashed line and solid line, respectively). (c, d) Resonance features in the magnetic field dependences of $\mathrm{d}R/\mathrm{d}B$ for (c) cadmium and (d) zinc observed in the experiments in [85, 86]. The magnetic field is directed along the $[11\bar{2}0]$ axis coinciding with the symmetry axis of the electron lens located in the third Brillouin zone (fig. c, curve *1*) and tilted to that axis at the angle $\Phi = 8°$ (curve *2*). The recording of curve *2* is amplified as 10:1 with respect to that of curve *1*.

Our analysis thus shows that the local geometry of the FS can give rise to cyclotron resonance in the normal magnetic field in good metals. A simple explanation of the effect is that the contributions from flattened or nearly cylindrical segments of the FS to the electron DOS can be significantly greater than those from the rest of the surface. Therefore, the number of effective electrons moving inside the skin layer increases, which provides favorable conditions for the resonance to occur. Because inflection lines as well as points of flattening exist on the FSs of most metals, we can expect cyclotron resonance in a normal magnetic field to be manifested there for suitable magnetic field directions. There is experimental evidence that 'necks' connecting quasispherical pieces of the FS of copper include nearly cylindrical belts [5]. When the magnetic field is directed along the axis of a neck (for instance, along the [111] direction in the quasimomentum space), the extremal cross section of the neck can be expected to run along the nearly cylindrical strip where the FS curvature vanishes. It is also likely that the FS of gold has the same geometric features, because it closely resembles that of copper. As was already mentioned, the cyclotron resonance in the normal field was observed in organic metals of the α-BEDT-$TTF_2MHg(SCN)_4$ group and some other layered conductors. Hence, the cyclotron resonance in a normal magnetic field may occur in conventional metals and Q2D conductors, and the nature and origin of this effect in both kinds of materials are identical.

### 3.3 Fermi-liquid and Fermi surface geometry effects in the propagation of low-frequency electromagnetic waves through thin metal films

As discussed above, electromagnetic waves incident on the surface of a metal cannot penetrate inside the metal deeper than a thin surface layer (skin layer). This happens due to the damping effect of conduction electrons absorbing the wave energy via the dissipationless Landau damping mechanism [4]. A strong magnetic field $\mathbf{B} = (0, 0, B)$ applied to the metal restricts the motion of electrons in the $x, y$ plane, creating 'transparency windows.' These windows are regions in the $q, \omega$ plane where the Landau damping cannot appear. As a result, in the presence of an external magnetic field, various weakly attenuated electromagnetic waves, such as helicoidal, cyclotron, and magnetohydrodynamic waves, can propagate in the electron liquid of a metal [51, 96].

Fermi-liquid correlations of conduction electrons induce changes in the wave spectra. Also, new collective modes can appear in metals due to FL interactions between the electrons. These modes solely occur owing to the FL interactions, and are absent in a gas of charge carriers. One such mode is the FL cyclotron wave first predicted by Silin [64] and observed in alkali metals [53]. In a metal with a nearly spherical FS, this mode is a transverse circularly polarized wave propagating along the external magnetic field whose dispersion in the collisionless limit ($\tau \to \infty$) has the form [97]

$$\frac{\omega}{\omega_0} = 1 + \frac{8}{35}\frac{1}{\alpha}(qR)^2 , \qquad (47)$$

where $R = v_0/\Omega$, $v_0$ is the maximum value of the electron velocity component along the magnetic field (for the spherical FS, $v_0$ equals the Fermi velocity $v_F$), and the dimensionless parameter $\alpha$ characterizes the FL interactions of conduction electrons. The difference between the frequency $\omega_0 = \omega(0)$ and the cyclotron frequency is determined by the value of the FL parameter $\alpha$, $\omega_0 = \Omega(1 + \alpha)$. Depending on whether $\alpha$ takes a positive/negative value, $\omega_0$ is greater/less than $\Omega$. In what follows, we assume for definiteness that $\alpha < 0$. When $qR \ll 1$, the dispersion curve of this FL cyclotron wave is situated in the transparency window whose boundary is given by the relation $\omega = \Omega - qv_0$, which corresponds to the Doppler-shifted cyclotron resonance for the conduction electrons. This is shown in Fig. 8a. But the dispersion curve meets the boundary of the transparency region at $q = q_m \approx 5|\alpha|/3R$ [97], and at this value of $q$ it terminates [98]. Therefore, for reasonably weak FL interactions ($|\alpha| \sim 0.1$), the FL cyclotron wave can appear only at $qR \ll 1$, and its frequency remains close to the cyclotron frequency for the whole spectrum [99]. Similar conclusions were made using some other models to mimic the FS shape, such as an ellipsoid, a nearly ellipsoidal surface, and a lens made out of two spherical segments [100, 101].

It is clear that the leading contribution to the formation of a weakly attenuated collective mode near the boundary of the transparency region comes from those electrons that move with the greatest possible speed along the magnetic field $\mathbf{B}$. The greater the relative number of such electrons is, the more favorable conditions develop for the wave to emerge and to exist at comparatively low frequencies $\tau^{-1} \ll \omega \ll \Omega$. The relative number of such 'efficient' electrons is determined by the FS shape, and the best conditions are reached when the FS includes a lens made out of two paraboloidal cups. This model was used in some previous works to study transverse collective modes occurring in a gas of charge carriers near the Doppler-shifted cyclotron resonance, which are known as dopplerons [102–104].

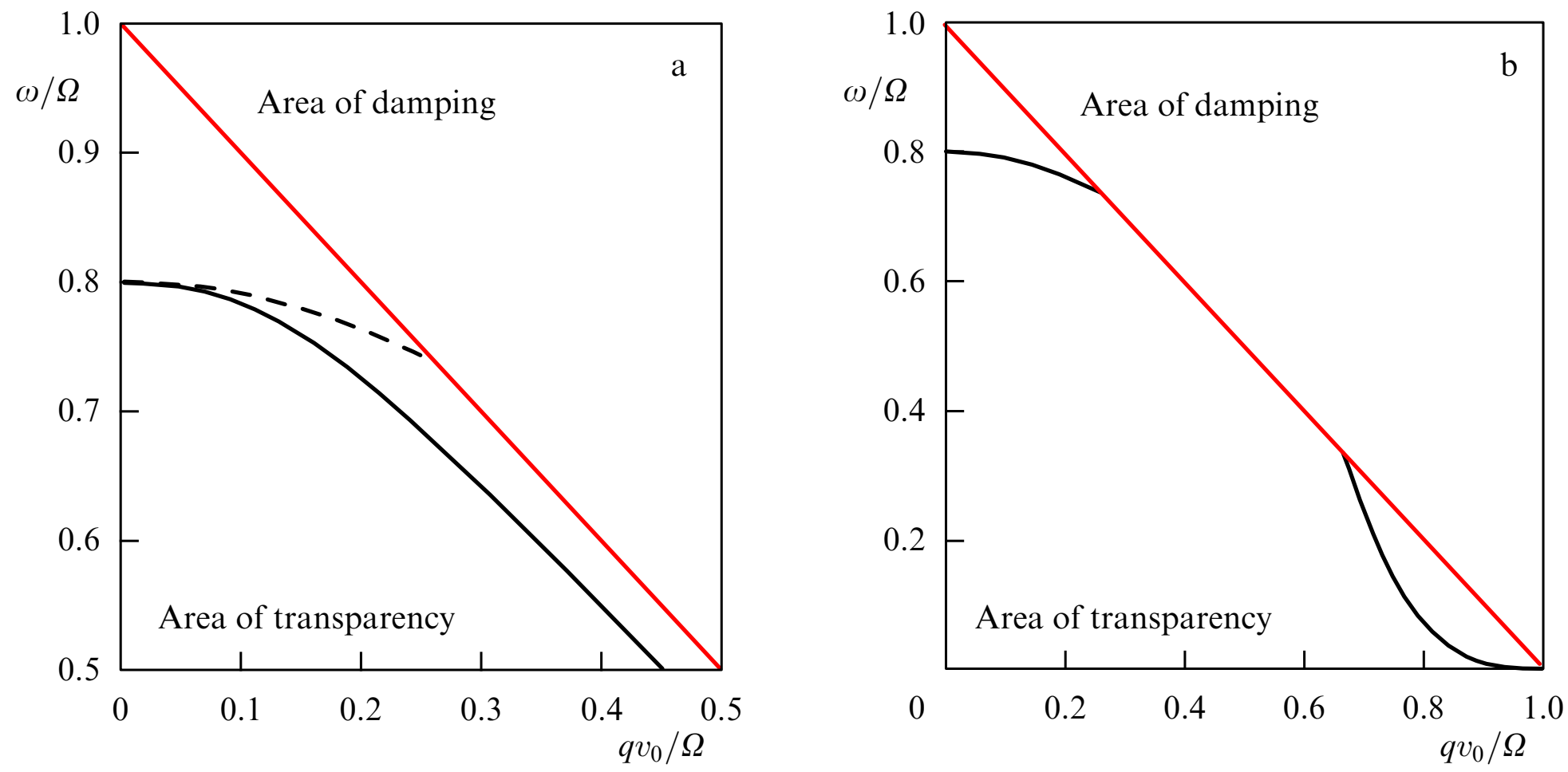

**Figure 8.** (a) Dispersion of the transverse FL cyclotron wave traveling along the external magnetic field for spherical (dashed-dotted line) and paraboloidal (solid line) FSs. The curves are plotted using Eqns (47) and (48) with $\alpha = -0.2$. (b) A schematic plot of the dispersion of the transverse FL mode in a metal whose FS includes nearly paraboloidal segments. The low-frequency ($\omega \ll \Omega$) branch is shown along with the cyclotron wave. For both panels, the straight line corresponds to the Doppler-shifted cyclotron resonance.

It was shown in [66] that for negative values of the FL parameter $\alpha$, the dispersion of the transverse FL wave propagating along the magnetic field is ($\tau \to \infty$)

$$\frac{\omega}{\Omega} = 1 - \frac{1}{2}\big(qR + |\alpha|\big) - \frac{1}{2}\sqrt{\big(qR - |\alpha|\big)^2 + \frac{4}{3}\,\frac{|\alpha|(qR)^2}{qR + \sqrt{(qR)^2 + |\alpha|^2}}}\,. \qquad (48)$$

This result shows that for a paraboloidal FS, there are no bounds on the frequency of the FL cyclotron wave in the collisionless limit (Fig. 8a). The only restriction on the wave frequency is due to the increase in the wave attenuation due to electron scattering. Taking the electron scattering into account, we can prove that the wave is weakly attenuated up to a magnitude of the wave vector of the order of $\Omega(1 - 1/|\alpha|\Omega\tau)/v_0$. This value can be significantly greater than the value $q_{\rm m}$ for a spherical FS. Therefore, the frequency of FL cyclotron waves for negative $\alpha$ can be much smaller than $\Omega$ (remaining greater than $1/\tau$). Comparing the dispersion curves of the transverse FL cyclotron wave for spherical and paraboloidal FSs, we see that the FS geometry strongly affects the wave dispersion, and it can provide a weak attenuation of this mode at moderately low frequencies $\omega \ll \Omega$. We next consider favorable conditions for the occurrence of these low-frequency modes in practical metals.

We restrict our consideration to the case of an axially symmetric FS whose symmetry axis is parallel to the magnetic field. The response of the electron liquid of the metal to an electromagnetic disturbance can then be expressed in terms of the electron conductivity circular components. Using the main equations of the FL theory and suitable approximations for FL kernel (24), we can derive the following expressions for the conductivity components [66, 83]:

$$\sigma_{\pm} = \frac{2ie^2 A(0)}{(2\pi\hbar)^3 q} \times \frac{\Phi_0^{\pm}\left(1 - \dfrac{\alpha_2 u}{Q_2}\,\Phi_2^{\pm}\right) + \dfrac{\alpha_2 u}{Q_2}(\Phi_1^{\pm})^2}{\left(1 - \dfrac{\alpha_1 u}{Q_0}\,\Phi_0^{\pm}\right)\left(1 - \dfrac{\alpha_2 u}{Q_2}\,\Phi_2^{\pm}\right) + \dfrac{\alpha_1\alpha_2}{Q_0 Q_2}\,u^2(\Phi_1^{\pm})^2}\,. \qquad (49)$$

Here,

$$\Phi_n^{\pm} = \int_{-1}^{1} \frac{\bar a(x)\bar m_{\perp}(x)x^n\,{\rm d}x}{u\chi_{\pm} \mp \bar v(x)}\,, \qquad Q_n = \int_{-1}^{1} \bar a(x)\bar m_{\perp}(x)x^n\,{\rm d}x\,, \qquad (50)$$

$$\bar a(x) = \frac{A(x)}{A(0)}\,, \qquad \bar v(x) = \frac{v_z}{v_0}\,, \qquad \bar m_{\perp}(x) = \frac{m_{\perp}(x)}{m_{\perp}(0)}\,, \qquad x = \frac{p_z}{p_0}\,, \qquad (51)$$

$$\chi_{\pm} = 1 \mp \frac{\Omega}{\omega} + \frac{i}{\omega\tau}\,, \qquad u = \frac{\omega}{q v_0}\,,$$

where $p_0$ is the maximum value of the longitudinal component of the electron quasimomentum and $A(x)$ is the FS cross-sectional area. The dimensionless factors $\alpha_{1,2}$ in Eqn (49) are given by

$$\alpha_{1,2} = \frac{f_{1,2}}{1 + f_{1,2}}\,, \qquad (52)$$

and $f_{1,2}$ are related to the FL parameters $\varphi_{10}$ and $\varphi_{11}$.

Presently, we are interested in the transverse waves propagating along the magnetic field. The corresponding dispersion equation has the form

$$c^2 q^2 - 4\pi i\omega\sigma_{\pm}(\omega, \mathbf{q}) = 0\,. \qquad (53)$$

For the electron FL, this equation for '$-$' polarization has solutions corresponding to helicoidal waves and the transverse FL waves traveling along the magnetic field. When the relevant charge carriers are holes, the '$+$' polarization is to be chosen in Eqn (53).

In considering FL waves, we can simplify dispersion equation (53) by omitting the first term. Also, we can neglect corrections of the order of $c^2q^2/\omega_{\rm p}^2$ ($\omega_{\rm p}$ is the electron plasma frequency) in the expression for the conductivity. Then the FL parameter $\alpha_1$ drops out from the dispersion equation, and that equation becomes

$$\varDelta(u) = \frac{1}{\alpha_2}\,, \qquad (54)$$

where $\varDelta(u) = (u/Q_2)\big[\Phi_2^- - (\Phi_1^-)^2/\Phi_0^-\big]$.

Assuming the mass $m_{\perp}$ to be the same over the whole FS, we expand the integrals $\Phi_n^-$ in powers of $u^{-1}$. Keeping terms of the order of $u^{-2}$, we obtain the dispersion relation for the cyclotron mode at small $q$ ($u \gg 1$):

$$\omega = \Omega(1 + f_2)\left[1 + \frac{b^2}{f_2}\left(\frac{q v_0}{\Omega}\right)^2\right], \qquad (55)$$

where $b^2$ is a dimensionless positive constant of the order of unity. For an isotropic electron liquid, $b^2 = 8/35$ and expression (55) coincides with expression (47) where $\alpha = f_2$.

We next analyze the possibilities of the low-frequency ($\tau^{-1} \ll \omega \ll \Omega$) transverse FL mode emerging in realistic metals where the cyclotron mass depends on $p_z$. Such waves can appear near the Doppler-shifted cyclotron resonance. Assuming $\alpha_2 < 0$ and $\omega \ll \Omega$, we can describe the relevant boundary of the transparency region by the following equations, applicable for small $\omega$:

$$\Omega(p_z)\left(1 + \frac{cq}{2\pi|e|B}\,\frac{{\rm d}A}{{\rm d}p_z}\right) = 0\,, \qquad (56)$$

$$\frac{{\rm d}\Omega}{{\rm d}p_z}\left(1 + \frac{cq}{2\pi|e|B}\,\frac{{\rm d}A}{{\rm d}p_z}\right) + \frac{\Omega(p_z)cq}{2\pi|e|B}\,\frac{{\rm d}^2A}{{\rm d}p_z^2} = 0\,. \qquad (57)$$

It follows from these equations that the attenuation at the boundary for $\omega \ll \Omega$ is carried out by electrons belonging to neighborhoods of particular cross sections on the FS where extrema of ${\rm d}A/{\rm d}p_z$ are reached. These can be neighborhoods of limit points or inflection lines, as shown in Fig. 9.

In general, to study various effects in the response of the electron liquid of metal near the Doppler-shifted cyclotron resonance, we must take contributions from all segments of the FS into account; therefore, the expressions for conductivity components (49) are to be correspondingly generalized. However, in the considered case, it is possible to separate that particular segment of the FS where electrons producing the low-frequency FL wave belong. The contribution from the rest of the FS is small, and we can omit it, as is shown in Ref. [66].

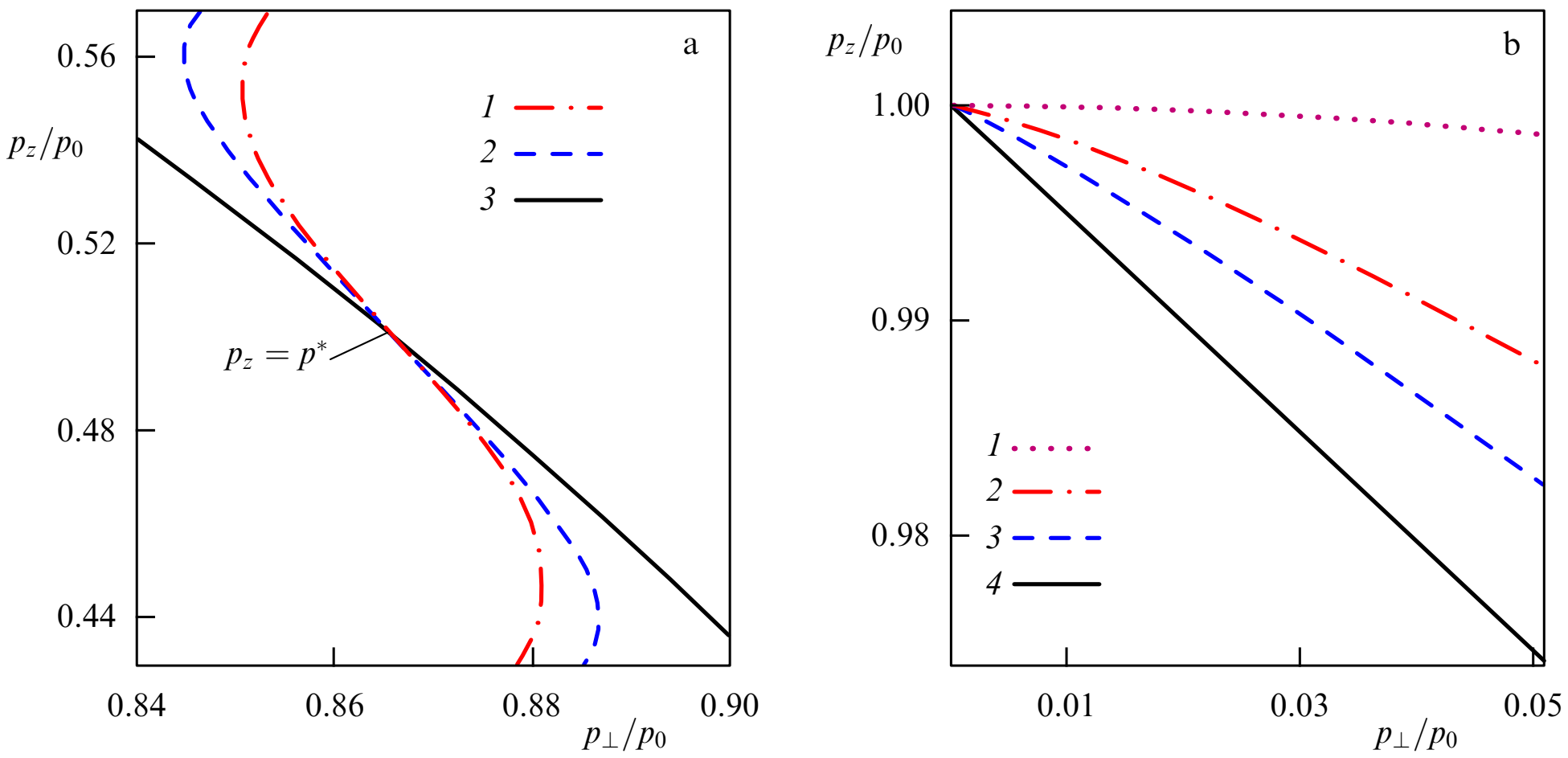

**Figure 9.** Schematic plots of the FS profiles in the vicinities of (a) inflection lines and (b) vertices. (a) The profiles are drawn in accordance with Eqn (58) assuming $p^* = 0.5p_0$, $|\mathrm{d}^s\bar{a}/\mathrm{d}x^s|_{x=x^*} = |\mathrm{d}\bar{a}/\mathrm{d}x|_{x=x^*} = 1$, and $s = 5$ (curve *2*), $s = 3$ (curve *1*). Curve *3* corresponds to a paraboloidal strip on the FS near $x^* = 0.5$ ($s \to \infty$). (b) The curves are plotted assuming $x^* = 1$ and $\bar{a}(1) = 0$. Curves *1* and *4* respectively correspond to spherical and paraboloidal FSs; curves *2* and *3* respectively represent nearly paraboloidal FSs with $s = 7$ and 9.

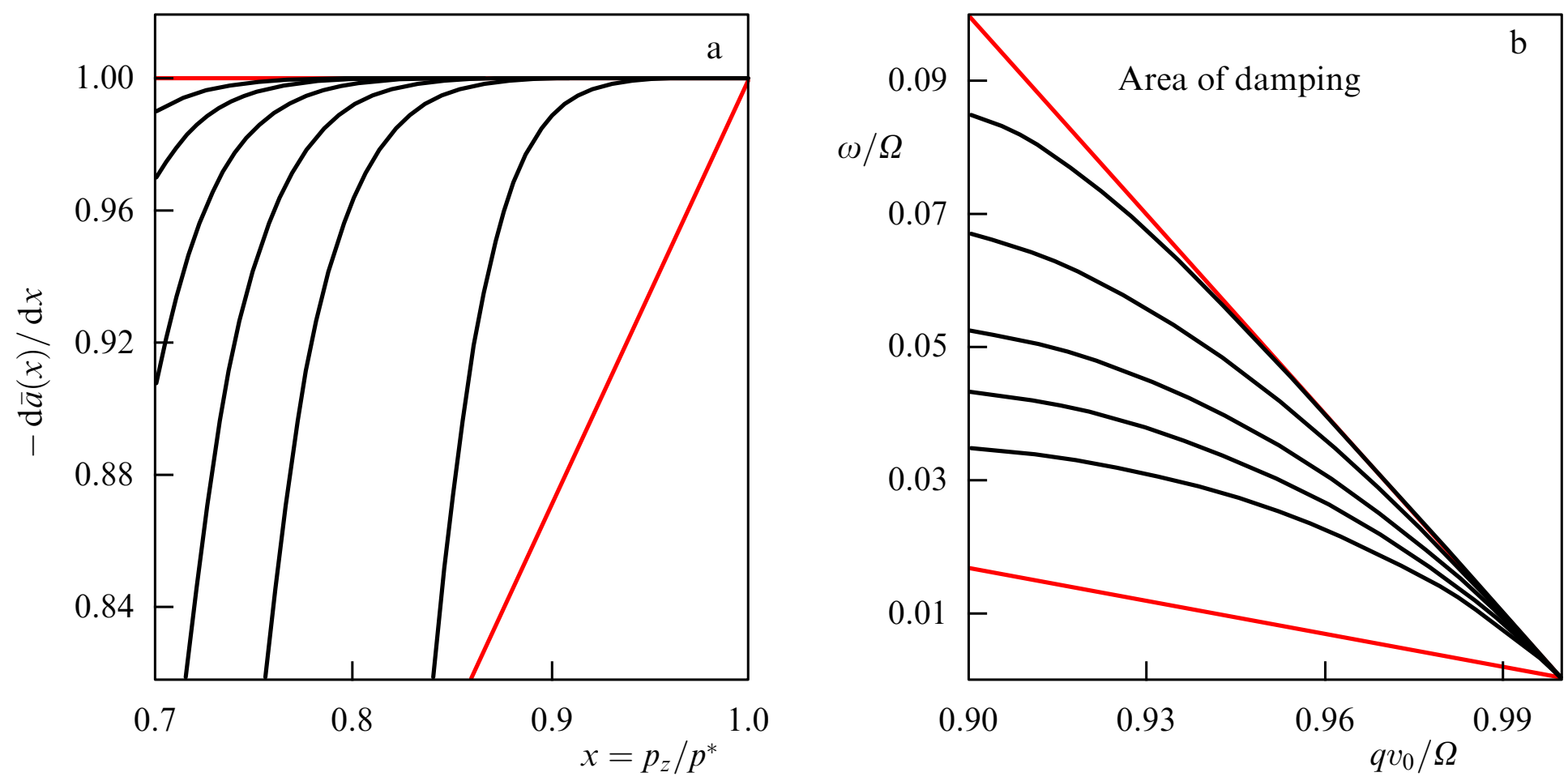

**Figure 10.** (a) Dependences of $\mathrm{d}\bar{a}/\mathrm{d}x$ on $x$ near the inflection line on the FS at $x = x^*$. The curves are plotted for $s = 4, 5, 6, 7, 8$, and 9 (from right to left). (b) Dispersion curves of the low-frequency transverse FL waves. The curves are plotted at $\alpha_2 = -0.2$; $s = 4, 5, 6, 7$, and 8, and $s \to \infty$ (from top down) in the collisionless limit assuming that $|\mathrm{d}^s\bar{a}/\mathrm{d}\bar{x}^s|_{x=x^*} = |\mathrm{d}\bar{a}/\mathrm{d}\bar{x}|_{x=x^*} = 1$.

In the analysis in what follows, we can therefore use dispersion equation (54) with the integrals $\Phi_n^{\pm}$ calculated for the appropriate segment of the FS. It follows from this equation that the dispersion curve of the cyclotron wave does not intersect the boundary of the transparency region if the function $\Delta(u)$ diverges there. A similar analysis was carried out in the theory of dopplerons [104]. It was proved that when the appropriate component of the conductivity [a $\Phi_0(u)$ integral] tends to infinity at the Doppler-shifted cyclotron resonance, the doppleron propagates without damping in a broad frequency range.

In what follows, we assume for definiteness that the extrema of $\mathrm{d}A/\mathrm{d}p_z$ are reached at the inflection lines $p_z = \pm p^*$. In the vicinities of these lines, we can use the approximation

$$\bar{a}(x) \approx \bar{a}(x^*) + \frac{\mathrm{d}\bar{a}}{\mathrm{d}x}\bigg|_{x=x^*}(x \mp x^*) \pm \frac{1}{s!}\frac{\mathrm{d}^s\bar{a}}{\mathrm{d}x^s}\bigg|_{x=x^*}(x \mp x^*)^s\,, \tag{58}$$

where $x^* = p_z/p^*$ and the parameter $s \geqslant 3$ characterizes the FS shape near the inflection lines at $x = \pm x^*$. The greater the value of $s$ is, the closer the FS near $p_z = \pm p^*$ is to a paraboloid (see Fig. 9). When $s = 2$, the FS has a spherical/ellipsoidal shape.

The dependences of the derivative $\mathrm{d}\bar{a}/\mathrm{d}x$ on $x$ near $x = x^*$ are presented in Fig. 10a. We can see that the greater the shape parameter $s$ is, the broader nearly paraboloidal strips are in the vicinities of the FS inflection lines. Consequently, the greater number of conduction electrons is associated with the nearly paraboloidal parts of the FS. This creates more favorable conditions for the wave to occur. A similar analysis can be carried out in the case where $\mathrm{d}A/\mathrm{d}p_z$ reaches its extremal values at the vertices of the FS. Again, to provide the emergence of the transverse low-frequency FL mode, the FS near $p_z = \pm p_0$ must be nearly paraboloidal in shape.

Using asymptotic expression (58), we can calculate the leading term of the function $\Delta(u)$. This term diverges at the boundary of the transparency region if $s \geqslant 3$. The low-frequency solutions of the dispersion equation in the collisionless limit are plotted in Fig. 10b. All dispersion curves are located between the boundary of the transparency window and the line corresponding to the limit $s \to \infty$ (a paraboloidal FS). The greater the value of $s$ is, the closer the dispersion curve is to this line. We see that the low-frequency ($\omega \ll \Omega$) transverse FL wave can appear in a metal exposed to a strong ($\Omega\tau \gg 1$) magnetic field. This can happen when the FS is close to a paraboloid near those cross sections where $\mathrm{d}A/\mathrm{d}p_z$ reaches its maxima/minima. Therefore, the possibility of this wave propagating in a metal is provided by the local geometry of the FS near its inflection lines or vertices.

When $\Omega$ depends on $p_z$ and $\omega$ increases, electrons associated with various cross sections of the FS participate in the formation of the wave. To ensure the divergence of the function $\Delta(u)$ near the Doppler-shifted cyclotron resonance, we have to require not only that narrow strips near lines of inflection or vicinities of limiting points be nearly paraboloidal but also that large segments of the FS be nearly so. This condition is too stringent for FSs of real metals. We can therefore expect that the dispersion curve of the low-frequency transverse FL wave intersects the boundary of the transparency region at rather small $\omega$, as shown in the right panel of Fig. 8.

To clarify possible manifestations of the considered FL wave in experiments, we calculate the contributions of these waves to the transmission coefficient of a metal film. We assume that the film occupies the region $0 \leqslant z \leqslant L$ in the presence of an applied magnetic field directed along the normal to the interfaces. An incident electromagnetic wave with the electric and magnetic components $\mathbf{E}(z,t)$ and $\mathbf{b}(z,t)$ propagates along the normal to the film. We also assume that the symmetry axis of the FS is parallel to the magnetic field ($z$ axis) and the interfaces reflect the conduction electrons in a similar manner. The Maxwell equations inside the metal then reduce to decoupled equations for circular components of the electric field $E_\pm(z)\exp(-\mathrm{i}\omega t)$. Expanding the incident electric field $E_\pm(z)$ and current density $j_\pm(z)$ inside the film in Fourier series, we arrive at the equations for the Fourier transforms

$$-\frac{c^2 q_n^2}{4\pi\mathrm{i}\omega} E_n^\pm + j_n^\pm = \mp\frac{\mathrm{i}c}{4\pi}\left[(-1)^n b_\pm(L) - b_\pm(0)\right], \qquad (59)$$

where $b_\pm(z)$ are the magnitudes of the magnetic component of the incident field and $q_n = \pi n/L$. Due to the high density of conduction electrons in good metals, the skin effect has a strongly anomalous character over the whole frequency range of the low-frequency FL mode $|\alpha_2|\tau^{-1} \ll \omega < \Omega$. We can therefore disregard the effects originating from the diffuse scattering of conduction electrons from the surfaces of the film as long as the film thickness is not too small. As is estimated in Ref. [105], $L$ must take values of the order of 100 μm or greater for the effects of the surface roughness on the transmission of the electromagnetic waves through the film to be negligible. For thinner metal films, the surface roughness can bring noticeable changes to the transmission. A detailed consideration of this case is given in Ref. [105] and in book [40]. Here, we assume the film surfaces to be sufficiently smooth, such that we can treat the electron reflection from the metal film surfaces as nearly specular.

Substituting the expressions for $j_n^\pm$ in Eqn (59) yields

$$E_n^\pm = \mp\frac{\omega}{c} F_\pm(\omega, q_n)\left[(-1)^n b_\pm(L) - b_\pm(0)\right], \qquad (60)$$

where we introduce the notation

$$F_\pm(\omega, q_n) = \left(q_n^2 - \frac{4\pi\mathrm{i}\omega}{c^2}\sigma_n^\pm\right)^{-1}. \qquad (61)$$

To obtain the expression for the transmission coefficient, which is determined by the ratio of the amplitudes of the transmitted field ($E_\mathrm{t}$) at $z = L$ and the incident field ($E_\mathrm{i}$) at $z = 0$, we use the Maxwell boundary conditions, which gives

$$T_\pm = \left|\frac{E_\mathrm{t}^\pm}{E_\mathrm{i}^\pm}\right| = \left|\frac{E_\pm(L) + b_\pm(L)}{E_\pm(0) + b_\pm(0)}\right|. \qquad (62)$$

Assuming that the transmission is small ($T_\pm \ll 1$), we obtain the asymptotic expression [106]

$$\frac{E_\mathrm{t}^\pm}{E_\mathrm{i}^\pm} \approx \frac{c}{4\pi} Z_\pm^{(1)}, \qquad (63)$$

where

$$Z_\pm^{(1)} = \pm\frac{8\pi\omega}{Lc^2}\sum_{n=0}\left(1 - \frac{1}{2}\delta_{n0}\right)(-1)^n F_\pm(\omega, q_n). \qquad (64)$$

Keeping the '$-$' polarization and using the Poisson summation formula, we obtain

$$T = \frac{2}{\pi}\frac{\omega}{c}\int_{-\infty}^{\infty} \operatorname{sign}(q)\operatorname{cosec}(Lq)\, F_-(\omega, q)\,\mathrm{d}q, \qquad (65)$$

where $\operatorname{sign}(q) = |q|/q$ is the sign function. An important contribution to integral (65) comes from the poles of the function $F_-(\omega, q)$, which are the roots of dispersion equation (53) for the relevant polarization. The contribution from the considered low-frequency mode to the transmission coefficient is equal to the residue at the appropriate pole of the integrand in (65). Size oscillations of the transmission caused by the low-frequency FL mode are shown in Fig. 11. For $B = 5$ T and $s = 3$ [$s$ is the shape parameter in Eqn (58)], the oscillations take peak values $\sim 10^{-8} - 10^{-9}$.

The values of that order can be measured in experiments on the transmission of electromagnetic waves through thin metal films. However, the oscillation magnitudes can reach significantly greater values when the shape parameter increases. As can be seen from Fig. 11, $T$ can reach values of the order of $10^{-6}$ when $s = 5$. Under the considered conditions, the transmission coefficient also contains a contribution $T'$ from electrons corresponding to the vicinities of those cross sections of the FS where the longitudinal component of their velocity vanishes. This contribution always exists under the anomalous skin effect. The explicit expression for $T'$ is given in Ref. [106]. In strong magnetic fields ($B \sim 5$ T) and for moderately thin films ($L\omega\xi > v_0$), the term $T$ exceeds $T'$. This creates favorable conditions for the observation of size oscillations arising due to the FL mode. Also, the transmission includes the term originating from branch points of the function $F_-(\omega, q)$ in the $q, \omega$ complex plane. These points cause the Gantmakher–Kaner size oscillations of the transmission coefficient [107]. But for $L\Omega > v_0$, these oscillations are weak and can be disregarded. Under these conditions, the contribution from the FL mode to the transmission coefficient dominates, determining the shape and size of the oscillations.

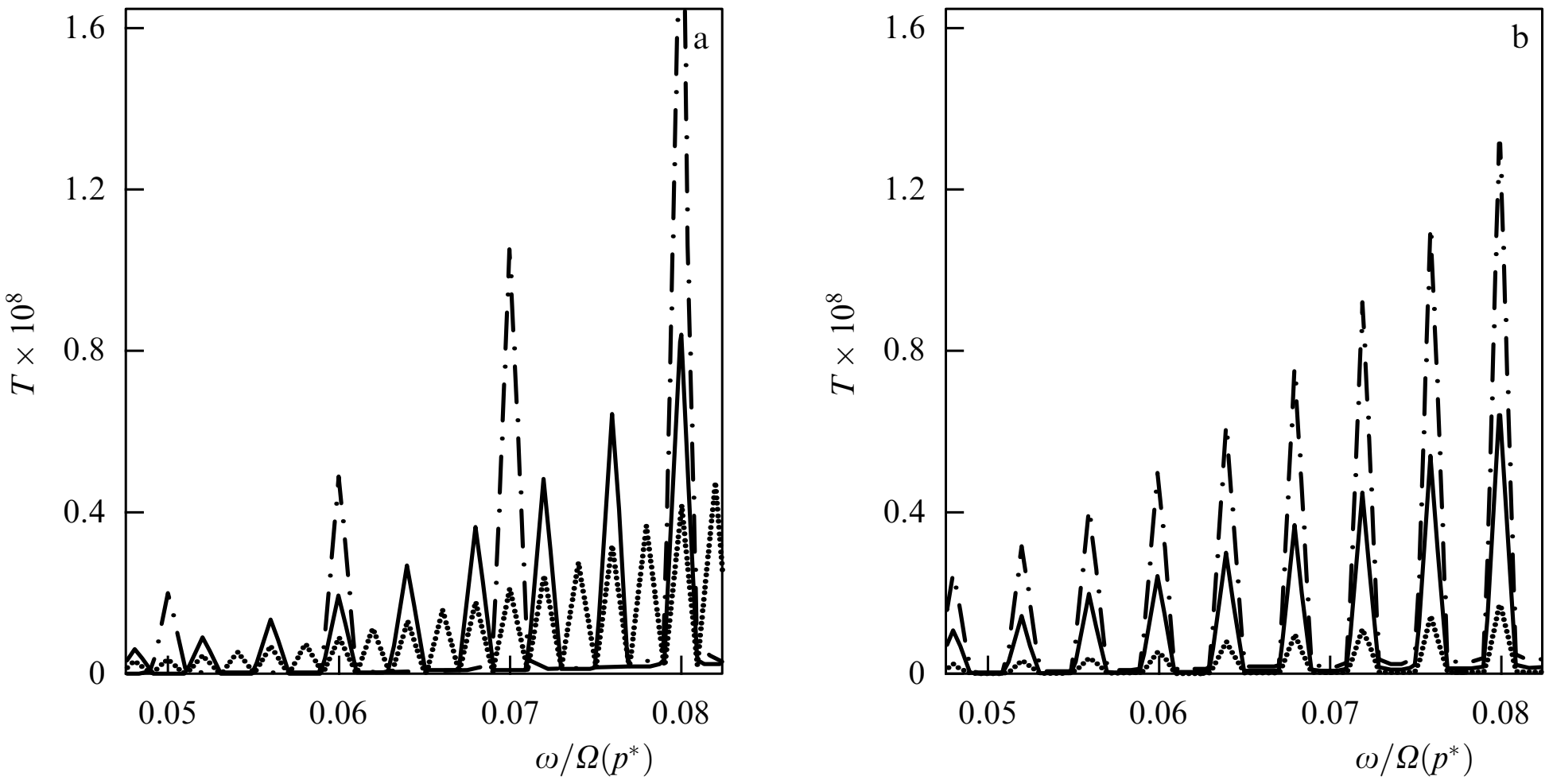

**Figure 11.** (a) Size oscillations in the transmission coefficient for a transverse electromagnetic wave traveling through a metal film and originating from the low-frequency FL mode. The curves are plotted at $\alpha_2 = -0.2$, $s = 3$, $\Omega\tau \sim 50$, $\xi = 10^3$, $L/l = 0.01$ (dashed-dotted line); 0.025 (solid line), and 0.05 (dashed line). (b) The dependence of the transmission on the FS shape near the inflection line. The curves are plotted for $s = 3$ (dashed line), 4 (solid line), and 5 (dashed-dotted line), $L/l = 0.025$. The other parameters coincide with those used to plot the curves in Fig. a.

It is known that most of the FSs of practical metals have inflection lines, and hence there are grounds to expect the low-frequency FL modes to appear in some metals for suitable directions of the applied magnetic fields. Especially promising are metals such as cadmium, tungsten, and molybdenum, where collective excitations near the Doppler-shifted cyclotron resonance (dopplerons) can occur. Another kind of interesting material is quasi-two-dimensional conductors. Assuming that the magnetic field is applied along the symmetry axis of the FS described by Eqn (7), it can be seen that the maximum longitudinal velocity of charge carriers is reached at the inflection lines where $d^2A/dp_z^2 = 0$. We can therefore expect low-frequency FL modes to appear in some of these materials [108].

## 4. Local geometry of the Fermi surface and magnetoacoustic response of a metal

### 4.1 Magnetoacoustic oscillations in metals with a nearly cylindrical Fermi surface

When an acoustic wave travels through a metal, the crystalline lattice is deformed periodically. These deformations give rise to an alternating electric field, which accompanies the wave. Conduction electrons are exposed to this electric field, and their response contributes to the metal elastic properties. Moreover, the periodic deformation of the lattice causes changes in the conduction electron spectra, which can be described by so-called deformation potentials. Usually, taking these deformation-induced terms into account in the energy–momentum relations for the conduction electrons does not qualitatively change the magnetoacoustic response, and we therefore omit them for brevity. When the mean free path of conduction electrons $l$ is greater than the sound wavelength $\lambda$, the electron response to the wave is determined by those electrons whose motion is consistent with the propagating perturbation. These electrons can strongly absorb the energy of the electric field. The efficient electrons are concentrated at small segments of the FS, and hence local anomalies of the FS curvature at such effective segments can noticeably affect the acoustic response of a metal. The influence of locally flattened or nearly cylindrical segments of the FS on the attenuation rate and the velocity shift of ultrasonic waves propagating in a metal has been analyzed before (see, e.g., Refs [1, 2, 4, 109]). Some results of this theoretical analysis were confirmed in experiments concerning the attenuation of ultrasonic waves in metals [36, 37].

Especially interesting effects are known to occur in the magtenoacoustic response of a metal in the case where the external magnetic field is moderately strong, such that the inequalities $\Omega\tau \gg 1$ and $qR \gg 1$ ($2R$ is the characteristic diameter of the cyclotron orbit) are simultaneously satisfied. Under these conditions, both sound velocity and attenuation oscillate as the magnetic field magnitude varies when the magnetic field $\mathbf{B}$ is directed perpendicularly to the wave vector $\mathbf{q}$ of the sound wave. These magnetoacoustic oscillations, which are also known as geometric resonances, are generated as a result of periodic reproduction of the most favorable conditions for the 'resonance' absorption of the acoustic wave energy by electrons moving along the wave front. The oscillations appear due to the commensurability of the cyclotron orbits of electrons with the wavelength of the sound wave. Their period is determined by the extremal diameter $2R_{ex}$ of the cyclotron orbit. The geometric oscillations exist in both low-frequency ($\omega\tau < 1$) and high-frequency ($\omega\tau > 1$) ranges. The main contribution to the oscillating corrections to the sound attenuation and velocity shift originates from the vicinities of so-called stationary points on the cyclotron orbits of the extremal diameter, where an electron moves in parallel to the wave front. This leads to a conjecture that the local geometry of the FS near these stationary points can strongly affect the geometric oscillations.

In what follows, we consider a longitudinal sound wave traveling along the $y$ axis of the chosen coordinate system ($\mathbf{q} = (0, q, 0)$), whereas the magnetic field is directed along the $z$ axis. We assume that the elastic displacement of the lattice $\mathbf{u}(\mathbf{r}, t)$ is proportional to $\exp(\mathrm{i}qy - \mathrm{i}\omega t)$. The force exerted by conduction electrons on the lattice is given by [40, 109]

$$F_{q\omega} = \mathrm{i}q\left(\gamma_\alpha - \frac{\mathrm{i}Ne}{q}\,\delta_{\alpha y}\right)E'^{\alpha}_{q\omega} + \mathrm{i}\omega q^2 \Lambda u_{q\omega}\,, \tag{66}$$

where $\mathbf{E}'_{q\omega} = \mathbf{E}_{q\omega} + (\mathrm{i}\omega/c)[\mathbf{u}_{q\omega} \times \mathbf{B}]$, and $\mathbf{E}_{q\omega}$ and $\mathbf{u}_{q\omega}$ are the amplitudes of the electric field accompanying the wave and of the lattice displacement.

The amplitude $\mathbf{E}_{q\omega}$ satisfies the Maxwell equations. Correspondingly, it can be expressed in terms of the total density of the current $\mathbf{j}_{q\omega}$ induced by the passage of an acoustic wave. The components of $\mathbf{j}_{q\omega}$ are

$$j^{\alpha}_{q\omega} = \sigma_{\alpha\beta} E'^{\beta}_{q\omega} + \omega q\left(\bar{\gamma}_\alpha - \frac{\mathrm{i}Ne}{q}\,\delta_{\alpha y}\right)u_{q\omega}\,. \tag{67}$$

The electron transport coefficients $\Lambda$ and $\sigma$ are

$$\Lambda = \frac{\mathrm{i}N^2}{2\pi^2\hbar^3 g^2}\int \mathrm{d}p_z\, m_\perp \sum_r \frac{n_{-r}(p_z, -q)\, n_r(p_z, q)}{\omega + \mathrm{i}/\tau - r\Omega}\,, \tag{68}$$

$$\sigma_{\alpha\beta} = \frac{\mathrm{i}e^2}{2\pi^2\hbar^3}\int \mathrm{d}p_z\, m_\perp \sum_r \frac{v^{\alpha}_{-r}(p_z, -q)\, v^{\beta}_r(p_z, q)}{\omega + \mathrm{i}/\tau - r\Omega}\,, \tag{69}$$

where $n_r(p_z, q)$ is the Fourier transform in the expansion with respect to the azimuthal angle specifying the position of an electron on the cyclotron orbit:

$$n_r(p_z, q) = \frac{1}{2\pi}\int_0^{2\pi} \exp\left[\mathrm{i}r\psi - \frac{\mathrm{i}q}{\Omega}\int_0^{\psi} v_y(p_z, \psi')\,\mathrm{d}\psi'\right]. \tag{70}$$

Here, $v_y(p_z, \psi)$ is the component of the electron velocity, $N$ is the electron concentration, and $g$ is the electron DOS on the FS. The Fourier transforms of the electron velocity components in the expansion in the angle $\psi$ are determined by relations similar to (70). For a multisheet FS, the integration with respect to $p_z$ in Eqns (69) and (69) must be supplemented with summation over all sheets of the FS. We can obtain the expression for the electroacoustic coefficient $\gamma_\alpha$ by replacing $n_{-r}(p_z, -q)$ by $ev^{\alpha}_{-r}(p_z, -q)$ in Eqn (68). To obtain the expression for $\bar{\gamma}_\alpha$, we have to replace $n_r(p_z, q)$ by $ev^{\alpha}_r(p_z, q)$. In the case of a multisheet FS, the term $\delta_{\alpha y}\mathrm{i}Ne/q$ in Eqn (67) is to be replaced by

$$\frac{\mathrm{i}e}{q}\sum_k N_k \frac{e_k}{|e|}\,,$$

where the summation is over all FS segments and the $e_k$ are the charges of the charge carriers (electrons/holes) associated with these segments.

To determine the wave vector of the acoustic wave propagating in a metal, we have to solve the equation for the amplitude of elastic displacement of the lattice together with the Maxwell equations. For small amplitudes of acoustic waves, the wave vector is given by

$$q = \frac{\omega}{s} + \Delta q\,, \tag{71}$$

where $s$ is the speed of sound. The increment $\Delta q$, which emerges as a result of interaction with electrons, is

$$\Delta q = \frac{\mathrm{i}q^2}{2\rho_{\mathrm{m}} s}\left(\Lambda + \frac{\gamma(\bar{\gamma} - Bcq/4\pi\omega)}{\sigma^* - c^2q^2/4\pi\mathrm{i}\omega}\right), \tag{72}$$

where $\sigma^* = \sigma_{xx} + \sigma^2_{yx}/\sigma_{yy}$ and $\rho_{\mathrm{m}}$ is the mass density of the metal.

In the case where $qR \gg 1$, the leading contribution to the integral over $\psi$ in (70) comes from the neighborhoods of stationary points on the cyclotron orbits. Accordingly, estimating the integrals by the stationary phase method, we can obtain the asymptotic expressions

$$n_{\pm r}(p_z, \pm q) = \frac{1}{\pi}\exp\left[\pm\mathrm{i}qR(p_z) \pm \mathrm{i}\pi\,\frac{r}{2}\right] \times\left\{\cos\left[qR(p_z) - \pi\,\frac{r}{2}\right]V(p_z) - \sin\left[qR(p_z) - \pi\,\frac{r}{2}\right]W(p_z)\right\}, \tag{73}$$

where $2R(p_z)$ is the diameter of the cyclotron orbit of electrons in the direction of propagation of the acoustic wave. The form of the functions $V(p_z)$ and $W(p_z)$ is determined by the energy–momentum relation for electrons in the vicinities of stationary points.

To analyze the effect of a nearly cylindrical cross section of the FS on geometric oscillations, we suppose that the metal FS includes a double-convex lens axially symmetric with respect to the $z$ axis and that the electron energy momentum relation has the form

$$E(\mathbf{p}) = \frac{p_1^2}{2m_1}\left(\frac{p_x^2 + p_y^2}{p_1^2}\right) + \frac{p_2^2}{2m_2}\left(\frac{p_z^2}{p_2^2}\right)^k, \tag{74}$$

where $p_1$ and $p_2$ are quantities having the dimensions of momentum and characterizing the diameter and thickness of the lens. For $k = 1$, the lens is ellipsoidal. If $k > 1$, there is a zero-curvature line coinciding with the central cross section of the lens by a plane perpendicular to its axis (Fig. 12). In the vicinity of this cross section, the shape of the lens surface is close to cylindrical, especially for $k \gg 1$.

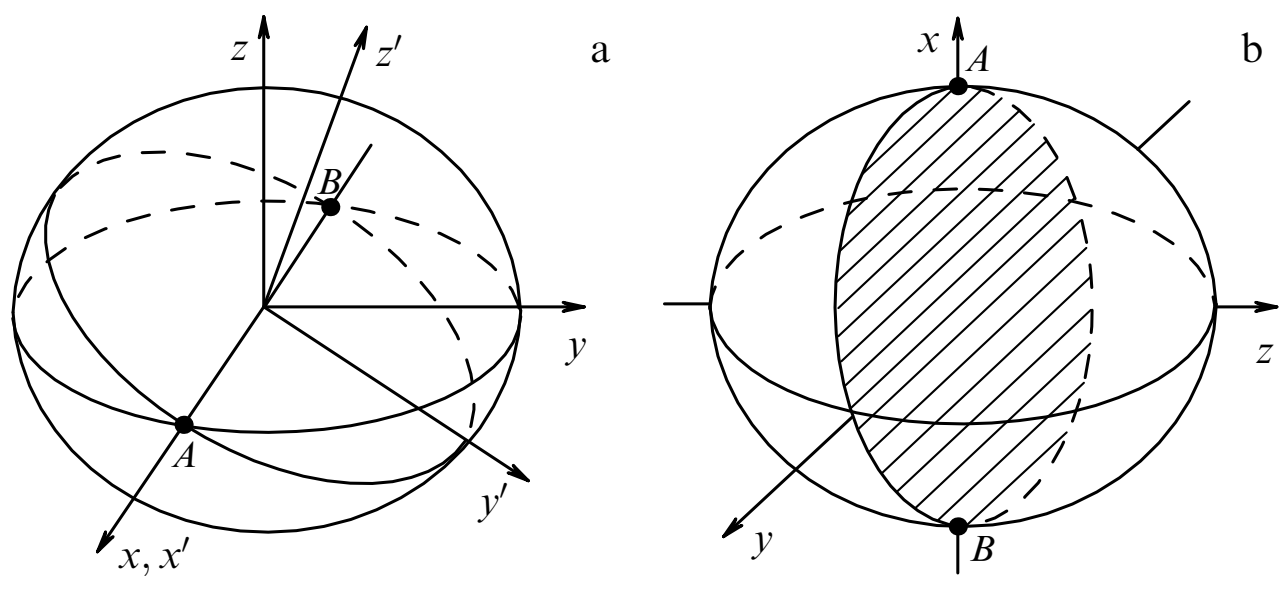

**Figure 12.** (a) The segment of a FS described by energy–momentum relation (74). For $k > 1$, the surface cross section by the plane $p_z = 0$ is a zero-curvature line. Points $A(p_1, 0, 0)$ and $B(-p_1, 0, 0)$ correspond to the stationary points on the cyclotron orbit. When the external magnetic field (directed along the $z'$ axis) is tilted to the symmetry axis of the lens ($z$ axis), the geometric oscillation amplitude decreases. (b) The segment of a FS associated with energy–momentum relation (33) for $k > 1$. The lens is flattened at the points $A(p_1, 0, 0)$ and $B(-p_1, 0, 0)$, which correspond to the stationary points.

The functions $V(p_z)$ and $W(p_z)$ for electrons of the lens have the form

$$V(p_z) = W(p_z) = \frac{1}{\sqrt{\pi q R_{\mathrm{ex}} f(p_z/p_2)}}\,, \tag{75}$$

where $f(x) = \sqrt{1-x^{2k}}$.

If the lens is ellipsoidal in shape ($k=1$), it can be proved that the second term in Eqn (72) for the dynamic correction $\Delta q$ is smaller than the first one, which allows approximately writing

$$\Delta q = -\frac{\mathrm{i}q^2}{2\rho_{\mathrm{m}} s}(\Lambda_1 + \Lambda_2) \equiv \Delta q_1 + \Delta q_2\,. \tag{76}$$

This approximation remains valid for the moderately manifested curvature anomaly at the lens edge, where the value of the shape parameter $k$ is not too large and $(qR_{\mathrm{ex}})^{-1/2k}$ remains small at $qR_{\mathrm{ex}} \gg 1$. The smooth ($\Lambda_1$) and oscillating ($\Lambda_2$) terms in Eqn (76) have the form [110]

$$\Lambda_1 = Q\,\frac{Np_1}{q}\,d\coth\left[\frac{\pi}{\Omega\tau}(1-\mathrm{i}\omega\tau)\right],$$

$$\Lambda_2 = \frac{Np_1}{q}\,bU_0^2\,\frac{\sin(2qR_{\mathrm{ex}} - \pi/4k)}{(qR_{\mathrm{ex}})^{1/2k}}\left\{\sinh\left[\frac{\pi}{\Omega\tau}(1-\mathrm{i}\omega\tau)\right]\right\}^{-1}, \tag{77}$$

where

$$U_0 = \frac{2\sqrt{m_1 m_2}}{p_1 p_2}\,, \qquad Q = \int_0^1 \frac{U_0^2\,\mathrm{d}t}{f(t)}\,,$$

and $b$ and $d$ are dimensionless constants of the order of unity. For $k=1$, $m_1 = m_2 = m$, and $p_1 = p_2 = p_{\mathrm{F}}$, Eqns (76) and (77) give the well-known result for a spherical FS.

When the shape of the lens in the vicinity of its central cross section is very close to cylindrical ($k \gg 1$), the parameter $(qR_{\mathrm{ex}})^{-1/2k}$ can no longer be regarded as small. In this case, the relation between the first and second terms in (76) depends on the relative number of charge carriers associated with the lens. When the contributions from all parts of the FS are taken into account, the smooth components of the transport coefficients $\Lambda$ and $\sigma_{xx}$ are proportional to the total concentration $N_0$ of the charge carriers (for $\omega\tau < 1$), whereas their oscillating components are proportional to the concentration $N$ of electrons/holes associated with the lens. The lens contribution determines the leading terms of the asymptotic expressions for the coefficients $\gamma$ and $\bar{\gamma}$ in the region $qR \gg 1$; therefore, they are also proportional to $N$. When the major part of conduction electrons/holes in the considered metal is associated with the lens ($N/N_0 \sim 1$), the enhancement of geometric oscillations of transport coefficients makes the second term in (76) as significant as the first. For $N/N_0 \ll 1$, the second term in Eqn (76) remains much smaller than the first one.

The enhancement of geometric resonances in the magnetoacoustic response of a metal arising due to the FS geometry was analyzed in Ref. [111]. The model of the FS used in that work satisfies both conditions $k \gg 1$ and $N/N_0 \sim 1$. This model can be applied to layered conductors with a quasi-two-dimensional energy spectrum of charge carriers. In such compounds, the enhancement of geometric oscillations of electroacoustic coefficients leads to the resonance effect predicted in Ref. [111].

In conventional 3D metals, the relative number of conduction electrons associated with the nearly cylindrical FS segments is most probably small ($N/N_0 \ll 1$). But the enhancement of geometric oscillations of the electroacoustic coefficient $\Lambda$ due to local features of the FS geometry can lead to noticeable changes in the geometric oscillations of the dynamic correction $\Delta q$. Oscillations formed by the response of electrons belonging to nearly cylindrical FS segments are greater in amplitude by a factor of $(\sqrt{qR_{\mathrm{ex}}})^{1-1/k}$ than geometric resonances formed by electrons from other parts of the FS. For $k \gg 1$, the oscillation enhancement becomes significant.

The enhancement of geometric oscillations considered above can be manifested only for certain directions of the magnetic field with respect to the symmetry axes of the crystal lattice. Like many other effects associated with local anomalies in the shape of the FS, this effect must strongly depend on the direction of the applied magnetic field. This is illustrated in Fig. 12a. When the magnetic field is tilted to the symmetry axis of the lens shown in the figure, the effective cross section slips away from the line of zero curvature that runs along the lens rim. This results in a decrease in the geometric resonances.

### 4.2 Local flattening of the Fermi surface and magnetoacoustic oscillations in metals

An increase in the number of electrons participating in the absorption of acoustic energy under the conditions of the Pippard geometric resonance can occur due to the local flattening of the FS at those points corresponding to the stationary points of a cyclotron orbit. Here, we analyze the effect of local flattening of the FS on the geometric oscillations of sound velocity and attenuation.

We assume that the Fermi surface includes a segment shaped like a biconvex lens, whose symmetry axis is the $x$ axis of the chosen coordinate system. We write the dispersion relation for the electrons associated with the lens in form (33). If the parameter $k$ characterizing the shape of the lens takes values greater than unity, then the Gaussian curvature of the surface vanishes at the points $(\pm p_1, 0, 0)$, which coincide with the vertices of the lens as shown in Fig. 12b. The vertices are points where the surface of the lens is flattened. As before, we assume that $\mathbf{B}$ is parallel to the $z$ axis and a sound wave propagates along the $y$ axis of the coordinate system fixed in the lens; then the expression for the wave vector of the sound wave can be written in form (71). For longitudinal sound, the dynamic correction $\Delta q$ is described by Eqn (72), where the relevant transport coefficients are given by Eqns (68) and (69).

At $qR \gg 1$, the neighborhoods of stationary points on the cyclotron orbits make the leading contribution to the integrals over $\psi$ in the expressions for $n_{\pm r}(p_z, \pm q)$. Correspondingly, the asymptotic expressions for these quantities have form (73).

For small values of $p_z$, corresponding to the lens rim, the leading term in the asymptotic expansion of $V(p_z)$ in inverse powers of $qR$ at $k=2$ can be approximated as [112]

$$V(p_z) = \frac{\Gamma(1/4)}{4}\,\frac{\sqrt{m_1 m_2}}{m_\perp^{\mathrm{ex}}}\left(\frac{2}{qR_{\mathrm{ex}}}\right)^{1/4}\cos\frac{\pi}{8}\,, \tag{78}$$

where $\Gamma(x)$ is the gamma function, $m_\perp^{\mathrm{ex}} = m_\perp(0)$, and $R_{\mathrm{ex}} = R(0)$. For sufficiently large values of $p_z$, the approx-

imation

$$V(p_z) = \frac{1}{2}\sqrt{\frac{\pi m_1 m_2}{(m_\perp^{\rm ex})^2}\,\frac{p_2^2}{p_1^2}}\;\frac{\cos(\pi/4)}{\sqrt{qR(p_z)}} \tag{79}$$

can be used. The asymptotic expressions for $W(p_z)$ can be obtained from Eqns (78) and (79) by replacing the cosine by the sine of the same argument.

Correspondingly, in the calculation of the dynamic correction to the wave vector of the sound wave arising as a result of the interaction with the electrons of the lens, the range of integration over $p_z$ in expressions (68) and (69) must be divided into regions with small and large values of $p_z$. For frequencies that are not too high ($\omega\tau < 1$), the first term in Eqn (76) gives the smooth part of the contribution of the lens electrons to the attenuation and the velocity shift of the ultrasonic wave. The magnetoacoustic oscillations are described by the second term $\Delta q_2$, which can be represented as

$$\Delta q_2 = {\rm i}\gamma_0 U_0^2 b \cos\left(2qR_{\rm ex} + \frac{\pi}{4}\right)\left(\sinh\left[\pi\,\frac{1-{\rm i}\omega\tau}{\Omega_{\rm ex}\tau}\right]\right)^{-1}, \tag{80}$$

where $b$ has the order of unity and $\gamma_0$ has the order of the sound attenuation rate in the absence of the magnetic field.

The amplitude of oscillations described by expression (80) is of the same order as the above oscillating contribution to the wave vector of the ultrasound wave arising due to the interaction with conduction electrons. This is a direct consequence of the increase in the number of effective electrons originating from the flattening of the electron lens in the neighborhoods of its vertices. In a metal whose FS does not include locally flat segments, the oscillating correction to the sound absorption coefficient is small compared to the smooth part.

Equation (80) was derived for $k = 2$. For $k > 2$, the amplification of the oscillations is even more pronounced. However, as shown in [112], even a very well-pronounced flattening of the FS cannot result in predomination of $\Delta q_2$ over $\Delta q_1$. Both terms always have the same magnitude. It is worthwhile to note that the effect of the FS flattening on the geometric resonances in the ultrasound attenuation and velocity can be stronger than the effect of nearly cylindrical strips. The reason is that the increase in the number of electrons associated with the close vicinities of flattening points exceeds their increase in the vicinity of a point belonging to a nearly cylindrical cross section where only one of the principal curvature radii tends to infinity.

### 4.3 Acoustic cyclotron resonance and giant high-frequency magnetoacoustic oscillations in metals with a locally flattened Fermi surface

The enhancement of magnetoacoustic oscillations due to the local flattening of the FS can also exhibit itself in the high-frequency range ($\omega\tau > 1$). At high frequencies, magnetoacoustic oscillations can be superimposed over the acoustic cyclotron resonance. Keeping in mind that the largest contribution to the integrals over $p_z$ in expressions (68) and (69) originates from the range of small $p_z$, we can replace all smooth functions of $p_z$ in the integrands by their values at $p_z = 0$. For $qR \gg 1$, the leading contribution to the asymptotic expression for $\Lambda$ associated with the electrons of the lens has the form [113]

$$\Lambda = \frac{{\rm i}g}{\omega}\,\frac{\mu}{(qR_{\rm ex})^{1/k}}\,U_0^2 W(\omega)\,. \tag{81}$$

The frequency-dependent factor $W(\omega)$ in Eqn (81) is

$$W(\omega) = \int_{-1}^{1} Y(\omega, x)\,{\rm d}x\,, \tag{82}$$

where

$$Y(\omega,x) = -{\rm i}\pi\,\frac{\omega}{\Omega}\left\{\coth\left[\pi\,\frac{1-{\rm i}\omega\tau}{\Omega\tau}\right] + \cos\left(2qR + \frac{\pi}{2k}\right)\right.$$
$$\left.\times\left(\sinh\left[\pi\,\frac{1-{\rm i}\omega\tau}{\Omega\tau}\right]\right)^{-1}\right\}. \tag{83}$$

In the high-frequency range $\omega\tau \gg 1$, the function $Y(\omega, x)$ diverges at frequencies $\omega$ equal to the multiple cyclotron frequency $\Omega$. These divergences appear due to the acoustic cyclotron resonance, predicted and analyzed by Kaner [114, 115]. The second term in Eqn (83) also includes the factor $\cos(2qR + \pi/2k)$ describing geometric oscillations.

The asymptotic expression for the dynamic correction $\Delta q$ near the cyclotron resonance depends on the ratio of the parameters $2qR_{\rm ex}$ and $(\omega\tau)^{k/2}$. Under the considered conditions, both parameters are large compared to unity. We suppose that $2qR_{\rm ex} \gg (\omega\tau)^{k/2}$. Under conditions of the acoustic cyclotron resonance in normal metals, $qR_{\rm ex} \sim v_{\rm F}/s \sim 10^3$ ($v_{\rm F}$ is the Fermi velocity for the electrons associated with the lens). For $\Omega\tau \sim 10^2$, the above inequality can be satisfied when the lens is moderately flattened ($k < 2$). When $2qR_{\rm ex} \gg (\omega\tau)^{k/2}$, the dynamic correction $\Delta q$ near the acoustic cyclotron resonance remains small compared to the leading approximation for the ultrasound wave vector $\omega/s$.

The resonance contribution to the correction $\Delta q$ from the electrons associated with the neighborhood of the central cross section of lens (33) is given by

$$\Delta q_{\rm r} = \gamma_0\,\frac{1}{(qR_{\rm ex})^{1/k}}\,\frac{qR_{\rm ex}}{r}\,\frac{1}{\sqrt{1-\omega/r\Omega_{\rm ex} - {\rm i}/\omega\tau}}$$
$$\times\left[1 + \frac{b\cos(2qR_{\rm ex} + \pi r + \pi/4k)}{\sqrt{1-\omega/r\Omega_{\rm ex} - {\rm i}/\omega\tau}\,(qR_{\rm ex})^{1/2k}}\right], \tag{84}$$

where $b$ is a dimensionless constant.

For $k = 1$, the result for the attenuation rate determined by Eqn (84) coincides with the corresponding result in Ref. [114], which is obtained assuming that the FS of a metal has a finite and nonzero curvature everywhere. When $k = 1$, the magnitude of the resonance feature in the attenuation rate is of the order of $\gamma_0\sqrt{\omega\tau}/r$. In this case, the magnitude of the geometric oscillations is smaller than the magnitude of the resonance feature associated with the cyclotron resonance by the factor $\sqrt{\omega\tau/qR_{\rm ex}}$.

When $k > 1$, the effective strip on the FS passes through the flattened segments near the vertices of the lens, leading to the amplification of the acoustic cyclotron resonance. The magnetic field dependence of the ultrasound attenuation rate near the cyclotron resonance is shown in Fig. 13. Due to the FS local flattening, the resonance contribution to the ultrasonic attenuation coefficient increases $(qR_{\rm ex})^{(k-1)/k}$-fold over the case of an ellipsoidal FS. This amplification arises due to the increase in the number of electrons participating in the resonance absorption of the energy of an ultrasound wave. This increase in the number of efficient electrons also leads to the amplification of geometric oscillations. The corresponding term in Eqn (84) is $(qR_{ex})^{(k-1)/2}$ times larger than a similar term in the expression for $\Delta q_r$ in a simple metal. When the flattening of the FS becomes stronger, the magnitude of the geometric oscillations increases faster than the magnitude of the peak corresponding to the acoustic cyclotron resonance.

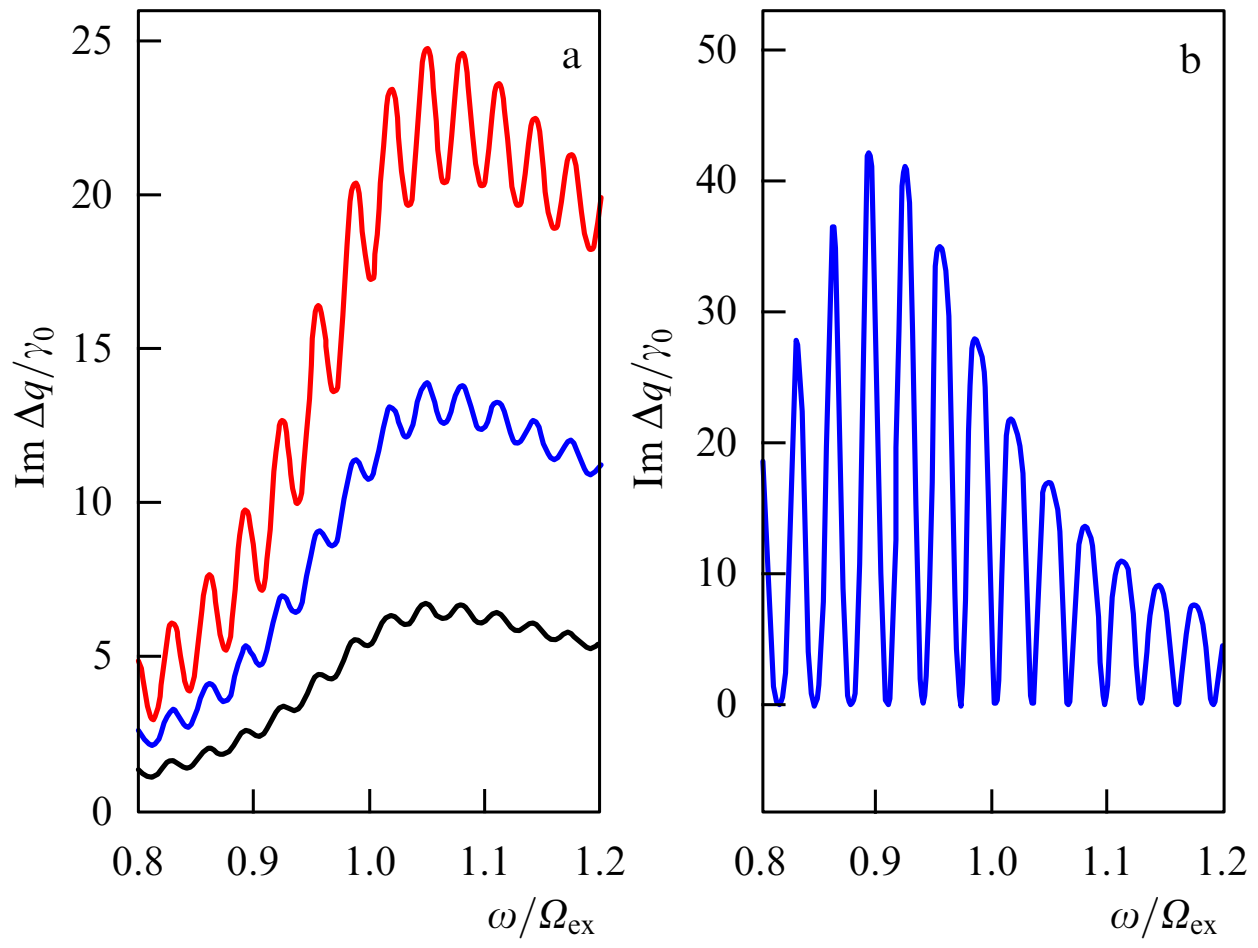

**Figure 13.** Attenuation of longitudinal ultrasound waves versus $\omega/\Omega_{ex}$ in the vicinity of the acoustic cyclotron resonance. The curves are plotted assuming that the magnetoacoustic response is mostly determined by the electrons associated with the vicinities of flattened vertices of a lens described by Eqn (33) at $qR_{ex} = 100$, $\omega\tau = 10$. The shape parameter $k$ takes the values (a) 1.25, 1.5, and 1.75 for the curves from bottom up, and (b) $k = 4$.

We can use expression (84) to describe the resonance part of the dynamic correction $\Delta q$ only for moderate flattening of the electron lens and moderately large $\omega\tau$. When the flattening of the lens near its vertices is strong, the quantity $(\omega\tau)^{k/2}$ exceeds the parameter $2qR_{ex}$. Under the conditions of acoustic cyclotron resonance in typical metals, the inequality $2qR_{ex} \ll (\omega\tau)^{k/2}$ can be satisfied for $k > 3$. In this case, the magnetic field dependence of the function $W(\omega)$ near the cyclotron resonance ($\omega \approx r\Omega_{ex}$) critically changes in such a way that $W(\omega)/(qR_{ex})^{1/k}$ reaches values of the order of unity at the oscillation peaks. Then the second term in Eqn (72) becomes significant, and the effects originating from the coupling of the sound wave to the electromagnetic cyclotron wave can appear.

The effects originating from the coupling of electromagnetic and ultrasound waves are well known. In particular, it has been shown that an ultrasound wave propagating perpendicularly to an external magnetic field can couple to shortwave cyclotron waves (see Refs [51, 96, 115]). In our geometry, longitudinal ultrasound waves couple to cyclotron waves whose dispersion relation is determined by the equation $\sigma_{yy} = 0$. The dispersion of such a mode near the frequency $r\Omega_{ex}$ can be written in the form [113]

$$\omega_1 = r\Omega_{ex}\left[1 + \frac{1}{(qR_{ex})^{2/k}}\, f^2(q)\right], \qquad (85)$$

where the function $f(q)$ includes an oscillating factor $\cos[qR_{ex} + \pi r/2 + \pi/4k]$. This cyclotron mode can appear in a metal under the condition $2qR_{ex} \ll (\omega\tau)^{k/2}$. The shape of the dispersion curve of the considered cyclotron wave depends on the local geometry of the FS. Longitudinal cyclotron waves similar to the mode described by Eqn (85) can occur in a metal with a spherical FS under the condition $qR_{ex} < \omega\tau$. Their dispersion relation has the form (see Ref. [96])

$$\omega_1 = r\Omega\left(1 + \frac{1}{2qR}\right). \qquad (86)$$

The difference between Eqns (85) and (86) describing the dispersion of the longitudinal cyclotron waves is completely caused by the local flattening of the considered FS.

For a very strong flattening of the vicinities of the vertices of an electron lens ($2qR_{ex} \ll (\omega\tau)^{k/2}$), we can write the expression for the resonance contribution to the dynamic correction $\Delta q_r$ as

$$\Delta q_r = \gamma_0 \frac{qR_{ex}}{r} \frac{f^2(q)}{(qR_{ex})^{2/k}} \frac{\omega}{\omega_1 - \omega - i/\tau}, \qquad (87)$$

where $\omega_1$ is the frequency of the longitudinal cyclotron wave described by Eqn (85). The frequency $\omega_1$ corresponds to the resonance rather than the cyclotron frequency $\Omega_{ex}$. The shift of the peak of the acoustic cyclotron resonance caused by the coupling of ultrasound to the cyclotron wave was already predicted for the spherical and ellipsoidal FSs [96]. When the effective segments of the FS are locally flat, this shift is more pronounced and more available for experimental observations. The factor $f^2(q)$ in Eqn (85) describes geometric oscillations superimposed on the peak corresponding to the acoustic cyclotron resonance. The amplitude of the geometric oscillations sharply increases near the resonance. In order of magnitude, it is determined by the height of the resonance peak. Therefore, geometric oscillations of the ultrasonic attenuation in metals with strongly flattened FSs can become giant near the acoustic cyclotron resonance. Figure 13b illustrates this conclusion.

Speaking of possible effects of the FS local geometry on the magnetoacoustic response of conduction electrons, it is worthwhile to briefly discuss an anomaly in the sound velocity shift experimentally observed in studying the transport properties of a two-dimensional electron gas (2DEG) in GaAs/AlGaAs heterostructures in a strong magnetic field corresponding to the half-filling of the lowest Landau level ($\nu = 1/2$) [116, 117]. In these experiments, the electron density was modulated along a chosen direction by an inhomogeneous static electric field, and the velocity shift and attenuation of the surface acoustic wave traveling over the system were recorded. The response revealed some unusual features. It was extremely anisotropic with respect to the mutual arrangement of the wave vector of the surface acoustic wave $\mathbf{q}$ and the electron density modulation wave vector $\mathbf{g}$. At $\mathbf{q} \parallel \mathbf{g}$, the response was similar to that repeatedly observed in the unmodulated quantum Hall system near $\nu = 1/2$ [118]. But at $\mathbf{q} \perp \mathbf{g}$, the minimum in the sound velocity at $\nu = 1/2$ was replaced by a maximum whose height increased as the magnitude of the modulating field increased.

As was first proposed by Halperin, Lee, and Read [119], the state of a quantum Hall system near half-filling of the lowest Landau level can be described by introducing new fermionic quasiparticles (so-called composite fermions), which are electrons decorated by attached quantum flux tubes. Near $\nu = 1/2$, the composite fermions experience the reduced effective magnetic field $B_{\text{eff}} = B - 4\pi\hbar cN/e$, which becomes zero at half filling. They form a Fermi sea. The corresponding 2D Fermi surface is a circle with the radius $p_{\text{F}} = (4\pi N\hbar^2)^{1/2}$.

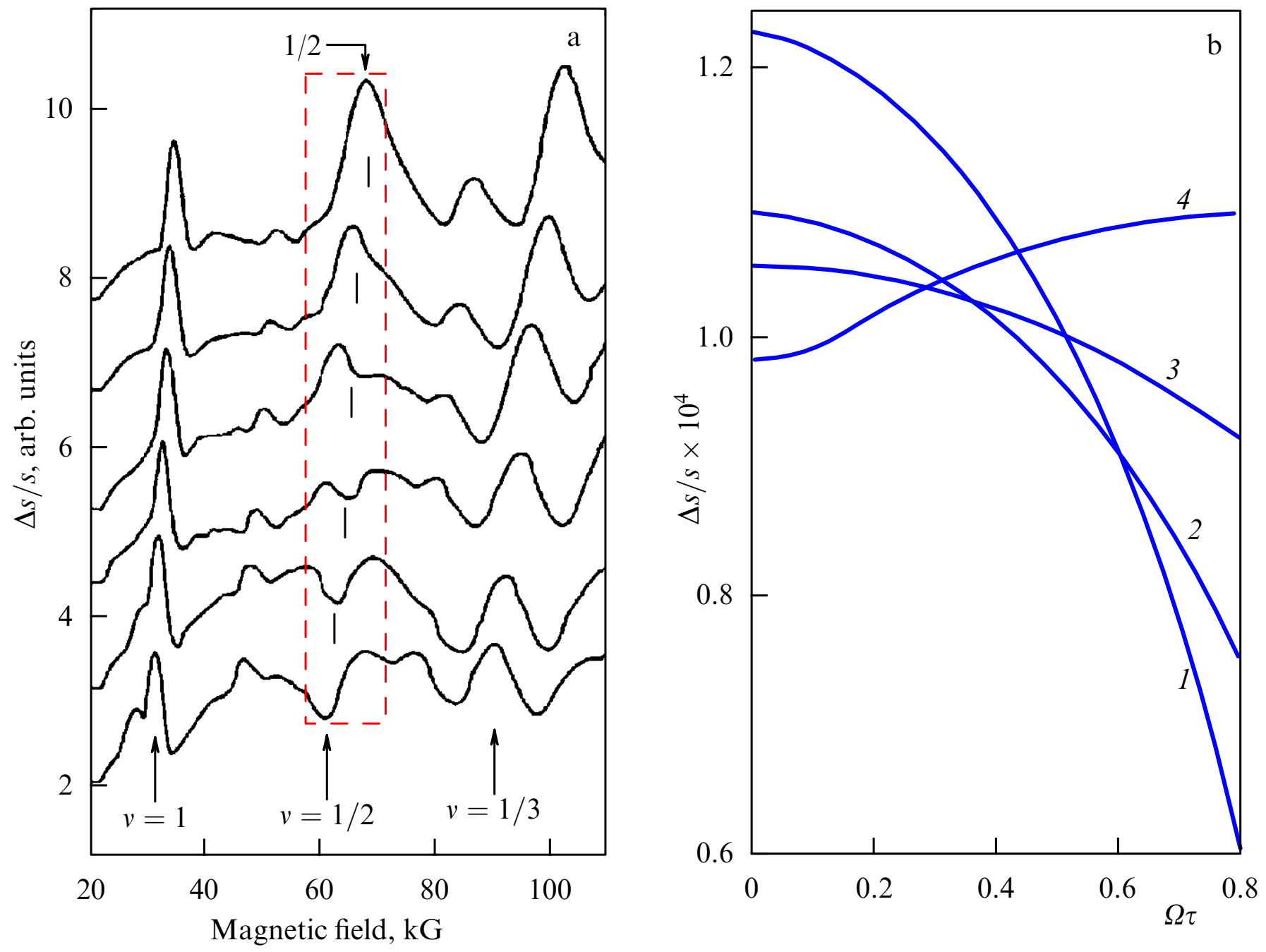

**Figure 14.** (a) The velocity shift of a surface acoustic wave traveling in the GaAs/AlGaAs heterostructure above the modulated 2DEG in the quantum Hall state near $\nu = 2$, as observed in experiment [116]. (b) Effective magnetic field dependences of the speed of a surface acoustic wave traveling above the modulated 2DEG in the quantum Hall state near $\nu = 1/2$. Curves *1–3* correspond to the deformed composite fermion FS (the assumed distortion increases from *3* to *1*) and exhibit maxima at $\nu = 1/2$. Curve *4* is plotted assuming that the FS is undeformed by the modulating potential. The curves are plotted at $N = 0.7 \times 10^{15}$ m$^{-2}$, $\tau = 2 \times 10^{-11}$ s, $q = 9 \times 10^6$ m$^{-1}$, $\alpha^2/2 = 3.2 \times 10^{-4}$, and $\sigma_m = 0.6 \times 10^4$ m s$^{-1}$.

Due to the piezoelectric properties of GaAs, a surface acoustic wave traveling over the surface of a heterostructure containing a 2DEG is influenced by the electron system. Assuming that the wave propagates along the $x$ axis of the chosen coordinate system, we can write [120, 121]

$$\frac{\Delta s}{s} - \frac{\mathrm{i}\Gamma}{q} = \frac{\alpha^2}{2}\left(1 + \frac{\sigma_{xx}}{\sigma_m}\right)^{-1}, \tag{88}$$

where $\alpha$ is the piezoelectric coupling constant, $\sigma_m = \epsilon s/2\pi$, and $\epsilon$ is the effective dielectric constant of the background. The electron conductivity component $\sigma_{xx}$ can be expressed in terms of the composite fermion conductivity components $\sigma_{\alpha\beta}^{\text{ef}}$ [119].

The electron density modulations induced by the electrostatic field affect the composite fermion system in two ways: through the changes in the effective magnetic field and through the direct effect of the modulating potential. The modulations of $B_{\text{eff}}$ play a crucial role in the modulated 2DEG response to the acoustic wave with a long wavelength, satisfying the condition $ql \ll 1$, where $l$ is the composite fermion mean free path [122, 123]. However, the experiments reported in Refs [116, 117] were carried out using acoustic waves with rather short wavelengths ($ql > 1$). In this 'nonlocal' regime, the direct effect of the electrostatic potential dominates [124].

The modulating potential deforms the originally circular composite fermion FS in the same way as crystalline fields shape the FSs of ordinary metals. But unlike the crystalline fields, the modulating electrostatic field in the considered case acts along a single direction indicated by its wave vector **g**. As shown in Ref. [124], small locally flat segments emerge on the distorted FS. These segments are located such that their contributions dominate in the 2DEG response to the acoustic wave when the wave propagates at a right angle to the modulating field ($\mathbf{q} \perp \mathbf{g}$). The specific FS geometry in these segments can be responsible for the appearance of the peak in the sound velocity shift observed in experiments, as is illustrated in Fig. 14. At the same time, when the sound wave travels along the modulating field ($\mathbf{q} \| \mathbf{g}$), the flattened segments of the FS only slightly contribute to the response, and the peak in the velocity shift disappears, being replaced by a dip typical for the unmodulated 2DEG response to the sound wave. These conclusions agree with experimental results.

## 5. Effect of the Fermi surface geometry on magnetic quantum oscillations

### 5.1 De Haas–van Alphen oscillations in quasi-two-dimensional conductors

Magnetic quantum oscillations are well known as a powerful tool repeatedly used in studies of the electronic properties of various conventional metals [5]. The theory of quantum oscillatory phenomena, such as de Haas–van Alphen oscillations in magnetization and Shubnikov–de Haas oscillations in the magnetoresistivity of conventional three-dimensional metals, was developed by Lifshitz and Kosevich (LK) in their well-known work [125]. This theory was successfully used to extract valuable information concerning electron band-structure parameters from experimentally measured magnetic quantum oscillations. In the last two decades, magnetic quantum oscillations have frequently been used as a tool to study the electron characteristics of various Q2D conductors with metallic-like conductivity [6–8, 10, 12–19]. A theory of magnetic oscillations in Q2D materials was proposed in several studies (see, e.g., Refs [43, 44, 126–128]). Significant progress is already being made in developing the theory, but some significant points are still not taken into account. One such point is the effect of the FS curvature on the amplitude and shape of the oscillations. The FSs of Q2D metals are known to include systems of weakly rippled cylinders. Accordingly, the current theory adopts the tight-binding approximation to describe the energy–momentum relation for the charge carriers.

Neglecting the anisotropies of the charge-carrier energy spectrum in the conducting layer planes for simplicity, we can write the energy–momentum relation in the form

$$E(\mathbf{p}) = \frac{\mathbf{p}_\perp^2}{2m_\perp} - 2wE_\parallel\left(\frac{p_z d}{\hbar}\right), \tag{89}$$

where

$$E_\parallel\left(\frac{p_z d}{\hbar}\right) = \sum_{n=1}^{\infty} \epsilon_n \cos\frac{np_z d}{\hbar}, \tag{90}$$

with $\epsilon_n = -E_n/2w$. It follows from this expression that $E_\parallel(p_z d/\hbar)$ is an even periodic function of $p_z$ whose period equals $2\pi\hbar/d$. Omitting all terms with $n > 1$ and setting $E_1 = 2w$, we can reduce our energy–momentum relation (89) to simple from (7). Introducing this expression opens up the possibilities to describe Q2D FSs of various profiles (see Fig. 3) and to analyze the influence of their fine geometric features on de Haas–van Alphen oscillations. These studies lead to some nontrivial results that cannot be obtained within the simple approximation in (7).

To analyze de Haas–van Alphen oscillations, we start from the standard expression for the longitudinal magnetization

$$M_\parallel(B, T, \zeta) \equiv M_z(B, T, \zeta) = -\left(\frac{\partial\bar{\Omega}}{\partial B}\right)_{T,\zeta}, \tag{91}$$

where the magnetization depends on the temperature $T$ and the chemical potential of charge carriers $\zeta$. The chemical potential itself is a function of the magnetic field and temperature, and oscillates in strong magnetic fields [5]. The expression for the thermodynamic potential $\bar{\Omega}$ can be written standardly

$$\bar{\Omega}(B, T, \zeta) = -T\sum_n \ln\left(1 + \exp\frac{\zeta - E}{kT}\right), \tag{92}$$

where the summation ranges over all possible states of quasiparticles. When a strong magnetic field is applied, the quasiparticles have the Landau energy spectrum of the form

$$E_{n,\sigma}(p_z) = \hbar\Omega\left(n + \frac{1}{2}\right) + \sigma\hbar\omega_0 - 2wE_\parallel\left(\frac{p_z d}{\hbar}\right), \tag{93}$$

where $\Omega$ is the cyclotron frequency, $\omega_0 = \beta B$, $\beta$ is the Bohr magneton, the quantum number $n$ labels Landau levels, and $\sigma$ is the spin quantum number. Using the Poisson summation formula, the expression for the thermodynamic potential can be represented as the sum of a monotonic term $\bar{\Omega}_0$ and an oscillating correction $\Delta\bar{\Omega}$:

$$\Delta\bar{\Omega} = \frac{\mathrm{i}}{4\pi^2\hbar\bar{\lambda}^2}\sum_\sigma\sum_{r=1}^{\infty}\frac{(-1)^r}{r}\int_0^\infty \frac{I(E_\sigma)\,\mathrm{d}E}{1 + \exp\left[(E_\sigma - \zeta)/kT\right]}, \tag{94}$$

where $\bar{\lambda}$ is the magnetic length and the function $I(E_\sigma)$ is given by

$$I(E_\sigma) = 2\mathrm{i}\int \exp\left[\mathrm{i}r\,\frac{\bar{\lambda}^2}{\hbar^2}\,A(E_\sigma, p_z)\right]\mathrm{d}p_z, \tag{95}$$

where $A(E_\sigma, p_z)$ is the cross-sectional area.

Until this point, we have followed the LK theory in deriving the expression for $\Delta\bar{\Omega}$. As a result, we arrived at Eqns (94) and (95), which are valid for conventional 3D metals and for Q2D and perfectly 2D conductors. Diversities in the expressions for $\Delta\bar{\Omega}$ appear in the course of calculations of the function $I(E_\sigma)$. These calculations yield different results for different FS geometries. In deriving the standard LK formula, it is assumed that the FS curvature is nonzero at the effective cross sections with the extremal areas and $I(E_\sigma, p_z)$ is approximated using the stationary phase method. For 2D metals, the calculation of $I(E_\sigma)$ is trivial because the FS is a cylinder and the cross-sectional area $A$ is independent of $p_z$. Obviously, the FS curvature is then everywhere zero. Correspondingly, we arrive at the following result for the oscillating part of the longitudinal magnetization for a 2D conductor [126]:

$$\Delta M_\parallel = -2N\beta\,\frac{\Omega}{\omega_0}\sum_{r=1}^{\infty}\frac{(-1)^r}{\pi r}\,R(r)\sin\left(2\pi r\,\frac{F}{B}\right). \tag{96}$$

Here, $F = cA/2\pi\hbar e$, $N$ is the density of charge carriers, and $R(r) = R_\mathrm{T}(r)R_\mathrm{S}(r)R_\mathrm{D}(r)$, where $R_\mathrm{T}(r)$ and $R_\mathrm{S}(r)$ describe the effects of temperature and spin splitting. Also, the scattering of electrons deteriorates magnetic quantum oscillations because it causes energy-level broadening. The simplest way to account for the effects of electron scattering on the oscillation amplitudes is to introduce an extra damping factor $R_\mathrm{D}(r)$ (Dingle factor). The usual approximation for that factor is $R_\mathrm{D}(r) = \exp(-2\pi r/\Omega\tau)$ [5]. In further calculations, we adopt this simple form for $R_\mathrm{D}(r)$, because more sophisticated expressions are irrelevant to the main point of our subject.

We can expect the effect of the FS curvature on the magnetization oscillations to appear when the FS warping is distinct ($w > \hbar\Omega$). To analyze these effects, we return to our generalized energy–momentum relation (89). The curvature of the corresponding FS near its cross sections with minimum/maximum areas is given by

$$K = -\frac{1}{2\pi p_\perp^2}\,\frac{\mathrm{d}^2 A}{\mathrm{d}p_z^2}. \tag{97}$$

For definiteness, we assume that the FS curvature vanishes at the effective cross section at $p_z = \pm p^*$. Then $\mathrm{d}^2A/\mathrm{d}p_z^2$ must vanish at $p_z = \pm p^*$. Accordingly, we set $\mathrm{d}\bar{a}/\mathrm{d}x|_{x=x^*} = 0$ in approximation (58) for the cross-sectional area and use this expression in what follows. The very essence of the model is that at $s > 2$, it describes an axially symmetric FS whose curvature vanishes at $p_z = \pm p^*$. Approximation (58) is necessarily applicable to each nearly cylindrical strip on any such FS. Otherwise, the strip has a nonzero curvature. We therefore see that the chosen model gives the general expression for the cross-sectional area of any nearly cylindrical segment of an axially symmetric FS. Using this approximation, we can derive the following expression for the contribution to the oscillating part of $M_\parallel$ from a nearly cylindrical cross section:

$$\Delta M_\parallel = -2N\xi_s\beta\left(\frac{\hbar\Omega}{w}\right)^{1/s}\sum_{r=1}^{\infty}\frac{(-1)^r R(r)}{(\pi r)^\rho}\sin\left[2\pi r\,\frac{F_{ex}}{B}\pm\frac{\pi}{2s}\right], \tag{98}$$

where $\rho = 1 + 1/s$. In the limit $s \to \infty$, the dimensionless factor $\xi$ tends to 0.5 and Eqn (98) passes into expression (96) describing magnetization oscillations in 2D conductors times 1/2. This extra factor appears because Eqn (98) describes the contribution from a single nearly cylindrical cross section of the FS. When the shape parameter for both effective cross sections tends to infinity, their contributions to the magnetization oscillations become identical and, putting them together, we arrive at expression (96).

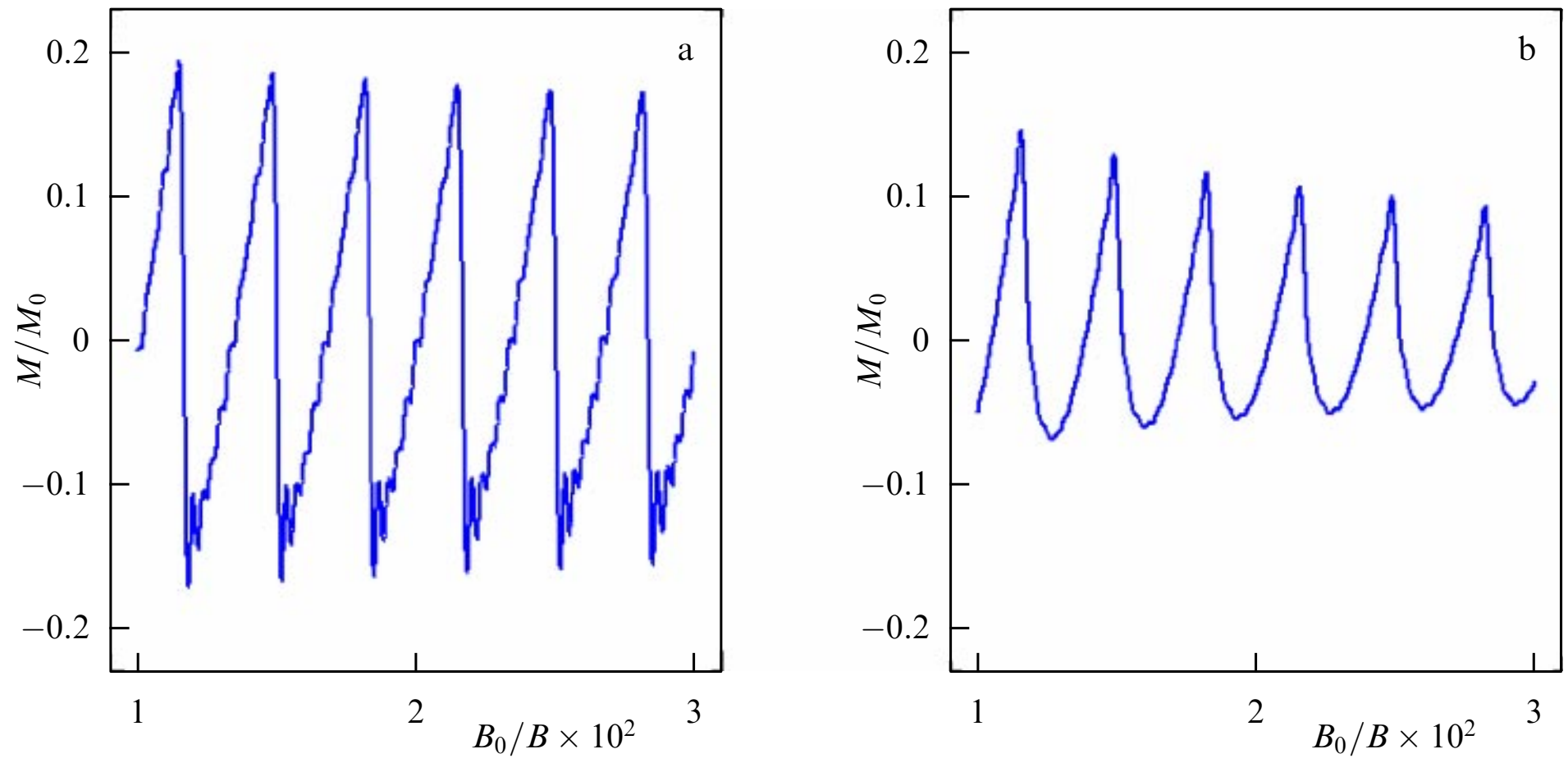

**Figure 15.** De Haas–van Alphen oscillations described by Eqn (98) for (a) $s = 8$ and (b) $s = 2$. Calculations are carried out for $T = T_D = 0.5$ K, $B_0 = 10$ T, $F/B_0 = 300$, and $M_0 = 2N\beta\Omega/\omega_0$.

Oscillations in magnetization described by expression (98) vary in magnitude, shape, and phase, depending on the value of the shape parameter $s$ that determines the local geometry of the FS near the effective cross section. This is illustrated in Fig. 15. As is shown in this figure, in the case of close proximity of the FS near $p_z = p^*$ to a cylinder, the oscillations are sawtoothed and resemble those occuring in 2D metals [127] or originating from cylindrical segments of FSs in conventional 3D metals [129]. When the FS curvature has a nonzero value at $p_z = p^*$ $(s = 2)$, the oscillations are similar to those in conventional metals. The present result (98) shows that the oscillation shape and phase can be determined not by the value of $w$ itself but by the form of the function $E_\parallel(p_z d/\hbar)$ specifying the FS profile [128]. The sawtoothed magnetization oscillations can occur at $w \sim \hbar\Omega$, when the FS curvature vanishes at an effective cross section. To simplify the interpretation of this point, we can imagine a FS shaped like a step-like cylinder [130]. The curvature of such a FS is everywhere zero, and oscillations from both kinds of cross sections (with minimum and maximum cross-sectional areas) should be similar to those in 2D metals. Nevertheless, the difference in the cross-sectional areas (the FS crimping) can be well pronounced, and a beat effect can be manifested. Obviously, this effect is absent when $w \ll \hbar\Omega$ and the FS warping is negligible.

Also, it may happen that the FS curvature vanishes at some effective cross sections and remains nonzero at the others. Then the contributions from zero-curvature cross sections $(s > 2)$ would exceed those originating from the ordinary cross sections $(s = 2)$. This follows from expression (98), where the factor $\xi_s(\hbar\Omega/w)^{1/s} \sim (B/F)^{1/s}$ $(B/F \ll 1)$ is included. Depending on the value of the shape parameter $s$, this factor takes values between $(B/F)^{1/2}$ $(s = 2)$ and 1 $(s \to \infty)$. Therefore, when there is a close proximity of the FS to a perfect cylinder at some extremal cross sections, the contributions from these cross sections are dominant, and they determine the shape and amplitude of the magnetization oscillations as a whole.

The effect of the FS curvature on the quantum oscillations in the magnetization is expected to be very sensitive to the geometry of the experiments. The reason is that the effective FS cross sections (with the minimum/maximum cross-sectional areas) run along zero-curvature lines (if any of these exist) only at certain directions of the magnetic field. When the magnetic field is tilted to that direction by an angle $\theta$, the extremal cross section slips from the nearly cylindrical strip on the FS containing a zero-curvature line. This results in a decrease in the oscillations amplitude. The phase of the oscillations also changes. These angular dependences of the oscillation amplitudes and phases radically differ in origin from the effect first described by Yamaji [131].

The Yamaji effect occurs due to the coincidence of the FS extremal areas $A_{max}$ and $A_{min}$ at certain inclination angles of the magnetic field with respect to the FS symmetry axis. At such angles, all the cross sections on the FS have the same area, and hence the amplitude of the de Haas–van Alphen oscillations increases. The Yamaji effect originates from the periodicity of the $p_z$-dependent contribution to the charge-carrier energy spectrum, and it is not related to the presence/absence of zero-curvature lines on the relevant FS. Also, there is a crucial difference in the manifestations of the two effects. The angular dependence originating from the effects of the FS curvature reveals itself at very small values of $\theta$, whereas the first maximum due to the Yamaji effect usually appears at $\theta \sim 10°$ or even greater. To further clarify the difference between the two effects, we analyze the angular dependence of de Haas–van Alphen oscillation amplitudes, assuming that the FS curvature vanishes at the extremal cross section $p_z = 0$ when the magnetic field is directed along the FS symmetry axis.

We suppose that the magnetic field is inclined to the FS symmetry axis by an angle $\theta$ within the $xz$ plane, and we use the coordinate system whose $z'$ axis is directed along the magnetic field. We use the energy–momentum relation given by Eqns (89) and (90) and rewrite them in terms of new coordinates $p'_z$, $p_y$, $p'_z$ ($p'_x = p_x \cos\theta + p_z \sin\theta$; $p'_z = p_z \cos\theta - p_x \sin\theta$). Then we can present the FS cross-sectional area in the form

$$\Delta M_{\parallel} = -2N\beta \frac{\Omega}{\omega_0} \sum_{r=1}^{\infty} \frac{(-1)^r}{\pi r} \sin\left[\frac{2\pi r F(\theta)}{B} + \Phi_r(\theta)\right] Y_r(\theta)\,, \tag{99}$$

where $\Phi_r(\theta)$ and $Y_r(\theta)$ describe the angular dependence of the magnetization. Both functions depend on the angle $\theta$ via their dependences on the cross-sectional area

$$A(p'_z, \cos\theta) = \frac{A}{\cos\theta} + \Delta A(p'_z, \cos\theta)\,, \tag{100}$$

where $A \equiv \pi p_0^2$ is the cross-sectional area of the unwarped FS ($w \to 0$) in the case where the cutting plane is perpendicular to the FS symmetry axis and $\Delta A(p'_z, \cos\theta)$ is given by

$$\Delta A(p'_z, \cos\theta) = 4\pi m_{\perp} w \sum_{n=1}^{\infty} \epsilon_n \cos\left(\frac{n p'_z d}{\hbar} \cos\theta\right) \times J_0\left(\frac{n p_0 d}{\hbar} \tan\theta\right). \tag{101}$$

The first term in this expansion coincides with the corresponding result in [131]. It was obtained assuming the simple cosine warping of the FS.

Requiring that $(\mathrm{d}^2 A/\mathrm{d}p_z^2)_{p_z=0} = 0$ and keeping only the first two terms in expansion (101), we obtain $\epsilon_1 = 1$ and $\epsilon_2 = -1/4$. To describe FSs possessing closer proximity to a perfect cylinder near a certain extremal cross section, we must keep more terms in expansion (101). For instance, setting $\epsilon_1 = 1$, $\epsilon_2 = -2/5$, $\epsilon_3 = 1/15$, and $\epsilon_n = 0$ ($n > 3$), we ensure that both $\mathrm{d}^2 A/\mathrm{d}p_z^2$ and $\mathrm{d}^4 A/\mathrm{d}p_z^4$ vanish at $p_z = 0$, which corresponds to $s = 6$. Similarly, at $\epsilon_1 = 1$, $\epsilon_2 = -1/2$, $\epsilon_3 = 1/7$, $\epsilon_4 = -1/56$, and $\epsilon_n = 0$ ($n > 4$), we obtain $s = 8$, and so on. Substituting these numbers in the general expression (100) for $\Delta A(p'_z, \cos\theta)$, we can finally calculate the functions $F(\theta)$, $Y_r(\theta)$, and $\Phi_r(\theta)$ that describe the desired angular dependences of the oscillation amplitudes.

Here, we carry out the calculations assuming that $w/\hbar\Omega = 0.5$ and $p_0 d/\hbar = 4\pi$ and keeping only the first term in the sum over $r$ in the expression for $\Delta M_{\parallel}$. The resulting curves are presented in Fig. 16. The solid line in this figure is associated with the energy spectrum of form (7). The corresponding FS has a cosine warping and a nonzero curvature at the extremal cross sections. The high peak at $\theta = 0.185$ ($10.6°$) corresponds to the first Yamaji maximum. The position of this peak is in agreement with the equation $(p_0 d/\hbar)\tan\theta = 3\pi/4$ (see [131]). Two preceding zeros originate from the beats. The remaining curves represent FSs whose curvature vanishes at their maximum cross sections at $\theta = 0$. We see that the closer the FS shape is to that of a perfect cylinder in the vicinities of these cross sections (the greater the value of $s$), the greater the oscillation amplitude near $\theta = 0$. At $s = 6$, the Yamaji maximum is approximately two times higher than the maximum at $\theta = 0$, whereas at $s = 2$, the ratio of the heights takes a value close to 4. We can expect that at very close proximity of the FS to a cylinder near the extremal cross section ($s \sim 10$), the amplitude maximum at $\theta = 0$ would exceed the Yamaji peak.

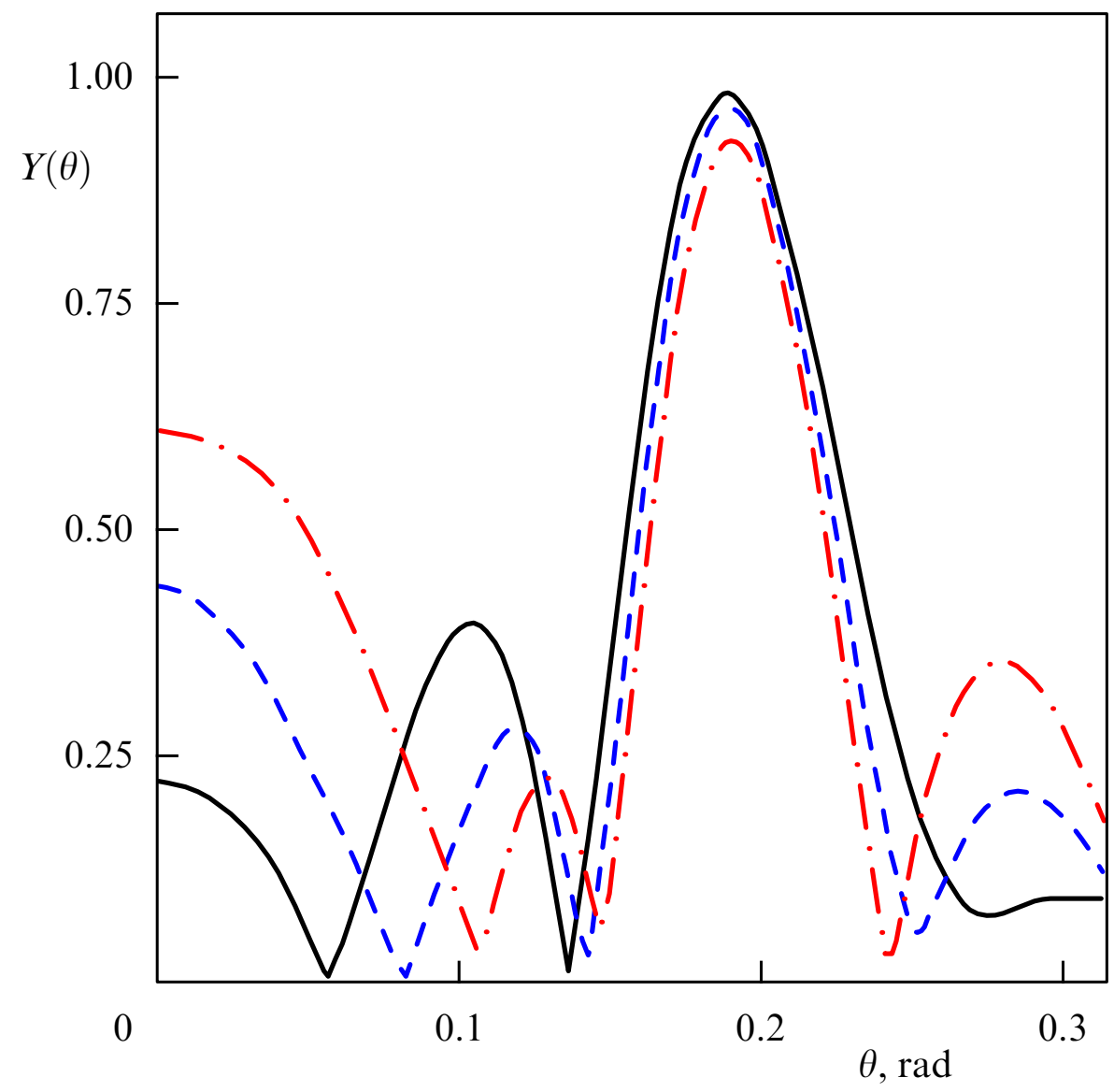

**Figure 16.** Angular dependences of the magnetization oscillations amplitudes. The curves are plotted assuming that $w/\hbar\omega = 0.5$ and $p_0 d/\hbar = 4\pi$. The shape parameter $s$ takes the values $s = 2$ (solid line), $s = 4$ (dashed line), and $s = 6$ (dashed-dotted line).

The angular dependence of the magnetization oscillation amplitude resembling that in Fig. 16 was reported to be observed in experiments on the Q2D organic metal $\alpha$-$(BETS)_2TlHg(SeCN)_4$ [132]. In these experiments, a high peak in the amplitude was observed when the magnetic field was directed along the axis of the corrugated cylinder, which is part of the FS. When the field was tilted to this axis by an angle $\theta$, the amplitude was rapidly reduced and reached approximately half the initial value at $\theta \sim 5°$. A further increase in the angle $\theta$ resulted in small variations in the amplitude until another peak was reached at $\theta \sim 18°$. Identifying this second peak with the first Yamaji maximum, we can conjecture that the higher peak at $\theta = 0$ arises due to the presence of the FS extremal cross sections of zero curvature. The relation between the heights of the peaks reported in [132] gives grounds to expect that the nearly cylindrical segments of the $\alpha$-$(BETS)_2TlHg(SeCN)_4$ Fermi surface (where the FS curvature vanishes) are very close to perfect cylinders. Recently, magnetic quantum oscillations were observed in several doped cuprates [10, 12–16], and the results confirm the existence of Q2D segments in the FSs of these materials. It can be expected that the present theoretical analysis will be useful in further explorations of the FS geometries in cuprates.

### 5.2 Local flattening of the Fermi surface and quantum oscillations in the magnetoacoustic response of a metal

The effect of the local flattening of the FS on the amplitudes of the magnetic quantum oscillations has not been thoroughly analyzed to date. But we can expect this effect to be revealed when magnetic quantum oscillations are modulated by commensurability oscillations (geometric resonances). Following Ref. [133], we consider the effect of FS flattening on the quantum oscillations of the velocity of ultrasound waves traveling in a metal perpendicularly to the external magnetic field. We adopt the same model for the FS as that corresponding to energy–momentum relation (33), such that the flattening points at the lens vertices are situated on the cross section with the maximum area. As before, the wave vector of the sound wave is given by (71), where the correction $\Delta q$ determines the velocity shift $\Delta s$ and the attenuation rate $\Gamma$. This correction is the sum of two terms. The first term, $\Delta q_1$, describes geometric oscillations in the ultrasonic attenuation rate and the velocity shift. Such oscillations are very well known in conventional metals (see Ref. [5]) as well as in two-dimensional electron systems [6, 8]. The effect is controlled by classical magnetotransport mechanisms.

The other term, $\Delta q_2$, originates from the quantization of the orbital motion of electrons in strong magnetic fields, and describes quantum oscillations in the velocity shift. Assuming that the cyclotron quantum $\hbar\Omega$ is small compared to the chemical potential of electrons $\zeta$ ($\gamma^{-1} = (\hbar\Omega/\zeta)^{1/2} \ll 1$), we obtain

$$\frac{\Delta q_2}{q} = -\frac{1}{2\rho_{\mathrm{m}} s^2}\,\frac{N^2}{g}\, n(-q) n(q) \varDelta\,, \tag{102}$$

where $g$ is the electron DOS at the FS in the absence of the magnetic field, $n(q) = n_0(0, q), n_0(0, q)$ is determined by Eqn (70) with $r = 0$, $p_z = 0$, and the function $\varDelta$ gives the contribution of the electron lens to the quantum oscillations of the electron DOS:

$$\varDelta = \frac{1}{\gamma}\sum_{r=1}^{\infty}\frac{(-1)^r}{\sqrt{r}}\, R(r)\cos\left(\frac{2\pi r F_{\mathrm{ex}}}{B} - \frac{\pi}{4}\right). \tag{103}$$

At very strong magnetic fields such that the characteristic diameter of the cyclotron orbit $2R_{\mathrm{ex}}$ is smaller than the wavelength of sound ($qR_{\mathrm{ex}} < 1$), we can go to the limit $q \to 0$ in expression (70), which yields $n_{\pm r}(p_z, \pm q)|_{q=0} = \delta_{r0}$. The semiclassical correction $\Delta q_1$ then becomes independent of the magnetic field, and magnetic oscillations are fully described by Eqn (102). These are ordinary quantum oscillations originating from the electron DOS oscillations.

In moderately strong but still quantizing magnetic fields ($qR_{\mathrm{ex}} > 1$), magnetic oscillations reveal a more complex structure. In this case, both $\Delta q_1$ and $\Delta q_2$ are to be included in consideration, and this leads to the result [133]

$$\frac{\Delta q}{q} = \frac{\Delta q_1}{q} + \frac{\Delta q_2}{q} = \frac{1}{4\rho_{\mathrm{m}} s^2}\,\frac{N^2}{g}\, Y_k(q, \omega)\,, \tag{104}$$

where we introduce the function

$$Y_k(q, \omega) = \frac{a^2(k)}{(qR_{\mathrm{ex}})^{1/k}}\left[V(\omega) + W(\omega)\cos\left(2qR_{\mathrm{ex}} + \frac{\pi}{2k}\right) + 2\cos^2\left(qR_{\mathrm{ex}} + \frac{\pi}{4k}\right)\varDelta\right] \equiv Y_{k1}(q, \omega) + Y_{k2}(q, \omega)\,. \tag{105}$$

Two oscillating terms are included in expression (105) for $Y_k(q, \omega)$. The first, $Y_{k1}$, originates from the semiclassical dynamical correction $\Delta q_1$ and represents commensurability oscillations. The second term, $Y_{k2}$, gives quantum oscillations superimposed on the geometric oscillations. The superposition of these two kinds of magnetic oscillations was studied for two-dimensional electron systems [134, 135]. Here, we showed that the same effect occurs in conventional 3D metals.

It follows from (105) that when the FS is flattened in the neighborhoods of the points corresponding to stationary points at the cyclotron orbit, the magnetic oscillations are noticeably amplified. In particular, when an ultrasound wave travels across the external magnetic field, the magnitude of quantum oscillations in the velocity shift $\Delta s/s$ is usually small compared to that of DOS oscillations due to the small factor $(qR_{\mathrm{ex}})^{-1}$, as shown in the top panels of Fig. 17. But when the FS includes locally flattened segments, the modulated quantum oscillations can reach the same order of magnitude as the DOS oscillations (Fig. 17c, d). Therefore, the increase in the number of effective electrons originating from the FS local flattening is too small to directly change the DOS oscillations. Nevertheless, it can have an effect on quantum oscillations in the observables by means of amplification of geometric oscillations modulating the latter. Unlike the direct effect of nearly cylindrical segments [128], the effect of points of flattening on the FS occurs due to the amplification of commensurability oscillations modulating DOS quantum oscillations. The effect could be observed in experiments for some particular directions of the magnetic field if the external disturbance propagates across the field. When revealed, this effect could be helpful in discovering the locations of flattened segments on the FSs.

### 5.3 Quantum oscillations of elastic constants and softening of the phonon modes in metals

As is known and as we have discussed, the effect of conduction electrons on the crystalline lattice arises due to a self-consistent electric field that appears under deformation. Lattice deformation also gives rise to an additional inhomogeneous magnetic field $\mathbf{b}(\mathbf{r})$. Usually, the effect of the field $\mathbf{b}(\mathbf{r})$ is small, and it can be omitted from consideration. However, in the presence of a quantizing external magnetic field $\mathbf{B}$ and at low temperatures ($\theta < 1$, where $\theta = 2\pi^2 kT/\hbar\Omega$), the field $\mathbf{b}(\mathbf{r})$ must be included in the consideration aiming at establishing the magnetic field dependences of the elastic constants [58–60, 136, 137]. The emergence of the electric and magnetic fields accompanying lattice deformation leads to a redistribution of the electron density $N$. The local change in the electron density $\delta N(\mathbf{r})$ is

$$\delta N(\mathbf{r}) = -\frac{\partial N}{\partial\zeta}\, e\varPhi(\mathbf{r}) + \frac{\partial N}{\partial B}\,\mathbf{b}(\mathbf{r}) \equiv -N_{\zeta}^{*}\left(e\varPhi(\mathbf{r}) + \frac{\partial\zeta}{\partial B}\,\mathbf{b}(\mathbf{r})\right). \tag{106}$$

The magnetic field $\mathbf{b}(\mathbf{r})$ satisfies the equation

$$\left[\nabla\times\mathbf{b}(\mathbf{r})\right] = 4\pi\left[\nabla\times\mathbf{M}(\mathbf{r})\right] = 0\,, \tag{107}$$

where $\mathbf{M}$ is the magnetization vector, $\zeta$ is the chemical potential of the charge carriers, and $\varPhi(\mathbf{r})$ is the potential of the electric field arising due to the deformation. The quantity $N_{\zeta}^{*}$ included in Eqn (106) is closely related to the electron density of states (DOS) on the FS $N_{\zeta}$. The difference between the two originates from correlations in the electron system. In the framework of the phenomenological FL theory, the renormalized DOS $N_{\zeta}^{*}$ has the form (see Ref. [40])

$$N_{\zeta}^{*} = -\sum_{\nu,\nu'}\frac{f_{\nu} - f_{\nu'}}{E_{\nu} - E_{\nu'}}\, n_{\nu\nu'}^{*}(-\mathbf{q})\, n_{\nu'\nu}(\mathbf{q})\bigg|_{q=0}\,, \tag{108}$$

where $n_{\nu'\nu}(\mathbf{q})$ is the Fourier transform of the electron density operator. The renormalized electron density operator $n_{\nu\nu'}^{*}(-\mathbf{q})$ is related to the 'bare' operator $n_{\nu\nu'}(-\mathbf{q})$ as

$$n_{\nu\nu'}^{*}(-\mathbf{q}) = n_{\nu\nu'}(-\mathbf{q}) + \sum_{\nu_1,\nu_2}\frac{f_{\nu_1} - f_{\nu_2}}{E_{\nu_1} - E_{\nu_2}}\, F_{\nu\nu'}^{\nu_1\nu_2}\, n_{\nu_1\nu_2}^{*}(-\mathbf{q})\,, \tag{109}$$

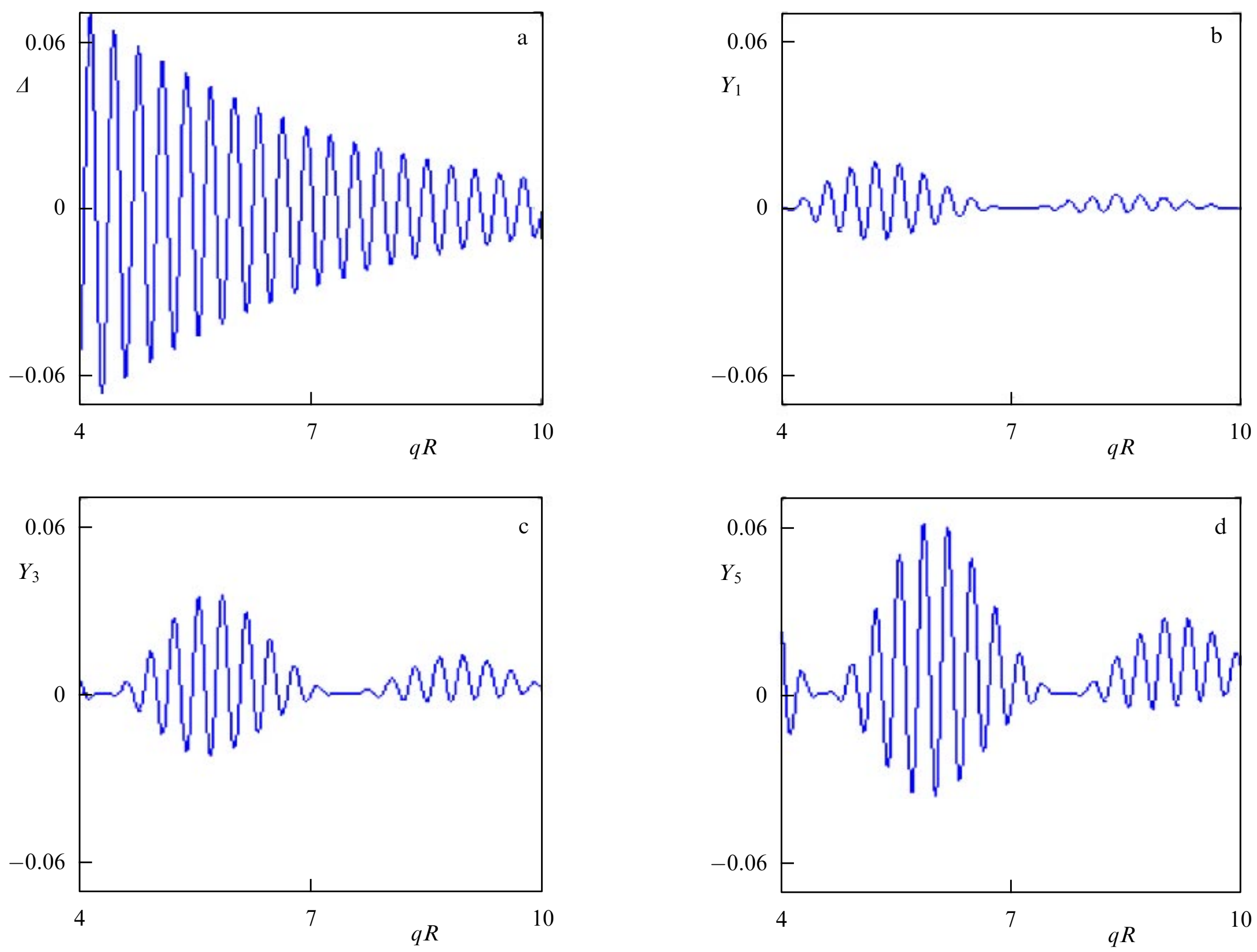

**Figure 17.** (a) Magnetic oscillations in the electron DOS and the response function $Y_{k2}(q, \omega)$ associated with the electron lens. The curves are plotted for (b) ellipsoidal and (c, d) flattened lenses. In plotting the curves, it was assumed that $\gamma = 10$, $kT/\zeta = 2 \times 10^{-3}$, and $\omega\tau = 0.1$.

where $F_{\nu\nu'}^{\nu_1\nu_2}$ are matrix elements of the FL kernel. Relations (106) and (107) must be supplemented by the condition of electrical neutrality of the system.

The set of these simultaneous equations was first presented in [58]. We use these equations to eliminate $\mathbf{b}(\mathbf{r})$ and to express the potential $\Phi(\mathbf{r})$ in terms of the lattice displacement vector. As a result, we arrive at the expression for the electron force $\mathbf{F}(\mathbf{r})$ acting on the lattice under its displacement by a vector $\mathbf{u}(\mathbf{r})$:

$$\mathbf{F}(\mathbf{r}) = \lambda_0 \mathbf{b}_0 \left[\mathbf{b}_0 \nabla\left(\nabla \mathbf{u}(\mathbf{r})\right)\right] + \lambda \Big[\mathbf{b}_0 \times \left[\nabla\left(\nabla \mathbf{u}(\mathbf{r})\right) \times \mathbf{b}_0\right]\Big] . \quad (110)$$

Here, $\mathbf{b}_0$ is the unit vector directed along $\mathbf{B}$. The result in (110) proves that the constants $\lambda_0$ and $\lambda_1$ represent electron contributions to the elastic constants corresponding to the deformation of the lattice along the external magnetic field ($\lambda_0$) and across this field ($\lambda$). In the chosen geometry, these constants equal the electron terms in the compression elastic moduli $c_{33}$ and $c_{11} = c_{22}$ (in the Voigt notation). Based on these equations, we can derive expressions for the elastic constants [58, 59]

$$\lambda_0 = \frac{N^2}{N_\zeta^*} , \quad (111)$$

$$\lambda = \lambda_0 \left(1 + \frac{4\pi\chi_\zeta}{1 - 4\pi\chi_\parallel}\right) , \quad (112)$$

where $\chi_\parallel = \partial M_z/\partial B + (\partial M_z/\partial\zeta)(\partial\zeta/\partial B)$ is the longitudinal part of the magnetic susceptibility and $\chi_\zeta = (\partial M_z/\partial\zeta)(\partial\zeta/\partial B)$. As follows from Eqn (111), $\lambda_0$ coincides with the compression modulus of the electron liquid. The structure of $\lambda$ is more complicated. In addition to the electron compression contribution, $\lambda$ also contains a contribution of a different origin. This extra term appears due to the inhomogeneous magnetic field $\mathbf{b}(\mathbf{r})$ produced by the lattice deformation. This field arises due to the change in the magnetization of electrons caused by the deformation. Hence, the appearance of the second term in (112) signifies a magnetostriction effect.

A strong magnetic field applied to a metal gives rise to quantum oscillations in the electron DOS, which in turn causes quantum oscillations in observables, including the magnetic susceptibility $\chi_\parallel$. At low temperatures, the magnitude of quantum oscillations increases so much that the oscillating term can dominate at the peaks of the oscillations. It was shown before (within the simple model of an isotropic electron liquid) that under such conditions, both $N_\zeta^*$ and $1 - 4\pi\chi_\parallel$ can go to zero near the oscillations peaks, producing magnetic instability of the metal and softening of some acoustic modes [58, 129, 136]. This is illustrated in Fig. 18a. Here, the magnetic fields $B_1'$ and $B_2'$ label thresholds of the magnetic instability region, and the differential magnetic susceptibility diverges at these points. Singularities in the longitudinal susceptibility $\chi_\parallel$ appear significantly closer to the field $B_0$, indicating the position of the oscillation peak. At the same time, the structural instability thresholds $B_1$ and $B_2$ are located respectively farther from $B_0$ than $B_1'$ and $B_2'$. However, in conventional 3D metals, these effects could be revealed in experiments only at extremely low temperatures of the order of 10 mK or lower. The stringent temperature requirements explain why the softening of the phonon

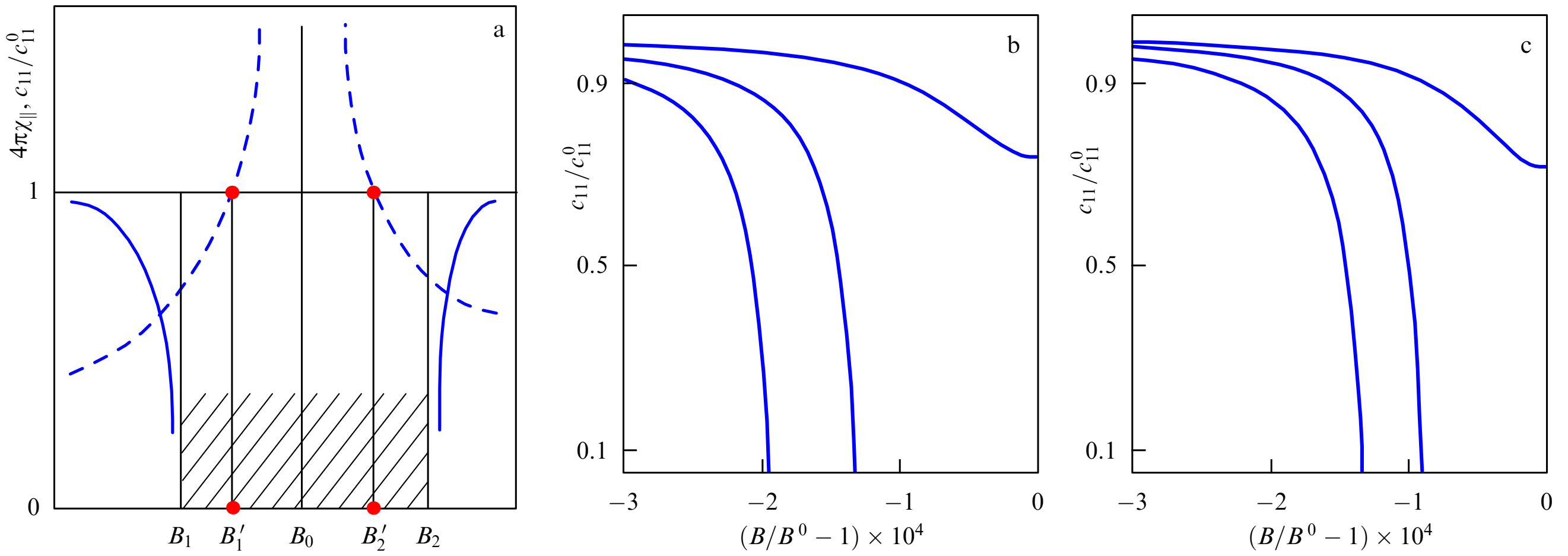

**Figure 18.** (a) Schematic plot of the magnetic field dependence of $4\pi\chi_{\parallel}$ (dashed lines) and $c_{11}/c_{11}^0$ (solid lines) near a peak of quantum oscillations at $B = B_0$ $(T = 0)$. The range of magnetic fields corresponding to the structural $(B_1 < B < B_2)$ and/or magnetic $(B_1' < B < B_2')$ instability is hatched. (b, c) Magnetic field dependences of the elastic constant $c_{11}$ near the diamagnetic phase transition. (b) The curves are plotted for $\theta = 1$, the shape parameter $s$ takes the values 16, 8, 4 from left to right; (c) $s = 8$, $\theta = 1$ takes the values 1, 2, 3 from left to right. For all curves, $N = 10^{27}$ m$^{-3}$, $\gamma^2 = 10^3$, and $B_0 = 10$ T.

modes at peaks of quantum oscillations has not yet been observed.

But these requirements can be noticeably moderated if the FS of a metal includes nearly cylindrical strips. This can influence anomalies of the elastic moduli and create much more favorable conditions for their observation in metals. To illustrate this statement, we consider a metal whose FS is axially symmetric in the vicinity of an extremal cross section at $p = p^*$ with the area $A_{\rm ex}$. We assume the magnetic field $\mathbf{B}$ to be directed along the symmetry axis and adopt the approximation given by Eqn (58) (assuming that $d\bar{a}/dx = 0$ at $x = x^*$) for the cross-sectional area around the extremal cross section.

Assuming that the cyclotron quantum $\hbar\Omega$ is small compared to $\zeta$ $(\gamma \gg 1)$ and using model (58), we arrive at the following expression for the contribution from the nearly cylindrical cross section to the electron DOS oscillations:

$$\varDelta = \frac{\eta_s}{\gamma^{2/s}} \sum_{r=1}^{\infty} \frac{(-1)^r}{r^{1/s}} R(r) \cos\left(2\pi r \frac{F_{\rm ex}}{B} - \frac{\pi}{2s}\right). \qquad (113)$$

Equation (113) agrees with the result obtained for a strictly cylindrical FS (see, e.g., Ref. [5]). We arrive at the corresponding result in the limit $s \to \infty$. Actual values of the shape parameter $s$ could be discovered in experiments where the FS local geometry is revealed. In the isotropic model, $s = 2$ and the oscillating function $\varDelta$ takes the well-known form (103).

Oscillations described by Eqns (113) and (103) differ in phase as well as amplitude. The amplitude of usual oscillations given by Eqn (103) is of the order of $\gamma^{-1}\theta^{-1/2}$, while Eqn (113) gives a magnitude of the order of $\gamma^{-2/s}\theta^{(1-s)/s}$. Therefore, the amplitude of oscillations related to the extremal zero-curvature section is approximately $(\gamma^{-1}\theta^{1/2})^{(2-s)/s}$ times greater than that of the usual quantum oscillations. As a result, the contribution from an extremal zero-curvature section can be considerably (more than tenfold) greater than contributions from other extremal sections, and the function $\varDelta$ can reach values around unity at oscillation peaks, even at $\theta \sim 1$. On these grounds, we conclude that the most favorable conditions for observation of softening of elastic moduli at peaks of quantum oscillations occur in metals whose FSs include nearly cylindrical segments. We consider such FSs in what follows.

Assuming $\gamma \ll 1$, we replace the matrix elements included in the FL kernel with their semiclassical analogs $\varphi(\mathbf{p}, \mathbf{p}')$ and $\psi(\mathbf{p}, \mathbf{p}')$, which depend on quasimomenta $\mathbf{p}$ and $\mathbf{p}'$ of interacting conduction electrons. For axially symmetric FSs, the FL functions can be approximated by Eqn (24). These approximations lead to the following expressions for the corrections to the elastic constants [60, 128]:

$$\tilde{c}_{11} = \tilde{c}_{22} = -\frac{N^2}{g} \frac{\varDelta}{1 + (1 + W - 4\pi\chi_0\gamma^4)\varDelta}, \qquad (114)$$

$$\tilde{c}_{33} = -\frac{N^2}{g} \frac{\varDelta}{1 + (1 + W)\varDelta}, \qquad (115)$$

where $\chi_0$ is related to the Landau diamagnetic susceptibility [which is equal to $-(1/3)\chi_0$] and the constant $W$ originates from the FL interactions. Hence, the elastic constants $c_{11}$ and $c_{22}$ appear to be affected by magnetostriction. At low temperatures $(\theta < 1)$, the denominator in Eqn (114) can vanish at the peaks of quantum oscillations. This indicates that the longitudinal magnetic susceptibility $\chi_{\parallel}$ diverges. The diamagnetic instability occurring at $1 - 4\pi\chi_{\parallel} = 0$ results in the structural instability of the metal.

As regards the elastic constant $c_{33}$, it can also be reduced at the peaks of low-temperature quantum oscillations, but this effect is not related to the magnetic instability. The expression for $c_{33}$ in (115) does not include the contribution due to magnetostriction. The possible softening of $c_{33}$ is a direct consequence of the behavior of the electron DOS under strong magnetic fields at low temperatures [138]. We have to remark here that the interactions between electrons significantly influence all the above effects. The value of the constant $W$ that accumulates the effects of electron–electron interactions in the framework of the FL theory can significantly influence the temperature range where both magnetic and lattice instabilities occur. With the function $\varDelta$ describing quantum oscillations given by Eqn (113), the amplitude of oscillations can become comparable to unity at moderately low temperatures if the FS shape reveals a fair proximity to a cylinder near the extremal cross section. For example, at $s = 6$, $\hbar\Omega/\zeta \sim 10^3$, and $B \sim 10$ T, the condition $\gamma^{-2/s}\theta^{(1-s)/s} \sim 1$ can be satisfied at temperatures of the order of 1 K.

To proceed with the analysis of the experimental feasibility of the effect, we numerically evaluate the decrease in the elastic constant $c_{11}$ using result (114). The results are shown in Fig. 18b, c. We see that the shape of the effective strip on the FS that is close enough to a cylinder gives rise to the structural instability near oscillation peaks at $\theta \sim 1$. Also, it is demonstrated that the effect is washed out as the temperature increases. Electronic contributions to the velocity of ultrasound waves propagating in metals are simply related to the elastic constants. It follows from the present results that the longitudinal sound velocity can depend on the propagation direction. Near the structural instability, the velocity of sound propagating perpendicularly to the magnetic field **B** can be noticeably reduced compared to the velocity of sound propagating along **B**. Again, we can expect this effect to appear in metals whose FSs include nearly cylindrical segments.

To summarize, it was demonstrated that structural instabilities can occur near the magnetic instabilities at the peaks of quantum oscillations. These magnetic instabilities have been analyzed in some studies (see, e.g., Refs [129, 137]). This effect can appear even in an isotropic metal. But it can be significantly strengthened when the immediate vicinities of some extremal cross sections of the FS are nearly cylindrical in shape, such that the FS curvature vanishes at these cross sections. The present analysis was carried out assuming the axial symmetry of the FS. The obtained results can be applied to practical metals, where the magnetic field is directed along a higher-order symmetry axis of the crystalline lattice. We can also expect phonon mode softening to occur in Q2D layered conductors, as was shown in Ref. [128].

## 6. Conclusion

The principal point that we tried to clarify here is that the fine characteristics of the FS shape can significantly affect the electron properties of conventional 3D metals and other materials such as Q2D layered conductors. The effect of the local geometric features of FSs (nearly cylindrical and/or nearly paraboloidal segments, flattened regions, and others) can be manifested under certain conditions, when the conduction electron response to an external disturbance is mostly determined by a small segment of the FS. This is the case with the anomalous skin effect and magnetic quantum oscillations of thermodynamic variables. In both cases, the response of a metal to an external disturbance is mainly formed by the charge carriers associated with narrow strips on the FS. When the FS curvature vanishes at these 'effective' strips, the frequency dependences of the surface impedance of a metal can noticeably change and the magnetic quantum oscillations can be enhanced. Local flattenings of the FS can significantly affect the commensurability magnetoacoustic oscillations described in Section 4. Observable effects arising due to the FS curvature anomalies at certain points can occur in truly low-dimensional conductors, such as the two-dimensional electron gas created in GaAs/AlGaAs superstructures, if the electron density is periodically modulated by applying an additional static electric field [116, 117].

The entire analysis presented in this review is based on the phenomenological models of the FS geometry over the relevant segments. The reason for the preference given to these models over the results of electron band structure computations is that even advanced computational methods cannot guarantee that such fine features as flattening points or narrow, nearly cylindrical strips are not missed in the process of FS reconstruction. In this paper, we concentrate on the analysis of possible experimental manifestations of such features. For instance, it was shown in Section 5 that the angular dependences of the magnitudes of magnetic quantum oscillations in Q2D metals predicted by Yamaji [131] radically differ from the angular dependences of similar characteristics that can appear due to the presence of nearly cylindrical strips on the relevant FSs. Phenomenological models remain extremely useful in these analyses because they allow relatively easily clarifying the nature and origin of various phenomena and identifying those that arise due to the FS local geometry.

We thank M V Sadovskii and I I Mazin for the helpful discussions. Also, the author is sincerely grateful to all the colleagues with whom she collaborated during the years given to the 'fermiological' studies, and to G M Zimbovsky for his help in preparing the manuscript.